\documentclass[a4paper,11pt]{article}
\usepackage{jcapppub} % for details on the use of the package, please see the JINST-author-manual
\usepackage{lineno}
\usepackage{amsmath}
\usepackage{amsfonts} 
\usepackage{graphicx}
\usepackage[normalem]{ulem}
\usepackage{url}
\PassOptionsToPackage{hyphens}{url}
\usepackage{hyperref}
\usepackage{subcaption}  
\usepackage{array}
\newcommand{\mycomment}[1]{}
\arxivnumber{ } % if you have one

\title{\boldmath Charged, rotating black holes 
in Einstein-Maxwell-dilaton theory}

\author{C. Herdeiro,}
\author{E. Radu} 
\author{and Etevaldo dos Santos Costa Filho} 
\affiliation{Departamento de Matem\'atica  da Universidade de Aveiro and
	Center for Research and Development in Mathematics and Applications (CIDMA)
	\\
    Campus de Santiago, 3810-183 Aveiro, Portugal}

\abstract{
The asymptotically flat, electrically charged, rotating black holes (BHs) 
in Einstein-Maxwell-dilaton (EMd) theory
are  known in closed form for \textit{only} two particular
values of the dilaton coupling constant $\gamma$: the Einstein-Maxwell coupling ($\gamma=0$), corresponding to the  %\textcolor{blue}
{Kerr-Newman} (KN)  solution, and the  Kaluza-Klein coupling ($\gamma=\sqrt{3}$). 
Rotating solutions with arbitrary $\gamma$ 
 are known only in the slow-rotation or weakly charged limits.
In this work,   
we numerically construct such EMd BHs with  arbitrary $\gamma$. 
 We present an overview of the parameter space of the solutions 
for illustrative values of
 $\gamma$
together with a study of their basic properties. 
The solutions are in general KN--like; 
there are however, new features.
The data
suggest that the spinning solutions with 
$0<\gamma<\sqrt{3}$
possess a zero temperature limit, which, albeit regular in terms of curvature invariants,
exhibits 
a $pp$-singularity.
A different limiting behaviour is found for
$\gamma>\sqrt{3}$, in which case, 
moreover, we have found hints of BH non-uniqueness for the same global charges.
}

\begin{document}
\maketitle
\flushbottom

\newpage
%%%%%%%%%%%%%%%%%%%%%%%%%%%%%%%%%%%%%%%%%%%%%%%%%%%%%%%%%%%%%%%%%%%
\section{Introduction}
%%%%%%%%%%%%%%%%%%%%%%%%%%%%%%%%%%%%%%%%%%%%%%%%%%%%%%%%%%%%%%%%%%%
 The 
Einstein-Maxwell-dilaton (EMd)  model
is one of the
simplest extensions of Einstein-Maxwell theory,
with an extra scalar field - the \textit{dilaton},
which features a specific (non-minimal) coupling to the Maxwell term.
This theory emerged in 
theoretical physics around a century ago,
occurring naturally in a Kaluza-Klein (KK) scenario,
when considering
the dimensional reduction of vacuum, five-dimensional gravity with a compact
extra dimension~\cite{kaluza,Klein:1926tv} (see also \cite{Ortin:2015hya,Duff:1986hr,Overduin:1997sri} and the bibliography therein).
Later, the same system also appeared in other contexts, most notably as
part of the low energy effective action of string theory~\cite{Gibbons:1987ps,Garfinkle:1990qj}.

In this work, we consider a 1-parameter family of EMd models, described by the action
(in units with $c=G=4\pi \epsilon_0=1$) 
\begin{equation}
\label{action}
S= \frac{1}{4 \pi}\int \left\{\dfrac{1}{4}R\, \boldsymbol{\epsilon}
-\dfrac{1}{2}\, d\Phi\wedge\star d \Phi
%- e^{2 \gamma \Phi} 
-\, \dfrac{f(\Phi)}{2}\, 
\mathcal{F}\wedge\star\mathcal{F}\right\} \ ,
\end{equation}
where $R$ is the Ricci scalar, $\boldsymbol{\epsilon}$ is the spacetime volume,
$\mathcal{F}=d\mathcal{A}$ is the Maxwell field strength 2-form,  $\mathcal{A}$ is the 1-form gauge potential and $\Phi$ is the scalar field/dilaton\footnote{In our conventions, we have $\mathcal{F}=\dfrac{1}{2}\mathcal{F_{\mu\nu}}\,dx^\mu\wedge dx^\nu, \boldsymbol{\epsilon}_{r\theta\varphi t}=\sqrt{-g}$. }. The non-minimal coupling function between the dilaton and the Maxwell field is taken as
\begin{eqnarray}
\label{dilaton}
    f(\Phi)=e^{-2 \gamma \Phi} \ ,
\end{eqnarray}
  $\gamma$ being a free parameter 
that governs the strength of the coupling of the dilaton 
field $\Phi$
to the Maxwell field.
It turns out that three  values of $\gamma$
are more relevant.
When $\gamma=0$, one can consistently set $\Phi=0$,
such that the action  (\ref{action}) reduces to the usual Einstein-Maxwell
theory.
When $\gamma=1$,
the action (\ref{action}) 
is part of the low energy action of string theory
\cite{Gibbons:1987ps,Garfinkle:1990qj}.
Finally, $\gamma=\sqrt{3}$ corresponds to 
compactified five-dimensional KK theory 
down to four dimensions. 

The black holes (BHs) of the EMd theory 
\eqref{action} with
\eqref{dilaton},
have been extensively studied over the last  three decades.
In what follows, we shall restrict to the case of asymptotically flat configurations
without a magnetic charge. Within this class of solutions, 
the static, spherically symmetric  (electrically charged)
dilatonic solutions were 
found by Gibbons and Maeda~\cite{Gibbons:1987ps} 
and Garfinkle, Horowitz and Strominger~\cite{Garfinkle:1990qj}
(from now on dubbed GMGHS BH). Their analysis reveals 
 certain qualitative features are independent of
$\gamma$.
For example, for a given mass, there is always 
a maximum electric charge $Q_e$
that can be carried by the BH. 
If $Q_e$ is less than this
extremal value, there is a regular event horizon. 
Other
qualitative features, however, depend crucially on $\gamma$;
for example,
  in the extremal limit, the surface gravity vanishes when
$0\leqslant\gamma<1$;
it reaches a finite limiting value when 
$\gamma=1$, and
it diverges when 
$\gamma>1$.

The case of rotating BHs is less studied.
Exact solutions exist only for 
$\gamma=0$ (the Kerr-Newman (KN) solution \cite{Newman:1965my})
and the KK case
$\gamma= \sqrt{3}$
\cite{Frolov:1987rj}.
The 
rotating solutions with generic $\gamma$
were studied in the slowly rotating 
\cite{Horne:1992zy,Shiraishi:1992np} and in the weakly charged approximation  
 \cite{Casadio:1996sj}.
To the best of our knowledge, no
rotating
non-perturbative solutions 
with $\gamma \neq (0,\sqrt{3})$ 
were reported in the literature\footnote{
Some properties of the (non-perturbative)  rotating dyonic BHs 
were studied  in \cite{Kleihaus:2003df}
within a numerical approach. When higher dimensions are considered, the model \eqref{action} and some of its generalizations have also been studied \cite{Deshpande:2024vbn,Deshpande:2024itz,Kunz:2006jd,Kleihaus:2016auo,Blazquez-Salcedo:2013wka}.
}.
 Notably, even in the absence of solutions,
 a uniqueness theorem has been established 
for values of 
$0\leq\gamma^2\leq3$, 
only \cite{Yazadjiev:2010bj}.
\medskip

The main goal of this paper is to propose
 a non-perturbative numerical approach for the construction of
 spinning electrically charged
 EMd BH solutions with arbitrary $\gamma$. 
 Also, a study of the basic physical properties  
of the solutions
is presented for several values of $\gamma$.
This, together with the knowledge based on the exact solutions
with $\gamma=(0,\sqrt{3})$,
would hopefully capture the pattern of the general EMd BHs.

Some highlights of our results are as follows.
 First,
all static EMd solutions 
  possess spinning generalizations, the domain of
existence being displayed for several
values of $\gamma$.
As with the static limit, some of their properties 
depend on the value of dilaton coupling constant $\gamma$.
For $0\leq \gamma \leq \sqrt{3}$,
rotation allows for  
an extremal limit 
with  nonzero area
and vanishing surface gravity;
this includes models with $\gamma\geq 1$, 
in which case the  static extremal limit has non-vanishing or divergent surface gravity (some understanding of this behaviour is provided by the exact solution with $\gamma=\sqrt{3}$ in Section
\ref{rotating-exact}). 

The situation is different for  
$ \gamma >\sqrt{3}$, 
in which case  
 the horizon area
of the maximally rotating
solutions
is still nonzero,
while 
the surface gravity
never vanishes  (and in fact  likely diverges,
while  although this limit cannot be approached numerically).

Also, there we have found some indication 
for the non-uniqueness of solutions,
with (at least) two distinct configurations
possessing the same global charges. 

This paper is organized as follows. In Section \ref{sec2}, we discuss the general framework
and
some relevant properties of the model. 
In Section \ref{sec3}, we provide a short review of the known (asymptotically flat) exact solutions of the EMd model.
In Section \ref{sec4}, we introduce the framework for the numerical
construction of electrically charged
spinning BHs with arbitrary $\gamma$, and discuss the Ansatz, boundary conditions, the physical quantities of
interest and the numerical procedure. 
The  numerical results are reported in 
 Section \ref{sec5},
where
we describe the spinning BH solutions, their domain
of existence, and the behaviour of different physical quantities. 
In Section \ref{Conclusions}, we present conclusions
and remarks.  
Appendices A-D contain some technical details on the solutions.

%%%%%%%%%%%%%%%%%%%%%%%%%%%%%%%%%%%%%%%%%%%%%%%%%%%%%%%%%%
\section{The model}
	\label{sec2}
%%%%%%%%%%%%%%%%%%%%%%%%%%%%%%%%%%%%%%%%%%%%%%%%%%%%%%%%%%

%%%%%%%%%%%%%%%%%%%%%%%%%%%%%%%%%%%%%%%%%%%%%%%%%%%%%%%%%%
\subsection{The equations and symmetries}
	\label{sec21}
%%%%%%%%%%%%%%%%%%%%%%%%%%%%%%%%%%%%%%%%%%%%%%%%%%%%%%%%%%

The field equations obtained by varying the action principle
	(\ref{action})
with respect
to the field variables $g_{\mu \nu}$, $\Phi$ and $\mathcal{A}$ are\footnote{Here, we define $f'(\Phi)=\dfrac{\mathrm{d}\, f}{\mathrm{d}\, \Phi}$.}
\begin{eqnarray}
\label{EME}
E_{\mu\nu}=R_{\mu\nu}-\dfrac{g_{\mu\nu}}{2}R-2T_{\mu\nu}&=&0\,,
\\
\label{ME}
d\left(f(\Phi)\star\mathcal{F}\right)&=&0
\,,
\\
\label{DE}
d\star d\Phi-\dfrac{f'(\Phi)}{2}\mathcal{F}\wedge\star\mathcal{F}&=&0 \,.
\end{eqnarray}
with the energy-momentum tensor
\begin{equation}
\label{tik}
T_{\mu\nu}=f(\Phi)
\left(
\mathcal{F}_\mu\,^\sigma\mathcal{F}_{\nu\sigma}-\frac{1}{4}g_{\mu\nu}\mathcal{F}_\sigma\,^\tau \mathcal{F}^\sigma\,_\tau
\right)
+ \nabla_\mu\Phi\nabla_\nu\Phi-\dfrac{1}{2}g_{\mu\nu}\nabla_\tau\Phi\nabla^{\tau}\Phi\,.
\end{equation}
	 
Let us remark that,
for a dilaton coupling (\ref{dilaton}),
the system possesses several symmetries \cite{Ashtekar:1999sn}:
\begin{itemize}
\item
First, there is the discrete duality rotation
{$(\mathcal{F}, \Phi) \rightarrow (e^{2 \gamma \Phi} \star\mathcal{F}, -\Phi)$},which is less relevant for the solutions with no net magnetic
charge in this work. 

\item
Second, the solutions are invariant under the simultaneous sign change 
$(\gamma, \Phi) \rightarrow -(\gamma, \Phi)$. 
As such, 
throughout this work, 
we shall consider $\gamma\ge0$ without any loss of generality.

\item
Finally, 
both the action and the equations of motion remain invariant under a constant dilaton shift,
 $\Phi \rightarrow \Phi + \Phi_0$,  accompanied by a simultaneous rescaling of the  $U(1)$ field, 
$\mathcal{A}_{\mu} \rightarrow e^{\gamma \Phi_0} \mathcal{A}_{\mu}$,
with $\Phi_0$ an arbitrary constant. 
%This implies that the presence of $\Phi_\infty$ is not necessary; hence, for the sake of %simplicity, we will proceed by setting $\Phi_\infty=0$.
This symmetry is fixed when imposing the
dilaton field to vanish asymptotically.
%$\Phi \to 0$ as $r \to  \infty$.
%\textcolor{blue}{
It also implies the existence of a conserved current, $d J =0$, given by 
\cite{Rakhmanov:1993yd}
\begin{equation}\label{noether}
    J= \star d \Phi+\gamma e^{-2\gamma\Phi}\mathcal{A}\land\star\mathcal{F}\,\,
\end{equation}
which remains conserved under gauge transformations. 
%}
 \end{itemize}

%%%%%%%%%%%%%%%%%%%%%%%%%%%%%%%%%%%%%%%%%%%%%%%%%%%%%%%%%%
\subsection{Circularity in EMd system and the Ansatz}
	\label{sec22}
%%%%%%%%%%%%%%%%%%%%%%%%%%%%%%%%%%%%%%%%%%%%%%%%%%%%%%%%%%

We focus on asymptotically flat, axisymmetric, and stationary solutions of the considered model.
These symmetries imply the existence of two commuting Killing vector fields \cite{Carter:1970ea}, which can be expressed in adapted coordinates as $\xi=\partial_t$ and  $\eta=\partial_\varphi$ where $t$, $\varphi$ are respectively the asymptotic time and the azimuthal angle.

Since the metric, $g_{\mu\nu}$, is invariant under infinitesimal diffeomorphisms generated by any linear combination $\kappa$ of $\xi$ and $\eta$, and given that both the field strength $\mathcal{F}$ and the dilaton field $\Phi$ are real, their Lie derivative along $\kappa$ must satisfy
\begin{equation}
L_{\kappa} g_{\mu \nu}=0\,,\qquad  L_\kappa \mathcal{F}=0\,,\qquad L_\kappa\Phi=0\,.
\end{equation}

The EMd theory also possesses the usual $U(1)$ gauge symmetry of electrodynamics  $\mathcal{A}\rightarrow \mathcal{A}+d\lambda$, where $\lambda$ is an arbitrary real function. This gauge freedom implies that the vector potential $\mathcal{A}$ does not necessarily remain invariant under the actions of $\kappa$. In order to see what the condition  $ L_\kappa \mathcal{F}=0$  becomes in terms of $A$,  we notice that  $ L_\kappa d \mathcal{A}=d L_\kappa A$,  which leads to the gauge-invariant condition $L_\kappa \mathcal{A}= d\lambda\,   $ \cite{heusler_1996,Forgacs:1979zs}.

 We now show that the circularity condition is an imposition of the EMd field equations for asymptotically flat, axisymmetric, stationary spacetimes, as in vacuum or electro-vacuum, rather than an ansatz choice\footnote{For a proof in a larger family of electromagnetic-scalar models, see \cite{Bokulic:2023oxw}.} \cite{Chinea:2002jz}. A circular spacetime is Ricci-circular \cite{heusler_1996,Kundt:1966zz,Carter:1969zz,Carter:2009nex}; therefore, in order to demonstrate circularity, we will use the equations of motion. A stationary and axisymmetric spacetime is said to be Ricci-circular if \cite{heusler_1996}
\begin{equation}
    \eta\land\xi \land R(\xi)=\xi\land\eta \land R(\eta)=0\,, \qquad\qquad(R(\kappa)_\mu=R_{\mu\nu}\kappa^\nu)\,.
\end{equation}

Consequently, using Einstein's equations, the circularity condition is met if the energy-momentum tensor fulfills the conditions
\begin{equation}
    \eta\land\xi \land T(\xi)=\xi\land\eta \land T(\eta)=0\,.
\end{equation}

The field strength, $\mathcal{F}= d \mathcal{A}$, satisfies
\begin{equation}\label{eomform}
    d \mathcal{F}=0\,,\qquad\qquad d(f(\Phi)\star \mathcal{F})=0\,.
\end{equation}

The first equality is a consequence of $\mathcal{F}$ being an exact form, and the second is the corresponding equation of motion. In order to make a parallelism with pure electromagnetism, we define the electric and magnetic 1-form fields \cite{Yazadjiev:2010bj,heusler_1996}
\begin{equation}
    E_\kappa=- \iota_\kappa \mathcal{F}=\star(\kappa\land\star\mathcal{F})\,,\qquad\qquad  B_\kappa= \iota_\kappa \left(f(\Phi)\star \mathcal{F}\right)=\star(\kappa\land f(\Phi) \mathcal{F})\,.
\end{equation}

It is straightforward to see that $E_\kappa$ and $B_\kappa$ are closed if the equations of motion hold. As a consequence, $ E_\kappa\,_{\mu}\kappa^\mu=B_\kappa\,_{\mu}\kappa^\mu=0$ and $L_\kappa E_\kappa=L_\kappa B_\kappa=0$.  Moreover, since $E_\kappa$ and $B_\kappa$ are closed, locally they can be written as: $E_\kappa=d\phi$ and $B_\kappa=d\psi$ \cite{Yazadjiev:2010bj}. Notice that $\phi$ and $\psi$ are directly defined from $\mathcal{F}$, and hence, gauge invariant \cite{Elgood:2020svt,Ortin:2022uxa}.

Additionally, regularity along the rotation axis ensures that everywhere in the domain \cite{Carter:1992dbj,Heusler:1996ft}
\begin{equation}
    E_\xi\,_{\nu}\eta^\nu= -\mathcal{F}_{\mu\nu}\xi^{\mu}\eta^{\nu}=0\,,\qquad\qquad   B_\xi\,_{\nu}\eta^\nu=(f(\Phi)\star\mathcal{F})_{\mu\nu}\xi^{\mu}\eta^{\nu}=0\,.
\end{equation}

We can now conclude that the EMd theory obeys the circularity condition. Using the above definitions, we have
\begin{equation}
    \xi\land T(\xi)=\star(E_\xi \land B_\xi)\,,
\end{equation}
\begin{equation}
    \eta\land T(\eta)=\star(E_\eta \land B_\eta)\,~,
\end{equation}
and
finally
\begin{equation}
   \eta\land \xi\land T(\xi)=\eta\land \star(E_\xi \land B_\xi)=\star\iota_\eta(E_\xi \land B_\xi)=0\,,
\end{equation}
\begin{equation}
   \xi\land \eta\land T(\eta)=\xi\land \star(E_\eta \land B_\eta)=\star\iota_\xi(E_\eta \land B_\eta)=0\,.
\end{equation}

Thus, a spacetime, which is asymptotically flat, axisymmetric, and stationary, sourced by real Maxwell-dilaton fields, is Ricci circular. Physically, this implies the existence of a reflection symmetry or a motion reversal $(t,\varphi)\rightarrow(-t,-\varphi)$.

Consequently, the metric can be expressed in a block-diagonal form \cite{heusler_1996,wald1984general}:
\begin{equation}
ds^2 = g_{\mu\nu}dx^\mu dx^\nu = \sigma_{ab}dx^a dx^b + \gamma_{ij} dx^i dx^j\,,
\end{equation}
where the metric $\sigma_{ab}$  is associated with the $(t, \varphi)$ manifold, while $\gamma_{ij}$ is associated with the $(x^1, x^2)$ manifold. 
Note that the two-dimensional metric 
$ds_2^2=\gamma_{ij}dx^i dx^j$ 
can always be brought to the diagonal form conformal to the $(x^1 - x^2)$-plane \cite{chandrasekhar1998mathematical}
\begin{equation}
ds_{2}^{2}=e^{2\mu}\left[\left(dx^{1}\right)^{2}+\Lambda(x^1,x^2)\left(dx^{2}\right)^{2}\right]\,.
\end{equation}

On the other hand, the metric $\sigma_{ab}$ can be written as \cite{heusler_1996}
\begin{equation}
    ds_1^2=- V\, dt^2+2 W\,dt\,d\varphi+ X\,d\varphi^2\, ,
\end{equation}
where $V=-\xi\cdot\xi$, $W=\xi\cdot\eta$, $X=\eta\cdot\eta$. A further simplification is achieved by introducing the new function  $\rho=\rho(x^1,x^2)$,  defined as
\begin{equation}
\rho \equiv \sqrt{-\sigma} = \sqrt{VX+W^2} \ ,
\end{equation}
where $\sigma$ denotes the determinant of the $\sigma_{ab}$ metric
\begin{equation}
    ds_1^2=- \dfrac{\rho^2}{X}\, dt^2+X (d\varphi+ A\,dt)^2\,,\qquad \text{with } A=\dfrac{W}{X}\,.
\end{equation}
The equation of motion for the function $\rho$ is \cite{Herdeiro:2019oqp,wald1984general,heusler_1996}
\begin{equation}
\frac{1}{\rho}\nabla^2_{(\gamma)}\rho = -\frac{1}{X}\text{tr}_\sigma \mathbf{R}.
\end{equation}
where
\begin{equation}
\text{tr}_\sigma \mathbf{R} = \sigma^{ab}R_{ab} =\frac{1}{\rho^2}[2 W R(\eta, \xi)-X R(\xi, \xi)+V R(\eta, \eta)] \ .
\label{harmonicc}
\end{equation}

When $\rho$ is a harmonic function, that is, $\text{tr}_\sigma \mathbf{R}=0$, it can be used as a coordinate alongside its harmonic conjugate $z$, and the metric function $\Lambda$ can be chosen to be 1 \cite{wald1984general,Carter:2009nex,heusler_1996,Herdeiro:2019oqp}. By using Einstein's equations, $R_{\mu\nu}=2\left(T_{\mu\nu}-\dfrac{g_{\mu\nu}}{2}T\right)$, we see that this is the case for the system we are dealing with. This leads to the most general metric for asymptotically flat, axisymmetric, stationary solutions in EMd theory
\begin{equation}
\label{m1}
d s^2=-\frac{\rho^2}{X(\rho, z)} d t^2+X(\rho, z)[d \varphi-w(\rho, z) d t]^2+\frac{e^{2 h(\rho, z)}}{X(\rho, z)}\left[d \rho^2+d z^2\right]\,.
\end{equation}

 %%%%%%%%%%%%%%%%%%%%%%%%%%%%%%%%%%%%%%%%%%%%%%%%%%%%%%%%%%%%%%%%%%%%%%%%%%%%%%%%%%%%%%%%%
\subsection{Effective field equations}
%%%%%%%%%%%%%%%%%%%%%%%%%%%%%%%%
%
One can introduce the twist vector $\boldsymbol{\omega}$, associated with a Killing vector $\kappa$, defined by $\boldsymbol{\omega}=\star(\kappa\land d\kappa)$.  The twist associated with $\kappa$ satisfies
\begin{equation}
d\boldsymbol{\omega}=-2\iota_\kappa\star R(\kappa) \ .
\end{equation}

In vacuum spacetimes, where $R_{\mu\nu} = 0$, the twist vector becomes exact, implying the existence of a scalar twist potential $\omega$ such that $\omega_\mu = \nabla_\mu \omega$. However, in non-vacuum spacetimes, this is generally not the case. Following \cite{Simon,10.21468/SciPostPhys.15.4.154}, the idea is to construct an improved twist, $\omega^I_\mu$, such that the total twist, $\omega^{\text{tot}}_\mu=\omega_\mu+\omega^I_\mu$, is curl free, $\nabla_{[\mu} \omega^{\text{tot}}_{v]}=0$. Following \cite{Yazadjiev:2010bj}, we have
\begin{equation}
    d\boldsymbol{\omega}=-2\iota_\kappa\star R(\kappa)=4 \iota_\kappa\mathcal{F}\land\iota_\kappa\left(f(\Phi)\star\mathcal{F}\right)\,.
\end{equation}

Notice that $\kappa^\mu\nabla_\mu\Phi=0$ implies that the contribution of the scalar field into the twist is through its interaction with the electromagnetic field. Hence, the improved twist takes the form
\begin{equation}
    \omega^I_\mu=2\phi \nabla_\mu\psi-2\psi\nabla_\mu\phi\,.
\end{equation}

As a result, the total twist becomes exact, $\boldsymbol{\omega}^{\text{tot}}=d\chi$, where $\chi$ is a scalar. 
Following the approach of \cite{Galtsov:1995zm,Yazadjiev:2010bj,Wells:1998gc},  we perform a dimensional reduction by choosing
$\kappa=\eta$, taken to be the spacelike axial Killing vector\footnote{In this context, dimensional reduction is performed with regard to the axial Killing vector to guarantee a positively definite metric in the target space \cite{Yazadjiev:2010bj}. Conversely, in \cite{Galtsov:1995zm}, the authors conduct this reduction with respect to the timelike Killing vector.}. In contrast to earlier treatments, however, here we keep a generic scalar coupling $f(\Phi)$.
%to the electromagnetic field - specifically, $f=e^{2\gamma\Phi}$ in the dilaton model.
Under this reduction, the equations of motion simplify to the system below
(with $\bar{\nabla}U \cdot \bar{\nabla}V=\partial_\rho U\partial_\rho V+\partial_z U\partial_z V$
and $\bar{\nabla}^2 U=\partial_{\rho\rho} U+\partial_{zz} U$):
\begin{subequations}
\begin{align}
    \label{eq-target}
& \rho^{-1} \bar{\nabla}\cdot\left(\rho \bar{\nabla} X\right)=X^{-1}   \left(\bar{\nabla}X\right)^2 -X^{-1}\left(\bar{\nabla} \chi+2 \phi \bar{\nabla} \psi-2 \psi \bar{\nabla} \phi\right)^2
\\\nonumber
&  -2 f \left(\bar{\nabla}\phi\right)^2-2 f^{-1} \left(\bar{\nabla} \psi\right)^2, \\[10pt]
& \bar{\nabla}\cdot\left[\rho X^{-2}\left(\bar{\nabla} \chi+2 \phi \bar{\nabla} \psi-2 \psi \bar{\nabla} \phi\right)\right]=0, \\[10pt]
& \rho^{-1} \bar{\nabla}\cdot\left(X^{-1} \rho f\bar{\nabla} \phi\right)=X^{-2} \bar{\nabla}\psi\cdot\left(\bar{\nabla} \chi+2 \phi \bar{\nabla} \psi-2 \psi \bar{\nabla} \phi\right), \\[10pt]
& \rho^{-1} \bar{\nabla}\cdot\left(X^{-1} \rho f^{-1} \bar{\nabla} \psi\right)=-X^{-2} \bar{\nabla}\phi\cdot\left(\bar{\nabla} \chi+2 \phi \bar{\nabla} \psi-2 \psi \bar{\nabla} \phi\right), \\[9pt]
& \rho^{-1} \bar{\nabla}\cdot\left(\rho \bar{\nabla} \Phi\right)=-X^{-1}\left[\dfrac{f'}{2} \left(\bar{\nabla}\phi \right)^2-\dfrac{f'}{2f^2}\left(\bar{\nabla} \psi\right)^2\right]. 
    \label{eq-targetf}
\end{align}
\end{subequations}

 As one can see, the coupling function  
enters nontrivially all equations (except the second one), which makes unlikely that the system with a generic
$f(\Phi)$ is integrable.
Also, let us remark that the case of a dilatonic
coupling
$f=e^{-2\gamma\Phi}$ 
does not lead to a significant simplification of the above equations.

%%%%%%%%%%%%%%%%%%%%%%%%%%%%%%%%%%%%%%%%%%%%%%%%%%%%%%%%%%%%%%%%%%%%%%%%%%%%%%%%%%%%%%%%%%%

\subsection{A mass formula and scalar charge}
 \label{sec6}

We aim at considering the Smarr formula for our model, which can serve different purposes
(in particular as a test of numerical accuracy). But before, let us comment on the generalized zeroth law. In thermodynamics, the zeroth law states that the constancy of the temperature is a condition for thermal equilibrium. Similarly, the zeroth law of BH thermodynamics asserts that the temperature and the ``chemical'' potentials must be constant at the horizon. Specifically for EMd BHs, this includes the constancy of
Hawking temperature $T_H$, event horizon angular velocity 
$\Omega_H$ and the electric and magnetic potentials $\phi$ and $\psi$  over the horizons. For completeness, here we will briefly comment on the constancy of the electric and magnetic potentials, only, since for the constancy of the
 $T_H$, $\Omega_H$
(and rigorous proofs), there exists a vast literature on the topic \cite{Ortin:2022uxa,Hajian:2022lgy,Bardeen:1973gs,heusler_1996,Gauntlett:1998fz,Copsey:2005se,Prabhu:2015vua}.

The approach for the EMd BHs is identical to the EM BHs and can be found in standards references \cite{heusler_1996,Frolov:1998wf,Carter:2009nex}. 
The argument goes as follows. Considering that the horizon of a BH is a Killing horizon and that the model obeys the null energy condition \cite{Maeda:2018hqu}, we have the following condition on the horizon
\begin{align}\label{Raychaudhuri}
	R_{\mu \nu}\chi^{\mu}\chi^{\nu} &= 0 \quad \text{where} \quad \chi^\mu=\xi^\mu+\Omega_H \eta^\mu\,.
\end{align}

By means of Einstein's equation, this will imply that both $E$ and $B$ are null on the horizon. Consequently, their contraction with any vector tangent to the horizon is zero, implying that $\phi$ and $\psi$ are constant over the horizon. In fact, consider \cite{Ashtekar:2000hw,Pacilio:2018kdk} the energy-momentum tensor 
can also be written as 
\begin{align}
\nonumber
%T_{\mu\nu}&=f(\Phi)F_\mu\,^\sigma F_{\nu\sigma}-\frac{f(\Phi)}{4}g_{\mu\nu}F_\sigma\,^\tau F^\sigma\,_\tau+ \nabla_\mu\Phi\nabla_\nu\Phi-\dfrac{1}{2}g_{\mu\nu}\nabla_\tau\Phi\nabla^{\tau}\Phi\, \\ 
T_{\mu\nu}&= f(\Phi)(\star F)_\mu\,^\sigma (\star F)_{\nu\sigma}-\frac{f(\Phi)}{4}g_{\mu\nu}(\star F)_\sigma\,^\tau (\star F)^\sigma\,_\tau+ \nabla_\mu\Phi\nabla_\nu\Phi-\dfrac{1}{2}g_{\mu\nu}\nabla_\tau\Phi\nabla^{\tau}\Phi\,.
\end{align}
where $(\star F)_{\mu\nu}=\epsilon_{\mu\nu\alpha\beta}F^{\alpha\beta}/2$. Hence, from Eq. \eqref{Raychaudhuri}, we have on the horizon
\begin{equation}
    E_{\chi\,\mu}E_{\chi}\,^{\mu}=0\,, \qquad\qquad B_{\chi\,\mu}B_{\chi}\,^{\mu}=0\,,
\end{equation}
meaning that the vector fields $E$ and $B$ are null on the horizon and, consequently, proportional to $\chi$ on the horizon. This leads to the following equality holding for any vector $s^\mu$ tangent to the event horizon
\begin{equation}
L_{s}\phi=\iota_s d\phi=\iota_s E=0\,,\qquad\qquad L_{s}\psi=\iota_s d\psi=\iota_s B=0\,.
\end{equation}
We conclude that the electric and the magnetic potentials assume constant values on the horizon \cite{heusler_1996,Ortin:2022uxa}.

\medskip

%%%%%%%%%%%%%%%%%%%%%%%%%%%%%%%%%%%%%%%%%%%%%%%%%%%%%%%%%%%%%%%%%

In an asymptotically flat, axially symmetric stationary spacetime, the Komar integral allows for the representation of the total mass and angular momentum through the 2-sphere at spacelike infinity, utilizing the Killing fields denoted by $\xi$ and $\eta$
\begin{equation}\label{mass_komar}
    M=- \dfrac{1}{8\pi}\int_{S^2_{\infty}}\star d\xi=- \dfrac{1}{8\pi}\int_{\mathcal{H}}\star d\xi - \dfrac{1}{4\pi}\int_{\Sigma}\star R(\xi)\,,
\end{equation}
\begin{equation}\label{ang_komar}
    J= \dfrac{1}{16\pi}\int_{S^2_{\infty}}\star d\eta= \dfrac{1}{16\pi}\int_{\mathcal{H}}\star d\eta + \dfrac{1}{8\pi}\int_{\Sigma}\star R(\eta)\, ,
\end{equation}
which yields  the generalized Smarr formula \cite{PhysRevLett.30.71}
\begin{equation}\label{generic_mass}
    M=2\Omega_H J + \dfrac{\kappa}{4\pi}A_H- \dfrac{1}{4\pi}\int_{\Sigma}\star R(\chi) \,.
\end{equation}
where
\begin{equation}
\kappa^2=-\frac{1}{2}(\nabla^a \chi^b)\left(\nabla_a \chi_b\right)\Big|_{\mathcal{H}}~,
\end{equation}
with $\kappa$ the surface gravity which determines the 
Hawking temperature $T_H=\kappa/(2\pi)$
and $A_H$
the event horizon area
(with the BH entropy in the EMd theory $S=A_H/4$).

Our goal now is to rewrite the last term in Eq. \eqref{generic_mass} in terms of physical quantities. To do so, we will make use of Einstein's equation, $ R_{\mu\nu}\,\chi^{\nu}-\dfrac{1}{2}R\, \chi_{\mu}=2  T_{\mu\nu}\,\chi^{\nu}$.  The energy-momentum tensor is expressed in terms of the matter Lagrangian through 
\begin{equation}
    T_{\mu\nu}=-2 \dfrac{\delta \mathcal{L}_{m}}{\delta g^{\mu\nu}}+g_{\mu\nu}\mathcal{L}_{m}\,.
\end{equation}

Following \cite{heusler_1996}, we denote $(\mathcal{L}_g)_{\mu\nu}=\dfrac{\delta \mathcal{L}_{m}}{\delta g^{\mu\nu}}$, so that we can rewrite equation \eqref{generic_mass} as \cite{Heusler:1993cj,Heusler:1993ke}
\begin{equation}
    M=2\Omega_H J + \dfrac{\kappa}{4\pi }A_H+ \dfrac{1}{\pi}\int_{\Sigma}\star \mathcal{L}_g(\chi)- \dfrac{1}{2\pi}\int_{\Sigma}\mathcal{L}\star\chi \,.
\end{equation}
Here, $\mathcal{L}$ is the total Lagrangian density of the system \eqref{action}. After straightforward calculations, we are able to express the mass formula in terms of the physical quantities of the system
\cite{Carter:2009nex,heusler_1996,RASHEED1995379,Kleihaus:2003sh,Compere:2007vx,Ortin:2022uxa} \footnote{To arrive in such formula, we have assumed the absence of magnetic monopoles, $Q_m$. Otherwise the formula would read $  M=2\Omega_H J + \dfrac{\kappa}{4\pi\,G}A+ \phi_{\mathcal{H}} Q_e +\psi_{\mathcal{H}} Q_m $. The dyonic solutions of
the EMd model
will be discussed elsewhere.}
\begin{equation}\label{mass_formula}
    M=2\Omega_H J + \dfrac{\kappa}{4\pi\,G}A_H+ \phi_\mathcal{H} Q_e \, ,
\end{equation}
where we have identified the electric charge
\begin{equation}
    Q_e= -\dfrac{1}{4\pi}\int_{\mathcal{H}}f(\Phi)\star\mathcal{F}\,.
\end{equation}

  Notice that the equations of motion 
  (\ref{eq-target})
  are left invariant under the transformation $\mathcal{A}\rightarrow -\mathcal{A}$. On the other hand, $\phi_\mathcal{H}$ and $Q_e$ assume the opposite sign. Hence, we will only consider $Q_e>0$.

%%%%%%%%%%%%%%%%%%%%%%%%%%%%%%%%%%%%%%%%%%%%%%%%%%%%%%%%%%%%%%%%%%%%%%

Regularity imposes that the scalar field 
assumes finite values at the horizon. 
By asymptotic flatness, the dilaton field asymptotes as 
\begin{equation}
\Phi=\Phi_{\infty}-\frac{D}{r}+\mathcal{O}\left(\frac{1}{r^2}\right) \ ,
\end{equation}
where $\Phi_{\infty}$ is a constant (assumed to be zero without loss of generality) and $D$ is the scalar monopole
\cite{Pacilio:2018gom} (see also \cite{Ballesteros:2023iqb}). 

Restricting to a dilaton coupling (\ref{dilaton}),
let us
consider the conserved current \eqref{noether}. By contracting the equation $dJ=0$
with the Killing vector $\kappa$ and using Cartan’s formula, $L_\kappa=d\, \iota_\kappa\; +\iota_\kappa d$ and that the fields obey $L_\kappa \Phi=L_\kappa \mathcal{A}=0$, we find
\begin{equation}
d\left[\iota_\kappa \star d \Phi + \gamma e^{-2 \gamma \Phi}\left(\iota_\kappa \mathcal{A}\right) \star \mathcal{F} - \gamma e^{-2 \gamma \Phi} \mathcal{A} \wedge \iota_\kappa \star \mathcal{F}\right]=0\,.
\end{equation}
Additionally, we have $d \phi = -\iota_\kappa \mathcal{F}$ and $d\psi=\iota_\kappa( e^{-2 \gamma \Phi}\star\mathcal{F})$. Here, we should remark that in \cite{Pacilio:2018gom}, it was used that $\iota_\kappa\star\mathcal{F}$ vanishes at the horizon, but one cannot, in general, neglect its contributions. Then,
\begin{equation}
d\left[\iota_\kappa \star d \Phi + \gamma e^{-2 \gamma \Phi} \phi\star \mathcal{F} - \gamma \psi \mathcal{F} \right]=0\, ,
\end{equation}
whose volume integral results in\footnote{Equation  \eqref{scalarcharge} was established in \cite{Pacilio:2018gom}; however, when a magnetic charge $Q_m$  is present, the relation becomes $    D= \gamma \phi_{\mathcal{H}} Q_e - \gamma \psi_H Q_m\,.
$}
\begin{equation}
\label{scalarcharge}
    D= \gamma \phi_{\mathcal{H}} Q_e. 
    %- \gamma \psi_H Q_m\,.
\end{equation}

Therefore, the electromagnetic field is sourcing the dilaton; if the former vanishes, the latter trivializes. 
Hence, the scalar hair is of the secondary type \cite{Coleman:1991ku}.

\medskip

Finally,  
let us remark that the equations 
(\ref{eq-target})-(\ref{eq-targetf})
of the model possesses a scaling symmetry $(\rho,z)\to \lambda (\rho,z)$,
with $\lambda$
an arbitrary positive parameter.
The quantities of interest 
scale accordingly, $i.e.$
$(Q_e,M)\to \lambda (Q_e,M)$,
$A_H \to \lambda^2 A_H $ ,
$T_H \to T_H/\lambda$ ,
$\Omega_H \to \Omega_H  /\lambda$ ,
$\phi_{\mathcal{H}} \to \phi_{\mathcal{H}}$.
This scaling symmetry is fixed 
by taking quantities measured in 
terms of ADM mass,
and
introducing the reduced quantities
\begin{eqnarray}
\label{scale1}
j\equiv \frac{J}{M^2}\ ,~~
q\equiv \frac{Q_e}{M}\ , ~~
a_H\equiv \frac{A_H}{16\pi M^2}\ , ~~
t_H\equiv 8\pi T_H M \ ,~~
\omega_H \equiv M \Omega_H~.
\end{eqnarray}
%
%%%%%%%%%%%%%%%%%%%%%%%%%%%%%%%%%%%%%%%%%%%%%%%%%%%%%%%%%%%%%%%%%
 %%%%%%%%%%%%%%%%%%%%%%%%%%%%%%%%%%%%%%%%%%%%%
 \section{Known limits}
 \label{sec3}
%%%%%%%%%%%%%%%%%%%%%%%%%%%%%%%%%%%%%%%%%%%%%
\subsection{Nonperturbative, exact solutions  }
%%%%%%%%%%%%%%%%%%%%%%%%%%%%%%%%%%%%%%%%%%%%%

\subsubsection{Static  BHs}
 \label{static}

The purely electric dilatonic solutions of~\eqref{action}   
were first considered by Gibbons and Maeda~\cite{Gibbons:1987ps} and Garfinkle, Horowitz and Strominger~\cite{Garfinkle:1990qj}. The  GMGHS BH is
spherically symmetric, with a line element 
\begin{eqnarray}
\label{GHS-solution}  
d s^2= && -\left(1-\frac{r_{+}}{r}\right)\left(1-\frac{r_{-}}{r}\right)^{\frac{1-\gamma^2}{1+\gamma^2}} d t^2+\frac{d r^2}{\left(1-\frac{r_{+}}{r}\right)\left(1-\frac{r_{-}}{r}\right)^{\frac{1-\gamma^2}{1+\gamma^2}}}
\\
&&{~~~~~~~~}
+r^2\left(1-\frac{r_{-}}{r}\right)^{\frac{2 \gamma^2}{1+\gamma^2}} (d\theta^2+\sin^2\theta d\varphi^2)  
\equiv ds_0^2
\end{eqnarray}
the Maxwell potential and dilaton fields being
\begin{equation}
\label{GHS-solution1}  
A=\frac{Q_e}{r}dt \ , \qquad 
e^{2\Phi}=\left(1-\frac{r_-}{r}\right)^{\frac{2\gamma}{1+\gamma^2}}\ .
\end{equation}
The two free parameters $r_{+}$, $r_{-}$ 
(with $r_-<r_+$)
 are related to the
ADM mass, $M$, and (total) electric charge, $Q_e$, by
\begin{eqnarray}   
  M = \frac{1}{2}
	\left[
	r_+ +\left(\frac{1-\gamma^2}{1+\gamma^2}\right)r_-
	\right] \ , \qquad 
	Q_e=\left(  
	\frac{r_-r_+}{1+\gamma^2}
	\right)^{\frac{1}{2}} \ .
\end{eqnarray}
For all $\gamma $, the
surfaces $r=  r_+$ is the location of the (outer) event horizon
and $r=  r_-$ is the inner horizon,
with
\begin{equation} 
A_H=4\pi r_+^2\left(1-\frac{r_-}{r_+} \right)^{\frac{2\gamma^2}{1+\gamma^2}},~
T_H=\frac{1}{4\pi}\frac{1}{r_+-r_-}
\left(1-\frac{r_-}{r_+}\right)^{\frac{2}{1+\gamma^2}}, ~
\Phi_\mathcal{H}=\frac{1}{\sqrt{1+\gamma^2} }\sqrt{\frac{r_-}{r_+}}~.~{~~}
\end{equation}
For $\gamma \neq 0$,
the extremal limit, which corresponds to the coincidence limit $r_- = r_+$, 
results in a singular solution (as can be seen $e.g.$   by evaluating the Kretschmann scalar).
In this limit, the event horizon's area goes to zero for
$ \gamma \neq 0$.
The Hawking temperature, however, only goes to zero in the extremal limit for $\gamma<1$,
while for $\gamma=1$ it
approaches a constant, and for $\gamma>1$ it diverges.

The reduced quantities~\eqref{scale1} have the following   expressions:
\begin{eqnarray}  
\label{GHS-solution-s1}  
\nonumber
q=\frac{2\sqrt{(1+\gamma^2)x}}{1+\gamma^2(1-x)+x},~~
 a_H=\frac{(1+\gamma^2)^2(1-x)^{\frac{2\gamma^2}{1+\gamma^2}}}
{(1+\gamma^2(1-x)+x)^2},~~
t_H=\frac{(1-x)^{\frac{1-\gamma^2}{1+\gamma^2}}(1+\gamma^2(1-x)+x)}
{1+\gamma^2}  \ ,
\end{eqnarray}
where
$0\leqslant x \leqslant 1$ (and $x\equiv r_-/r_{+}$).

%%%%%%%%%%%%%%%%%%%%%%%%%%%%%%%%%%%%% 
\subsubsection{ $\gamma=0,\sqrt{3}$ rotating  BHs}
 \label{rotating-exact}
%%%%%%%%%%%%%%%%%%%%%%%%%%%%%%%%%%%%% 

 The spinning BH solutions of the model~\eqref{action} 
 are known in closed form for
 two special values of the coupling constant
 $\gamma$, only.
 The $\gamma=0$
 case corresponds to the  KN 
solution, in which case the
dilaton field vanishes, while the 
metric and the vector field, as expressed in
 Boyer-Linquist coordinates, are 
\begin{eqnarray}  
\nonumber
 ds^2 & = &
 \Sigma \left(\frac{dr^2}{\Delta}+\Sigma d\theta^2 \right)
 +\left( \frac{ \left(r^2+a^2\right)^2-\Delta a^2\sin^2\theta}{\Sigma}
\right)\sin^2\theta d\phi^2 
 -\frac{ \left(\Delta -a^2\sin^2\theta\right)}{\Sigma}dt^2
\\
 & & 
 \label{KN}  
 - 2 a 
\sin^2\theta \frac{ \left(r^2+a^2-\Delta\right)}{\Sigma}dt\,d\phi ,
~~~~
A=  \frac{  Q_e r\left(dt-a\sin^2\theta d\phi\right) }{\Sigma}   \ ,
\end{eqnarray}
where
\begin{eqnarray}  
\label{s1KN}
\Sigma= r^2+a^2\cos^2\theta,~~
\Delta= r^2-2 \mu r+a^2+Q_e^2,
\end{eqnarray}
with $M=\mu$ and $Q_e$ are the mass and electric charge, respectively, while
$a=J/M$.
The KN BH possesses an (outer)
event horizon at $r=r_H$,
with $r_H>0$ the largest root of the equation $\Delta(r_H)=0$.
The expression of various relevant quantities, as written in terms
of $r_H,a,Q_e$ read
\begin{eqnarray}  
\label{s2KN}
&&
M=\frac{a^2+Q_e^2+r_H^2}{2r_H},~~
J=aM,~~A_H=4\pi (a^2+r_H^2),~~
\\
\nonumber
&&
T_H=\frac{1}{4\pi r_H}
\frac{r_H^2-a^2-Q_e^2}{r_H^2+a^2},~~
\phi_{\mathcal{H}}=\frac{Q_e r_H}{r_H^2+a^2},~~
\Omega_H=\frac{a}{a^2+r_H^2}~.
\end{eqnarray}
The reduced quantities~\eqref{scale1}
have a relatively simple expression
when taking
$a=r_H u \cos v$,
$Q_e=r_H u \sin v$  
(where $0\leq u\leq 1$, 
$0\leq v\leq  \pi/2$), with:
\begin{eqnarray}  
\label{KN1}  
&&
 j=\frac{2u \cos v}{1+u^2},~~
 q=\frac{2u \sin v}{1+u^2},~~
t_H=\frac{1-u^4}{1+u^2\cos^2 v},~~
\\
\nonumber
&& 
a_H=\frac{1+u^2\cos^2 v}{(1+u^2)^2},~~
w_H=\frac{(1+u^2)u \cos v}{2(1+u^2\cos^2 v)}~.
%~~\phi_{\mathcal{H}}=\frac{ u \sin v}{1+u^2\cos^2v}~.
\end{eqnarray}
The extremal limit is approached for $u\to 1$,
in which case $a_H>0$
and finite.

\medskip
 
The other special value  of the dilaton coupling constant
which
allows for an exact solution
is the KK one,
$\gamma= \sqrt{3}$.
This counterpart of the KN solution has been reported in Ref.
\cite{Frolov:1987rj}.
The corresponding expressions for 
metric and matter fields are
\begin{eqnarray}   
\nonumber
 ds^2 & = &
 \Sigma B \left(\frac{dr^2}{\Delta}+ d\theta^2 \right)
 +\left( B\left(r^2+a^2\right) +a^2\sin^2\theta\frac{Z}{B}
\right)\sin^2\theta d\phi^2 
 -\frac{ \left(\Delta -a^2\sin^2\theta\right)}{B\Sigma}dt^2
\\
\label{KK} 
 & &  
 - 2 a \sin^2\theta \frac{Z}{\sqrt{1-v^2}B}dt\,d\phi ,~~{\rm with}~
 Z=\frac{2\mu r}{\Sigma},~~
B=\sqrt{1+\frac{v^2 Z}{{1-v^2}}}~,
\end{eqnarray}
and  
\begin{eqnarray}  
\label{KK12}
\Phi=-\frac{\sqrt{3}}{2}\log B,~~
A=  \frac{  v Z  }{2\sqrt{1-v^2}B^2} 
\left(\frac{dt}{\sqrt{1-v^2}}-a\sin^2\theta d\phi\right) ~,
\end{eqnarray}
with $\Delta$ and $\Sigma$ given by (\ref{s1KN}) (with $Q_e=0$)
and   $0\leq v \leq 1$ a parameter.
The quantities of interest here are
\begin{eqnarray}  
\label{KKs1 }
&&
M=\frac{\mu (2-v^2)}{2(1-v^2)},~~
J=\frac{a (r_H^2+a^2)}{2r_H \sqrt{1-v^2}},~~Q_e=\frac{\mu v}{1-v^2},
\\
&&
\nonumber
A_H=\frac{4\pi (r_H^2+a^2)}{\sqrt{1-v^2}},~~
T_H=\frac{1}{4\pi r_H}\frac{ (r_H^2-a^2)\sqrt{1-v^2}}{r_H^2+a^2},~~
\Omega_H=\frac{a \sqrt{1-v^2}}{r_H^2+a^2},~~
\phi_{\mathcal{H}}=\frac{v}{2}~,
\end{eqnarray}
where $r_H = \mu + \sqrt{\mu^2 - a^2}$.
This results in the following expression of  reduced quantities~\eqref{scale1}:
\begin{eqnarray}  
\label{KK1}  
&&
 j=\frac{8x}{1+x^2}\frac{(1-v^2)^{3/2}}{(2-v^2)^2},~~
 q=\frac{2v}{2-v^2},~~
t_H=(1-x^2)\frac{2-v^2}{2\sqrt{1-v^2}},~~
\\
\nonumber
&& 
a_H=\frac{4}{1+x^2}\frac{(1-v^2)^{3/2}}{(2-v^2)^2},~~
w_H= \frac{(2-v^2)x}{4\sqrt{1-v^2}}, 
\end{eqnarray}
in terms of $v$ and $x$
(with $0\leq x \equiv \frac{a}{r_H}\leq 1$ 
 a parameter related to the BH rotation).

One notices the existence of a KN-like extremal limit,
with $t_H\to 0$
as $x\to 1$,
in which case all quantities are still finite.
However, as in the static case, the limit $v\to 1$
still
leads to a vanishing $a_H$ and (for $x\neq 1$)
a diverging $t_H$.

%%%%%%%%%%%%%%%%%%%%%%%%%%%%%%%%%%% 
\subsection{Perturbative solutions}

\subsubsection{Slowly rotating BHs}
 \label{slowly}

The slowly rotating solutions were first presented in \cite{Horne:1992zy}. For arbitrary $\gamma$, the metric is
\begin{equation}
d s^2= d s_0^2 -2 a f(r) \sin ^2 \theta d t d \phi,
\end{equation}
where $d s_0^2$ is the vacuum Kerr metric
($i.e.$ the limit $Q_e=0$ of (\ref{KN}))
and $f(r)$ is given by
\begin{equation}
f(r)= \frac{r^2\left(1+\gamma^2\right)^2\left(1-\frac{r_{-}}{r}\right)^{\frac{2 \gamma^2}{1+\gamma^2}}}{\left(1-\gamma^2\right)\left(1-3 \gamma^2\right) r_{-}^2} -\left(1-\frac{r_{-}}{r}\right)^{\frac{1-\gamma^2}{1+\gamma^2}}\left(1+\frac{\left(1+\gamma^2\right)^2 r^2}{\left(1-\gamma^2\right)\left(1-3 \gamma^2\right) r_{-}^2}+\frac{\left(1+\gamma^2\right) r}{\left(1-\gamma^2\right) r_{-}}-\frac{r_{+}}{r}\right) .
\nonumber
\end{equation}

The vector potential and dilaton field are
\begin{equation}
\Phi=\frac{\gamma}{1+\gamma^2} \log \left(1-\frac{r_{-}}{r}\right), \quad A_t=\frac{Q}{r}, \quad A_\phi=-a \sin ^2 \theta \frac{Q}{r} .
\end{equation}

As pointed out in \cite{Horne:1992zy}, the above solution agrees with the 
slowly rotating limit of
Kerr-Newman ($\gamma=0$) and KK ($\gamma=\sqrt{3}$)
solutions.
To this order in perturbation theory,
the angular momentum and gyromagnetic ratio
(as defined in Eq. (\ref{gyro}))
are
\begin{eqnarray}
 J=\frac{a}{2}
 \left(
r_+ 
+\frac{3-\gamma^2}{3(1+\gamma^2)}r_-
 \right),~~
 g=2-\frac{4\gamma^2}{(3-\gamma^2)r_-
 +3(3-\gamma^2)r_+}~,
\end{eqnarray} 
while
mass, electric charge, horizon area and Hawking temperature
do not change.

%%%%%%%%%%%%%%%%%%%%%%%%%%%%%%%%%%%%%%%%%%%%%%%%%%%%%%%%%%%%%%%%%%%%%%%%%%%%%
\subsubsection{Weakly charged BHs}

In the work \cite{Casadio:1996sj}, the authors consider perturbations around a Kerr BH, considering as perturbation parameter
the ratio $Q_e/M$. They do not require the BH to be slowly rotating. For arbitrary $\gamma$, the metric is
\begin{equation}
d s^2=  \widetilde{\Psi} d t^2+\frac{\rho^2}{\Delta}d r^2 +\rho^2 d \theta^2 +\left( \widetilde{\Psi}\widetilde{\omega}^2-\dfrac{\Delta \sin^2 \theta}{\widetilde{\Psi}}\right)d \phi^2 +2  \widetilde{\Psi}\widetilde{\omega}d t d \phi,
\end{equation}
with
\begin{eqnarray}
&& 
\nonumber
\widetilde{\Psi}=
-\frac{\Delta-a^2 \sin^2 \theta}{\rho^2},~~
\rho^2=\rho_0^2 e^{-\gamma \widetilde{\phi}/ 3},~~
\widetilde{\omega}=
%-a \delta\left[1+\widetilde{\Psi}^{-1}\right]\,.
-a \sin^2 \theta \left[1+\widetilde{\Psi}^{-1}\right],
~~
\Delta=\Delta_0+ (1-\dfrac{\gamma^2}{3}) Q_e^2~,
\\
&&
{\rm with}~~
\rho_0^2 = r^2+a^2 \cos^2\theta,~~
\Delta_0 = r^2-2 m r+a^2~.~~
\end{eqnarray}
The dilaton field is
\begin{equation}
\tilde{\phi}=-\frac{1}{\gamma} \ln b^{2 \alpha}\,,
\end{equation}
where
\begin{equation}
\alpha=\frac{2 \gamma^2}{1+\gamma^2},~~{\rm and}~~
 b^2=1+\frac{Q_e^2\left(1+\gamma^2\right) m r}{2 m^2 \rho_0^2} .
\end{equation}
To this order in perturbation theory,
the mass and angular momentum are
\begin{equation}
M  =m\left[1+\frac{\gamma^2 Q_e^2}{6 m^2}\right]\,, \qquad J=a m\left[1+\frac{\gamma^2 Q_e^2}{6 m^2}\right]\,.
\end{equation}

%%%%%%%%%%%%%%%%%%%%%%%%%%%%%%%%%%%%%%%%%
%\section{The results}
\section{The general solutions: a non-perturbative framework}
\label{sec4}
\subsection{The metric and boundary conditions}
%%%%%%%%%%%%%%%%%%%%%%%%%%%%%%%%%%%%%%%%%

We consider stationary, axially symmetric BH
spacetimes with Killing vector fields,
$\xi=\partial_t$, $\eta=\partial_\phi$. 
We consider a metric ansatz which has been employed in the past for the study of various generalizations\footnote{The Ansatz
(\ref{line_BH})
contains only three undetermined functions,
not four as for a generic matter content.
} of the Kerr BHs 
($e.g.$
\cite{Herdeiro:2014goa,Herdeiro:2016tmi,Delgado:2020rev}),
with a line element\footnote{
The Ansatz
(\ref{line_BH})
results from the generic form
(\ref{m1})
by taking 
\begin{eqnarray}
   \rho=\sqrt{r^2- r_H r } \sin \theta, ~~
   z=(r-r_H/2)\cos\theta~,
\end{eqnarray}
together with the function redefinition
\begin{eqnarray}
e^{2F_1}=\frac{e^{2h}}{X}(1-\frac{r_H}{r}+\frac{r_H^2}{4r^2}\sin^2 \theta) ,~~
 e^{2F_0}=r^2\sin^2 \theta/X,~~w=W.
\end{eqnarray}
 } 
\begin{equation}
\label{line_BH}
d s^2=-e^{2 F_0} N d t^2+e^{2 F_1}\left(\frac{d r^2}{N}+r^2 d \theta^2\right)+e^{-2 F_0} r^2 \sin ^2 \theta\left(d \varphi-W d t\right)^2\,,
\end{equation}
where
\begin{equation}
N \equiv 1-\dfrac{r_H}{r}\, ,
\end{equation}
and $(F_i, W)$ are functions of the spheroidal coordinates $(r, \theta)$; 
 $r_H>0$ is an input parameter again describing the location of the event horizon.
The coordinates $\theta,\varphi$ and $t$
possess the  usual range, while 
$r_H\leqslant r <\infty$.
The matter fields  are parametrized by
\begin{equation}
\label{matter}
\mathcal{A}_\mu d x^\mu=\left(
A_t- A_{\varphi} \sin\theta W\right)d t+A_{\varphi}\sin\theta d \varphi\,,\qquad\qquad 
\Phi \equiv \Phi(r,\theta)\,.
\end{equation}

Finding EMd solutions with the above
ansatz requires defining boundary behaviours. We have made the following choices. For the solutions to approach at spatial infinity ($r\rightarrow \infty$) a Minkowski spacetime we require
\begin{equation}
\label{inf}
		F_i=W=0,~~A_t=A_\varphi= \Phi=0 ~.
\end{equation}
The  regularity of the solutions
on the symmetry axis imposes
the following boundary conditions at $\theta=0,\pi$:
\begin{equation}
 \label{t0}
\partial_\theta F_i = \partial_\theta W = \partial_\theta \Phi = 0 \ ,~~
\partial_\theta A_t=A_\varphi=0.
\end{equation}
Moreover, the absence of conical singularities implies also that $
F_1=-F_0 
$
on the symmetry axis.

	We are looking for symmetric solutions concerning reflection symmetry with respect to the equatorial plane, $\theta=\pi/2$, such that we need to consider the solutions\footnote{As a numerical test, we have computed
    as well a number of solutions both for $0 \leq \theta \leq \pi/2$ and for $0 \leq \theta \leq \pi$ and verified that the results coincide.} only
for $0 \leqslant \theta \leqslant \pi/2$.   Therefore, we
impose
the subsequent boundary conditions at  $\theta=\pi/2$: 
	\begin{equation}
    \label{equat_plane}
		\partial_\theta F_i=\partial_\theta W=\partial_\theta A_t=A_\varphi=\partial_\theta \Phi=0 \ .
	\end{equation}

 For the metric ansatz~\eqref{line_BH}, the event horizon is located at a surface with constant radial variable, $r=r_H>0$.
The  horizon boundary conditions  and the numerical treatment of the problem greatly simplify  
by introducing a new radial coordinate 
\begin{equation}
x=\sqrt{r^2-r_H^2} ~.
\label{x}
\end{equation}
Then the boundary conditions we impose at the horizon are 
\begin{equation}
\partial_x F_i \big|_{x=0}= \partial_x \Phi  \big|_{x=0} =  0\ ,~~ W \big|_{x=0}=\Omega_H\ ,
A_t \big|_{x=0}=\phi_{\mathcal{H}},~A_\varphi \big|_{x=0}=0,
\label{bch1}
\end{equation}
where $\Omega_H $ is the horizon angular velocity, and 
the Killing vector $\chi =\xi+\Omega_H \eta$ is orthogonal and null on the horizon.
These conditions are consistent with a near-horizon solution on the form
\begin{eqnarray}
\label{rh}
{\cal F}_i(r,\theta)= {\cal F}_{i0}(\theta)+x^2 {\cal F}_{i2}(\theta)+\mathcal{O}(x^4)\ ,
\end{eqnarray}  
with ${\cal F}_i =\{F_0, F_1,  W; \Phi;A_\varphi,A_t\}$,
where the essential functions are
${\cal F}_{i0}$. We mention that $(F_0 -F_1)\big |_{r_H}=const.$,
as imposed by a constraint equation.

%%%%%%%%%%%%%%%%%%%%%%%%%%%%%%%%%%%%%%%%%%%%%%%%%%%%%%%%%%%%%%%%%%%%%
\subsection{Physical quantities}
\label{sec5o}

Most of the quantities of interest are 
encoded in the metric functions at the horizon or at infinity.
Considering first horizon quantities, the
Hawking temperature $T_H$,  
and the event horizon area $A_H$  
 are computed as
\begin{eqnarray}
\label{THAH}
&&
T_H=\frac{1}{4\pi r_H}e^{F_0(r_H,\theta)-F_1(r_H,\theta)} \ ,
\qquad 
A_H=2\pi r_H^2 \int_0^\pi d\theta \sin \theta~e^{F_1(r_H,\theta)-F_0(r_H,\theta)} \ .
\end{eqnarray}
The horizon angular velocity $\Omega_H$ is fixed by the horizon value of the metric function $W$,
\begin{eqnarray}
\label{OmegaH}
\Omega_H=-\frac{g_{\varphi t}}{g_{\varphi\varphi}}\bigg|_{r_H}=W \bigg|_{r_H}.
\end{eqnarray}
 % the entropy being $S=A_H/4G$.

As for Kerr BHs,  the solutions in this work have a topologically spherical horizon \cite{Hawking:1973uf}. Geometrically, however, the horizon is not a round sphere.
Its deformation can be seen by evaluating the ratio between the circumference of the horizon along the equator and the meridional circumference of the horizon (along the poles)
\begin{equation}
L_e=2 \pi r_H e^{-F_0\left(r_H, \pi / 2\right)}\,,\qquad\qquad 
L_p=2 r_H \int_0^\pi d \theta e^{F_1\left(r_H, \theta\right)}\,.
\end{equation}
As with other spinning BHs,
the ratio $L_e/L_p$
gives a
 measure for the deformation
of the horizon.

The total (ADM)  mass $M$, the angular momentum $J$ 
and the electric charge $Q_e$
of the BHs
are read off from the asymptotics of 
the metric components
$g_{tt}$, $g_{\varphi t}$
and electric potential $A_t$, respectively
\begin{eqnarray}
\label{asym}
g_{tt} =-1+\frac{2 M}{r}+\dots \ , \quad g_{\varphi t}=-\frac{2 J}{r}\sin^2\theta+\dots \ , \quad  A_t =- \frac{Q_e}{r}+\ldots\,.
\end{eqnarray}
 Of interest are also the asymptotics of magnetic potential and scalar field,
 \begin{equation}
 A_{\varphi} = \frac{\mu_m \sin \theta}{r}+\dots\,~~~ \Phi=- \dfrac{D}{r}+\cdots\, ,
\end{equation}
where $\mu_m$ is the magnetic dipole moment associated with the dipolar decay of the magnetic field. We can define the gyromagnetic ratio $g$ as
\begin{equation}
\label{gyro}
    \mu_m=g \dfrac{Q_e}{2M}J~,
\end{equation}
which assumes the value of 2 for the KN BH.

%%%%%%%%%%%%%%%%%%%%%%%%%%%%%%%%%%%%%%%%%%%%%%%%%%%%%%%%%%%%%%%%%%%%%
\subsection{The numerical approach}
\label{num}

In our approach, the field equations are reduced to a set of six 
coupled nonlinear elliptic partial differential equations for the functions 
${\cal F}_i =\{F_0, F_1,  W; \Phi;A_\varphi,A_t \}$,
which are found by plugging the Ansatz 
(\ref{line_BH}),
(\ref{matter})  
into fhe field eqs.~(\ref{EME}), (\ref{ME}), ~(\ref{DE}).
The set of equations which is solved in numerics
is given in Appendix A.

 	To perform the numerical calculations, we have utilized a professional software package \cite{SCHONAUER1989279,SCHONAUER1990279,SCHONAUER2001473}
  that employs a finite difference method with an arbitrary grid and arbitrary consistency order. The Newton-Raphson method is employed to iteratively solve the system, requiring an initial guess close to the solution for successful convergence. Further details about the solver are provided in \cite{Herdeiro_2015, Delgado:2022pwo}.

   After removing time and azimuthal dependencies, the equations reduce to two spatial dimensions. The equations are then discretized in  a  two-dimensional grid with $N = N_r \times N_\theta$ points. 	To enhance computational efficiency, we introduce a compactified radial coordinate: $ X=   x/(c +  x)$, 
   with $  x=\sqrt{r^2-r_H^2}$
    and $c$ being an input parameter typically set to one, but increased or decreased for better accuracy. This transformation maps the semi-infinite interval $[0, \infty)$ to $[0, 1]$, eliminating the need for a cutoff radius. Derivatives are modified as:
	\begin{equation}
		\mathcal{F}_{, r} \longrightarrow \frac{1}{c}(1-  X)^2 \mathcal{F}_{, X}, \quad \mathcal{F}_{, r r} \longrightarrow \frac{1}{c^2}(1-X)^4 \mathcal{F}_{, XX}-\frac{2}{c^2}(1-X)^3 \mathcal{F}_{, X} \ .
	\end{equation}
	
	We have utilized an equidistant grid which typically has 300 points in $X$, covering the integration region $0 \leq X \leq 1$, and 50 points in $\theta$ direction ($100$ points when working from $0 \leq \theta \leq \pi$). The choice of grid  has been made carefully, taking into consideration the trade-off between accuracy and computational cost. Furthermore, we apply a sixth-order finite difference scheme and a parallelized code. We have verified the convergence of the solutions with respect to the grid spacing, ensuring that our results are reliable and accurate.

In our numerical scheme, there are four input parameters: 
${\bf i)}$
 the dilaton coupling constant  $\gamma$ in the action 
 (\ref{action}), (\ref{dilaton}),
 ${\bf ii)}$
 the event horizon radius $r_H$ in the metric form (\ref{line_BH}),
  ${\bf iii)}$
the event horizon angular velocity $\Omega_H$ and
 ${\bf iv)}$
the value $\phi_{\mathcal{H}}$ of the electric  potential at the horizon 
in the boundary conditions (\ref{bch1}).
The quantities of interest are computed from the numerical output.
 For example, the mass the angular momentum 
 and the electric charge 
are extracted from the far field asymptotics
while the Hawking temperature, the entropy and the horizon area 
are obtained from the event horizon data.

%%%%%%%%%%%%%%%%%%%%%%%%%%%%%%%%%%%%%%%%%%%%%%%%%%%%%%%%%%%%%%%%%%%%%
\begin{figure}[ht!]
\begin{center}  
\includegraphics[height=.28\textheight, angle =0]{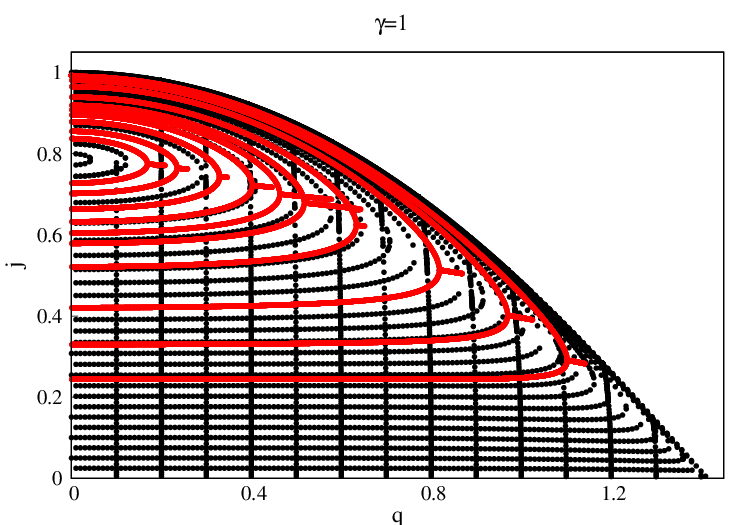} 
\end{center}
  \vspace{-0.5cm}
\caption{ Domain of existence of EMd spinning BHs
with
$\gamma=1$
is shown for an angular momentum 
$vs.$~electric charge diagram
(in units of ADM mass). This figure contains around 
fifteen thousands of  numerical solutions,
each one represented  as a  dot. 
Furthermore, some of the solutions
(shown with red dots)
were reobtained
by employing the spectral solver
described in \cite{Fernandes:2022gde}.
}
\label{dom0}
\end{figure}
%%%%%%%%%%%%%%%%%%%%%%%%%%%%%%%%%%%%%%%%%%%%%%%%%%%%

In Figure \ref{dom0} all data points
which were found numerically
for solutions with $\gamma=1$
are explicitly shown. 
The blue shaded regions in Figures
\ref{qj}-\ref{qj3}
with 
$\gamma\neq (0,\sqrt{3})$
are the extrapolation of such sets of data points into the continuum. 

In order to construct the domain of existence for a given  $\gamma$ , we have used
the scaling symmetry discussed in Subsection \ref{sec6}
%(\ref{scaling})
to fix the value of $r_H$
(typically $r_H=0.25$)
and considered a set of 
values for
$\Omega_H$ ($\phi_{\mathcal{H}}$);
subsequently, for each fixed value of  $\Omega_H$ ($\phi_{\mathcal{H}}$), we vary $\phi_{\mathcal{H}}$ ($\Omega_H$).

A large set of EMd configurations with
$\gamma =(1,3)$
were also constructed
independently, by using a spectral solver
introduced in  \cite{Fernandes:2022gde},
 with a resolution $N_r \times N_\theta=60 \times 12$ (see Figure \ref{dom0}). 
 However, a part of parameter space was not recovered with the spectral solver since that region becomes numerically challenging and would require significantly higher resolution, increasing computational time and resource demands. Nonetheless, for the considered solutions, we have found an overall very good agreement for the results obtained by these two different numerical schemes, the typical relative difference for various global quantities being around $10^{-6}$.

%%%%%%%%%%%%%%%%%%%%%%%%%%%%%%%%%%%%%%%%%%%%%%%%%%%%%%
\begin{figure}[h]
	\makebox[\linewidth][c]{%
		\begin{subfigure}[b]{8cm}
			\centering
\includegraphics[width=8cm]{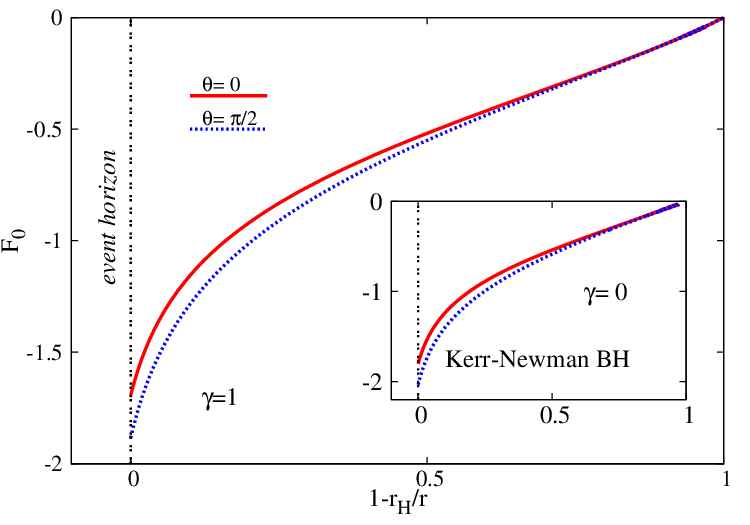}
		\end{subfigure}%
		\begin{subfigure}[b]{8cm}
			\centering
\includegraphics[width=8cm]{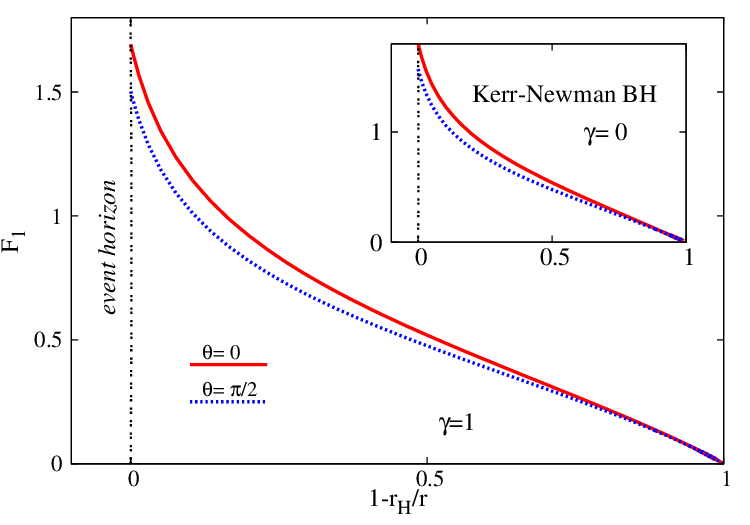}
		\end{subfigure}%
  } 
  \\
  \\
 \makebox[\linewidth][c]{%
		\begin{subfigure}[b]{8cm}
			\centering
\includegraphics[width=8cm]{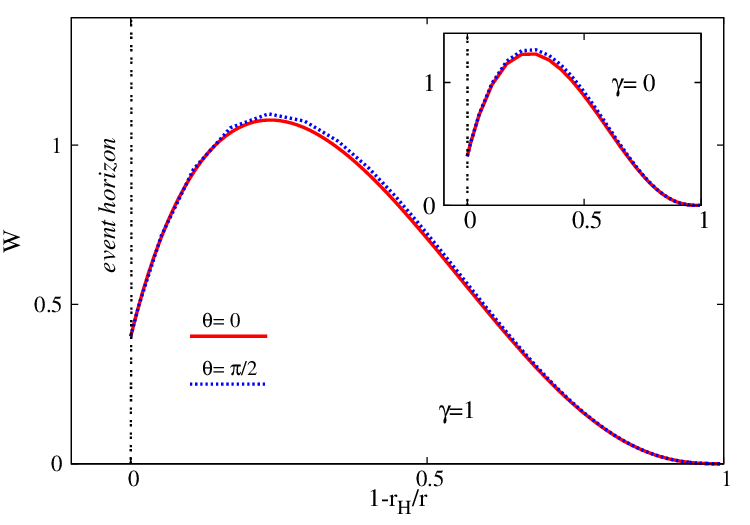}
		\end{subfigure}%
		\begin{subfigure}[b]{8cm}
			\centering
\includegraphics[width=8cm]{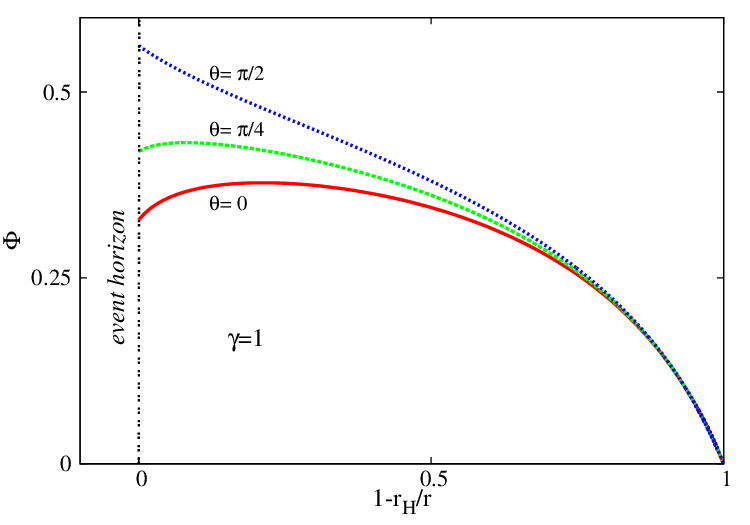}
		\end{subfigure}%
	}
  \\
  \\
 \makebox[\linewidth][c]{%
		\begin{subfigure}[b]{8cm}
			\centering
\includegraphics[width=8cm]{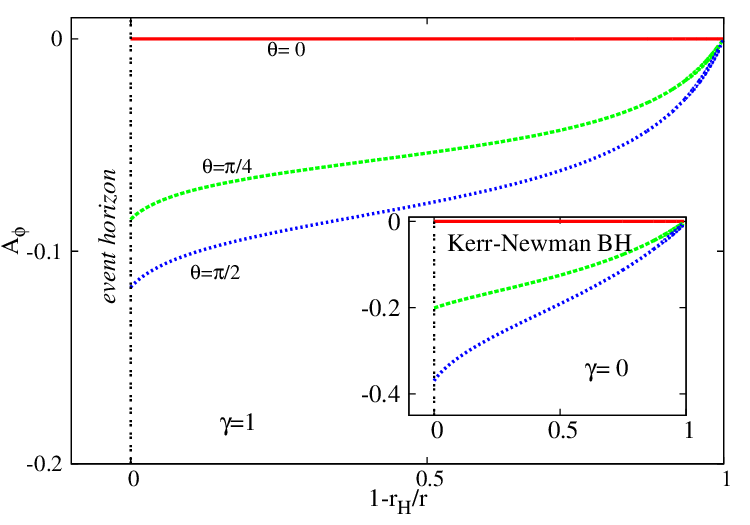}
		\end{subfigure}%
		\begin{subfigure}[b]{8cm}
			\centering
\includegraphics[width=8cm]{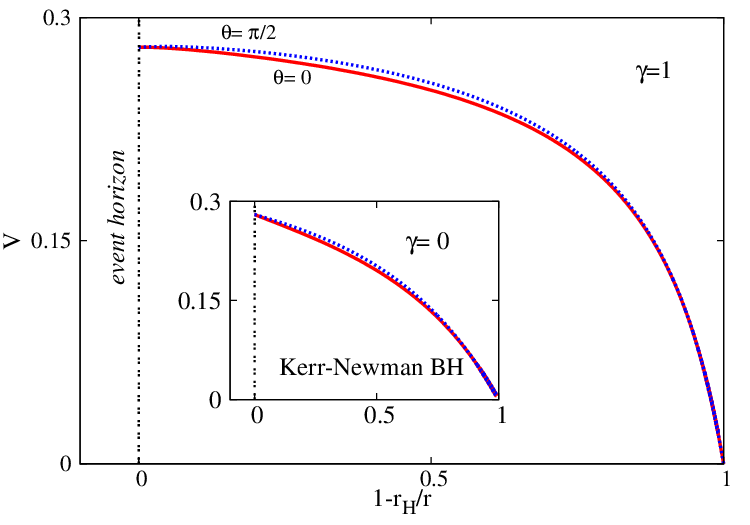}
		\end{subfigure}%
	} 
 \\
	\caption{
	{\small
Profile functions of a typical
 $\gamma=1$ solution 
with 
$r_H=0.25$,
$\Omega_H=0.4$,
$\phi_{\mathcal{H}}=0.28$, 
$vs.$ the compactified radial
coordinate $1-r_H/r$,  
 for several different polar angles $\theta$. The insets show the corresponding functions for a KN BH ($\gamma=0$) with the same
 input parameters $\{r_H,\Omega_H,\phi_{\mathcal{H}} \}$.
}
		\label{sol1}
  }
\end{figure}
%%%%%%%%%%%%%%%%%%%%%%%%%%%%%%%%%%%%%%%%%%%%%%%% 

%%%%%%%%%%%%%%%%%%%%%%%%%%%%%%%%%%%%%%%%%%%%%%%%%%%%%%
\begin{figure}[h]
	\makebox[\linewidth][c]{%
		\begin{subfigure}[b]{8cm}
			\centering
\includegraphics[width=8cm]{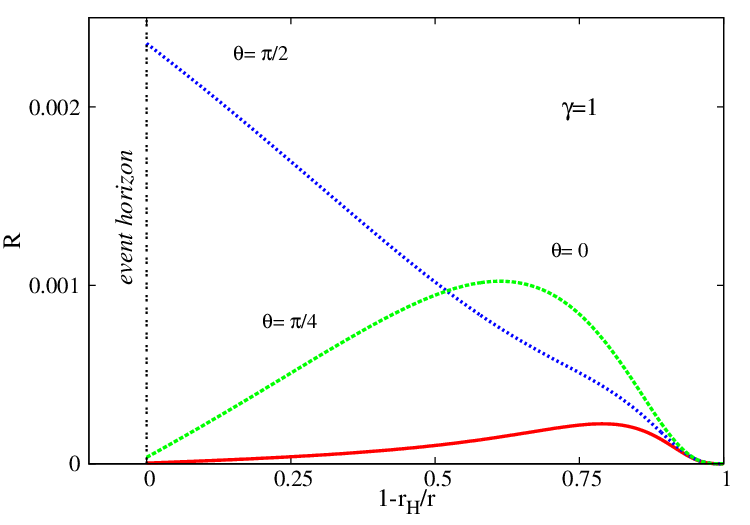}
		\end{subfigure}%
		\begin{subfigure}[b]{8cm}
			\centering
\includegraphics[width=8cm]{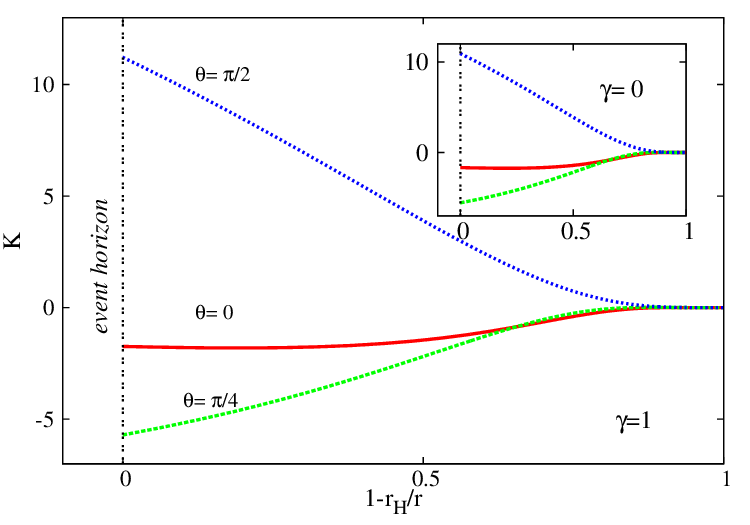}
		\end{subfigure}%
  } 
  \\
  \\
 \makebox[\linewidth][c]{%
		\begin{subfigure}[b]{8cm}
			\centering
\includegraphics[width=8cm]{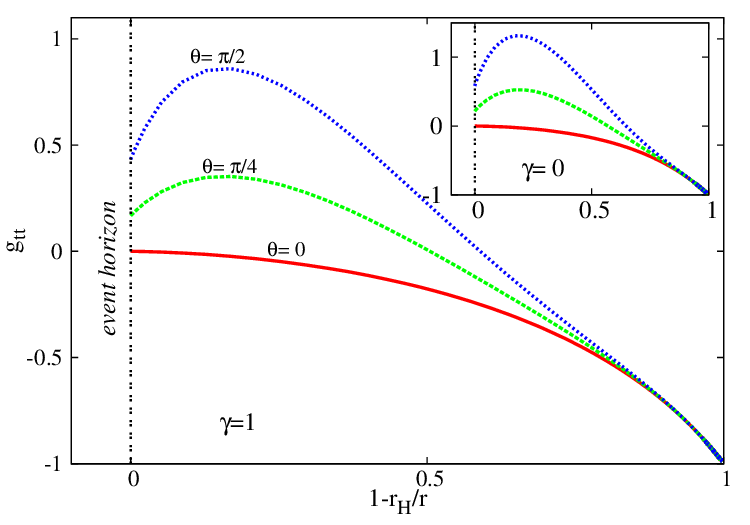}
		\end{subfigure}%
		\begin{subfigure}[b]{8cm}
			\centering
\includegraphics[width=8cm]{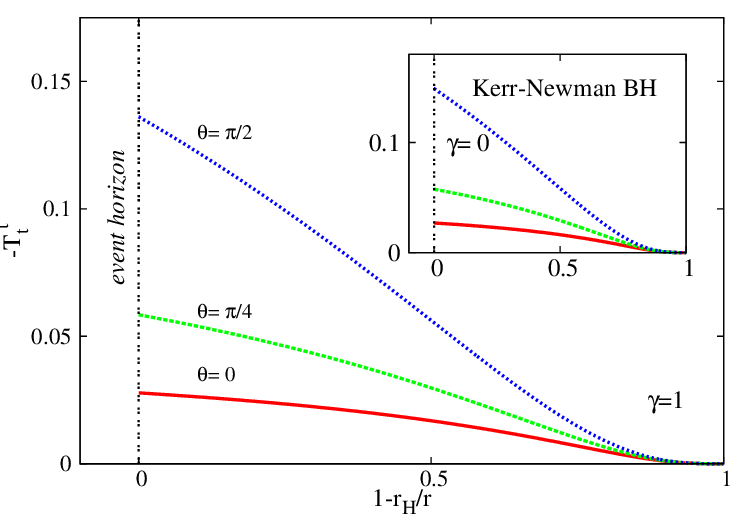}
		\end{subfigure}%
	} 
 \\
	\caption{
	{\small 
  The Ricci $R$ and Kretschmann $K$ scalars 
  are shown for
  the same solution as in Figure \ref{sol1}
  together with 
  the metric function $g_{tt}$ and the component $T_t^t$
  of the energy-momentum tensor.
 }
		\label{sol2}
  }
\end{figure}
%%%%%%%%%%%%%%%%%%%%%%%%%%%%%%%%%%%%%%%%%%%%%%%% 

The parameter $\gamma$ is special, since
it fixes the theory (rather than an integration constant of a given solution).
We have obtained the full parameter space of solutions for  $\gamma=\{ 0.5, 1, 1.5, 2,3\}$. 
Sets of solutions have been 
found also for various other values of $\gamma$, 
although not in a systematic way.
This includes the cases   $\gamma=\{ 0, \sqrt{3}\}$,
where an analytical solution is available.
%The numerical solutions here were cons
In Appendix \ref{appenB},
we present a comparison between 
the theory and numerical results for several quantities
of interest  as a function of
the event horizon velocity
for  EMd BHs with $\gamma= \sqrt{3}$
 (similar results were found for $\gamma=0$).
Figure \ref{errorKK} therein gives an overall estimate for the
numerical accuracy of the solutions, which is consistent with other diagnostics provided by the solver.
This supports the conclusion
that the proposed numerical scheme can be used in the construction of EMd BH solutions with an arbitrary $\gamma$.

 %%%%%%%%%%%%%%%%%%%%%%%%%%%%%%%%%%%%%%%%%%%%%%%%%%%%%%%%%%%%%%%%%%
\section{The new results}
\label{sec5}
 %%%%%%%%%%%%%%%%%%%%%%%%%%%%%%%%%%%%%%%%%%%%%%%%%%%%%%%%%%%%%%%%%%
\subsection{General properties}
%%%%%%%%%%%%%%%%%%%%%%%%%%%%%%%%%%%%%%%%%%%%%%%%%%%%%%%%%%%%%%%%%%

For all solutions we have found, the metric functions ${\cal F}_i$, together with their first and second derivatives with respect
to both $r$ and $\theta$ have smooth profiles. This leads to finite curvature invariants on the full domain of integration,
in particular at the event horizon.

The profile functions of a typical EMd solution 
with $\gamma=1$
are exhibited in Figure~\ref{sol1}. 
The insets shows for comparison  the same curves for a KN solution with the same
input parameters $(r_H$, $\Omega_H,\phi_{\mathcal{H}})$. 
The  Ricci and the Kretschmann scalars, $R$ and $K$, 
together with the metric function $g_{tt}$
  and the component $T_t^t$
  of the energy-momentum tensor are shown in 
 Figure \ref{sol2}.
In these plots,  the corresponding functions are shown in terms of the compactified radial coordinate
$1-r_H/r$
 for   different values\footnote{ 
 Since the  solutions possess  a $\mathbb{Z}_2$-reflection symmetry, in these plots we show the behaviour for  
 $0\leq \theta \leq \pi/2$, only.} of the
angular coordinate $\theta$.
As one can see,
the shape of the metric functions $F_0,F_1,W$ and
gauge potentials
$V,A_\varphi$ is similar to those  in the $\gamma = 0$ case. 
The maximal deviation from the
KN  profiles (with the same 
input parameters $r_H,\Omega_H,\phi_{\mathcal{H}}$)
is near the horizon.
At the same time, the dilaton field may possess a more complicated angular dependence.

%%%%%%%%%%%%%%%%%%%%%%%%%%%%%%%%%%%%%%%%%%%%%%%%%%%%%%
\begin{figure}[h]
	\makebox[\linewidth][c]{%
		\begin{subfigure}[b]{8cm}
			\centering 
\includegraphics[width=8cm]{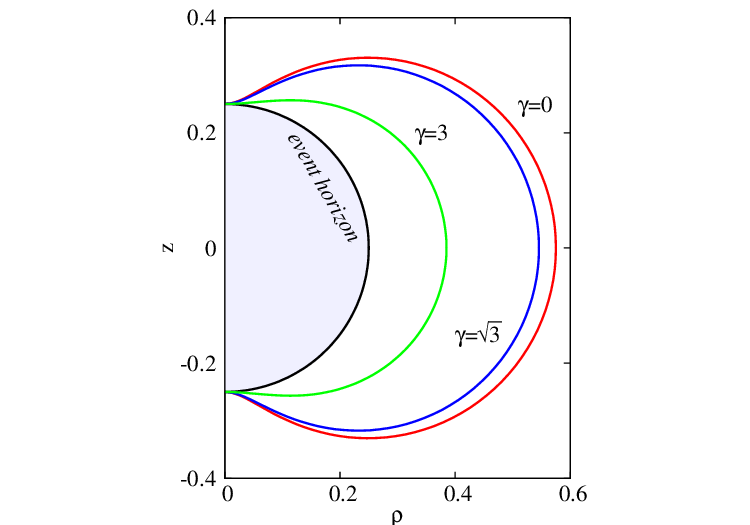}
		\end{subfigure}%
        		\begin{subfigure}[b]{8cm}
			\centering
\includegraphics[width=8cm]{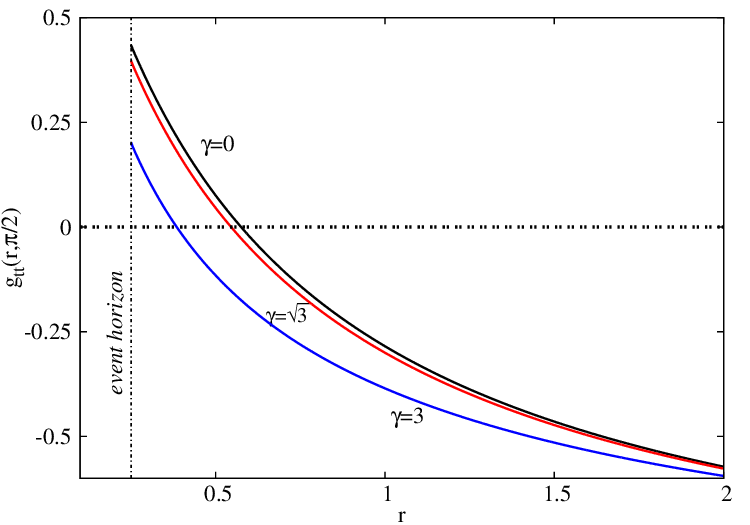}
		\end{subfigure}%
  }  
 \\
\caption{ 
A cross section of the ergo-sphere
 ({\it left panel})
along the
$\rho-z$ plane-- with $\rho=r\sin \theta$
and $z=r\cos\theta$, and the metric function 
$g_{tt}$ along the equatorial
plane ({\it right panel})
are shown 
for BHs solution
with
three value of $\gamma$
and the same input parameters
$r_H=0.25$, $\Omega_H=0.7$,
$\phi_{\mathcal{H}}=0.3$.
}
\label{ergoregion}
\end{figure}
%%%%%%%%%%%%%%%%%%%%%%%%%%%%%%%%%%%%%%%%%%

%%%%%%%%%%%%%%%%%%%%%%%%%%%%%%%%%%%%%%%%%%%%%%%%%%%%%%
\begin{figure}[h]
	\makebox[\linewidth][c]{%
		\begin{subfigure}[b]{8cm}
			\centering 
\includegraphics[width=8cm]{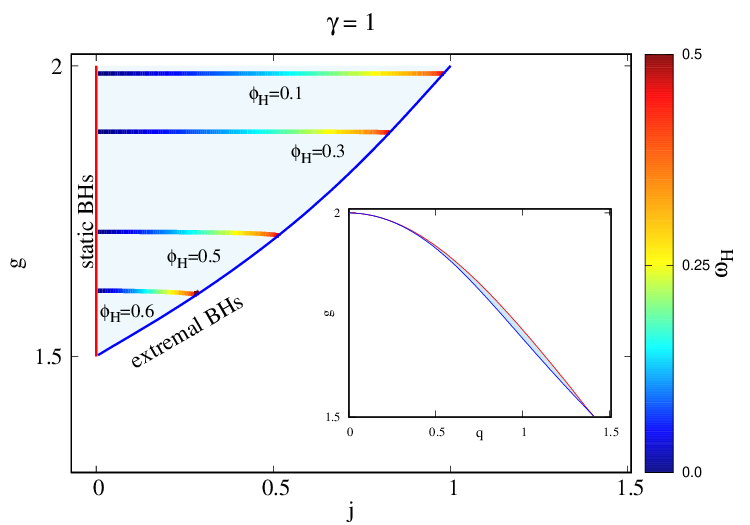}
		\end{subfigure}%
		\begin{subfigure}[b]{8cm}
			\centering 
\includegraphics[width=8cm]{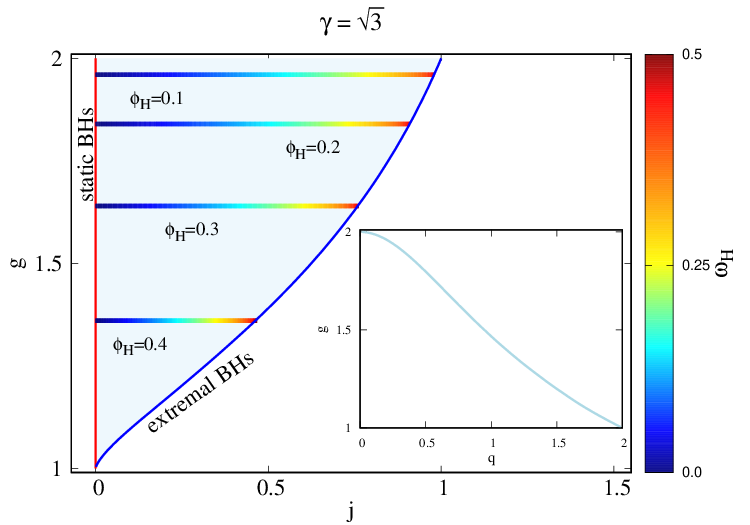}
		\end{subfigure}%
  } 
 \\
	\caption{
	{\small
The gyromagnetic ratio $g$ of 
EMd BHs is shown as
a function of reduced angular momentum $j$
and reduced electric charge $q$, for two values
of the dilaton coupling parameter $\gamma$.  
}
		\label{gyr}
  }
\end{figure}
%%%%%%%%%%%%%%%%%%%%%%%%%%%%%%%%%%%%%%%%%%%%%%%% 
%%%%%%%%%%%%%%%%%%%%%%%%%%%%%%%%%%%%%%%%%%%%%%%% 

%%%%%%%%%%%%%%%%%%%%%%%%%%%%%%%%%%%%%%%%%%%%%%%%%%%%%%
\begin{figure}[h]
	\makebox[\linewidth][c]{%
		\begin{subfigure}[b]{8cm}
			\centering 
\includegraphics[width=8cm]{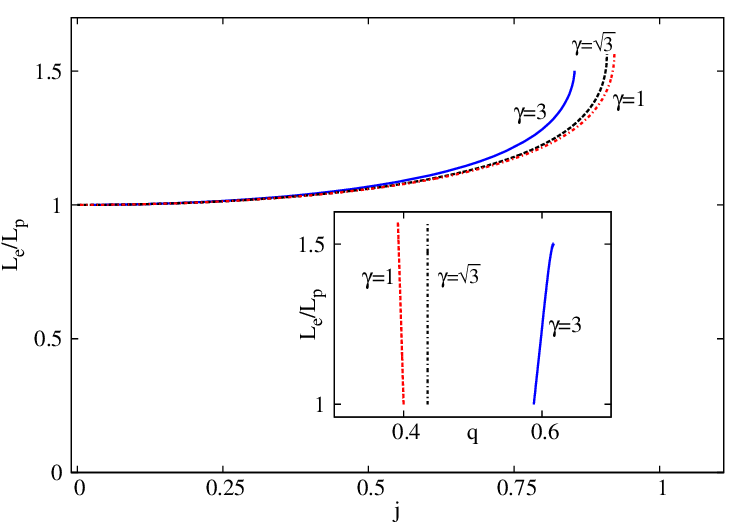}
		\end{subfigure}%
        }
 \\
	\caption{
	{\small
The horizon deformation (as given by the ratio $L_e/L_p$) 
EMd BHs with  $\gamma=1,\sqrt{3}$ and $3$ is shown as
a function of reduced angular momentum $j$
and reduced electric charge $q$ (inset), for three values
of the dilaton coupling parameter $\gamma$, for fixed $\phi_{\mathcal{H}}=0.3$.
}
		\label{LeLp}
  }
\end{figure}
%%%%%%%%%%%%%%%%%%%%%%%%%%%%%%%%%%%%%%%%%%%%%%%% 

 %%%%%%%%%%%%%%%%%%%%%%%%%%%%%%%%%%%%%%%%%%%%%%%%%%%%%%
\begin{figure}[h]
	\makebox[\linewidth][c]{%
		\begin{subfigure}[b]{8cm}
			\centering 
\includegraphics[width=8cm]{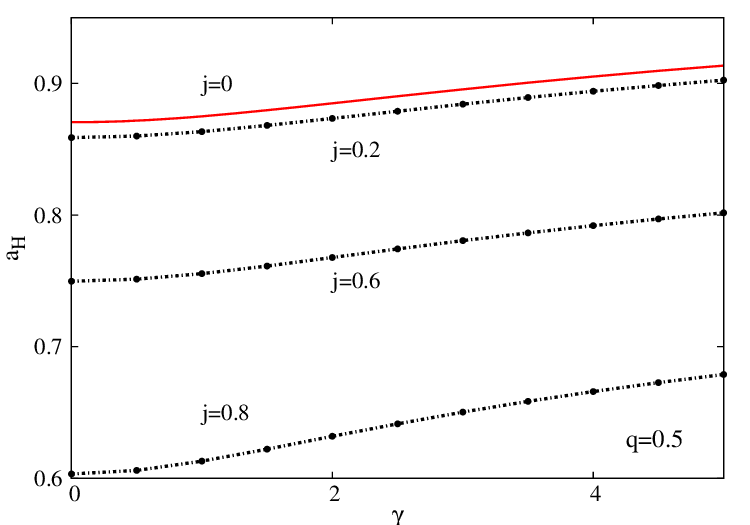}
		\end{subfigure}%
		\begin{subfigure}[b]{8cm}
			\centering 
\includegraphics[width=8cm]{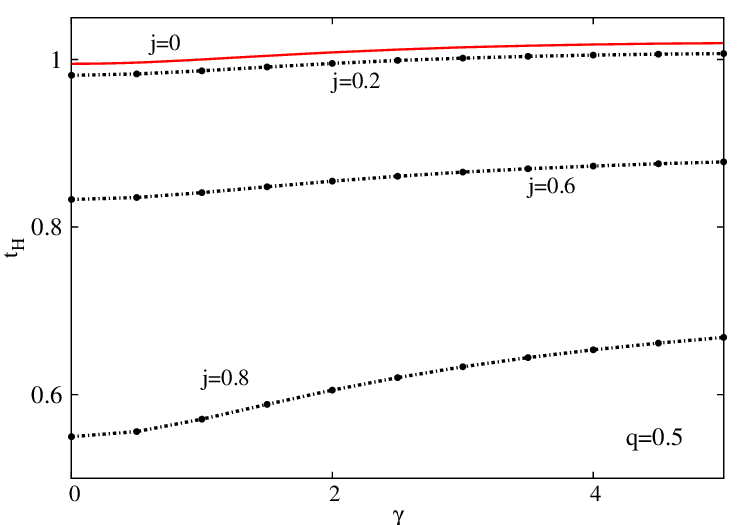}
		\end{subfigure}%
        }
 \\
	\caption{
	{\small
The event horizon area and the
Hawking temperature are shown as a function of the dilaton coupling constant $\gamma$
for several value of the  angular momentum 
for solutions with a fixed 
value of the  electric charge.
 Here and in Figures \ref{qj}-\ref{qj3} 
all quantities are given in units of the BH mass.
}
		\label{var-gamma}
  }
\end{figure}
%%%%%%%%%%%%%%%%%%%%%%%%%%%%%%%%%%%%%%%%%%%%%%%%
 All spinning EMd BHs  have an ergoregion, defined as the region where the norm of $\xi=\partial_t$ becomes positive outside the horizon. 
This region is bounded by the event horizon and by the surface where
\begin{equation}
 g_{tt}=-e^{2F_0} N+W^2e^{-2F_0}r^2 \sin^2\theta =0 \ .
\end{equation}
As with  the Kerr BH, this surface has a spherical topology and touches the horizon at the poles. 
As one can see in Figure \ref{ergoregion},
the size of the ergoregion decreases with increasing $\gamma$,
when keeping all other input parameters fixed. 
%\ch{
This hints at a possible decrease of the superradiant instability with increasing~
$\gamma$~\cite{Herdeiro:2014jaa}.
%}

\begin{figure}[h]
	\makebox[\linewidth][c]{%
		\begin{subfigure}[b]{8cm}
			\centering 
\includegraphics[width=8cm]{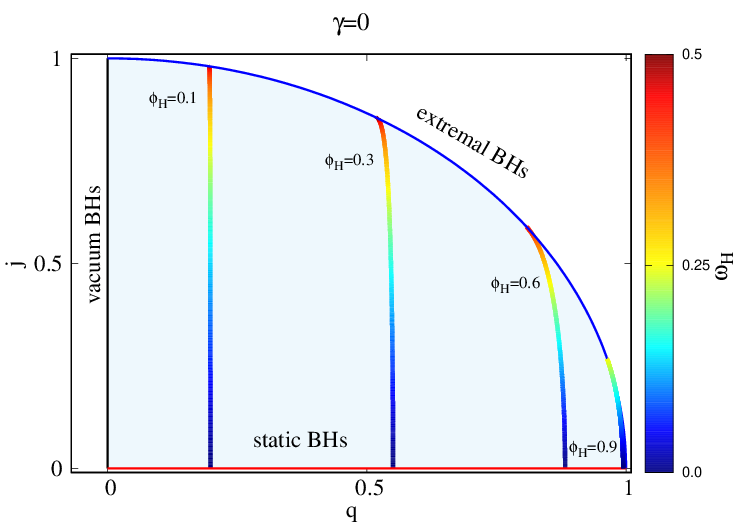}
		\end{subfigure}%
		\begin{subfigure}[b]{8cm}
			\centering 
\includegraphics[width=8cm]{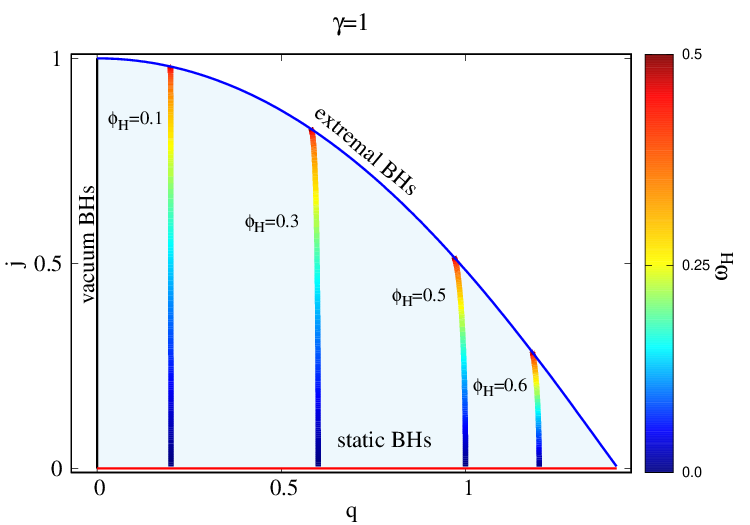} 
		\end{subfigure}%
  } 
  \\
  \\
 \makebox[\linewidth][c]{%
		\begin{subfigure}[b]{8cm}
			\centering 
\includegraphics[width=8cm]{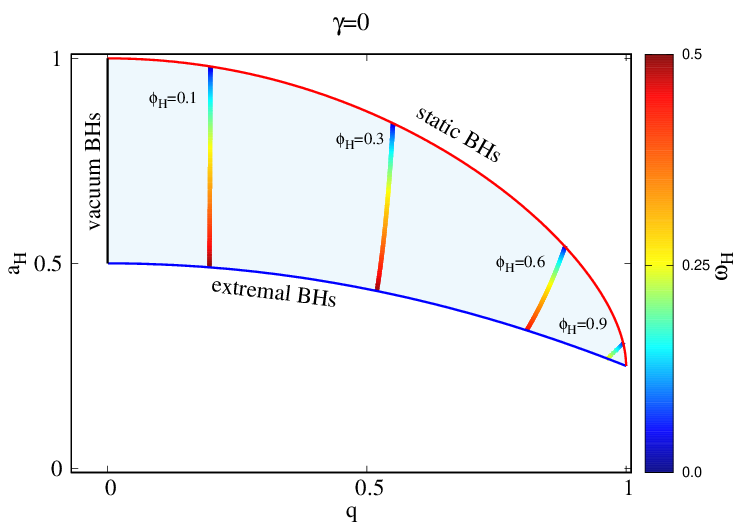}
		\end{subfigure}%
		\begin{subfigure}[b]{8cm}
			\centering
\includegraphics[width=8cm]{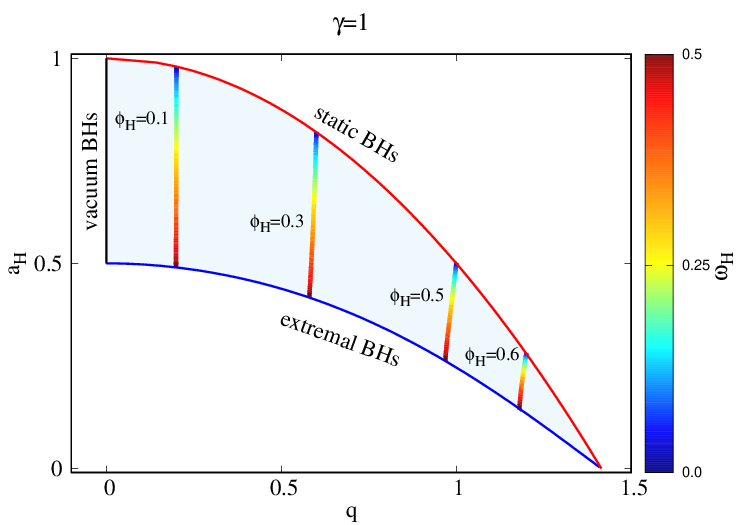}
		\end{subfigure}%
	}
      \\
  \\
 \makebox[\linewidth][c]{%
		\begin{subfigure}[b]{8cm}
			\centering
\includegraphics[width=8cm]{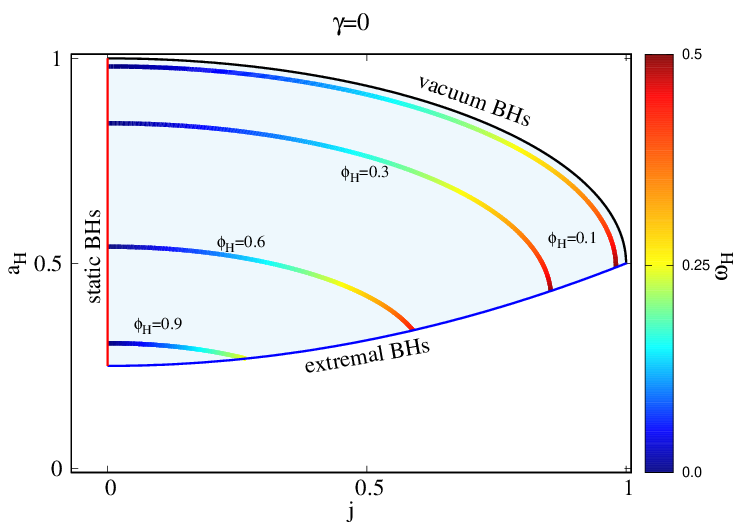}
		\end{subfigure}%
		\begin{subfigure}[b]{8cm}
			\centering
\includegraphics[width=8cm]{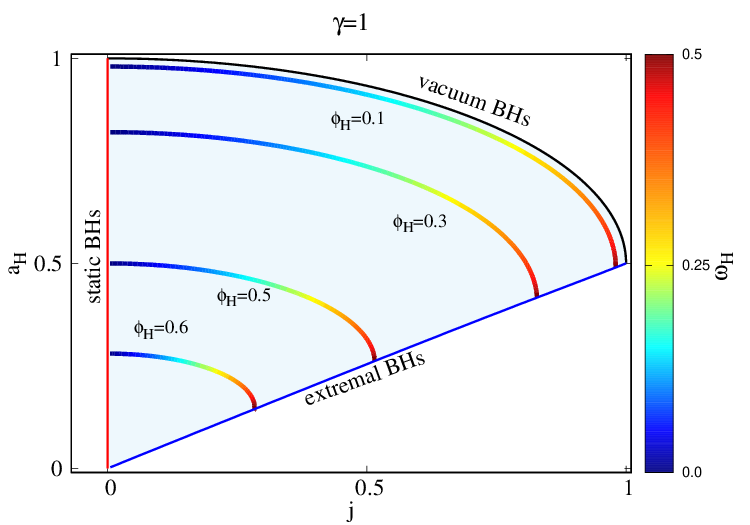}
		\end{subfigure}%
	}
 \\
	\caption{
	{\small
The domain of existence of KN BHs is shown together with
that of charged, rotating
EMd BHs with $\gamma=1$.
}
		\label{qj}
  }
\end{figure}

\begin{figure}[h]
	\makebox[\linewidth][c]{%
		\begin{subfigure}[b]{8cm}
			\centering
\includegraphics[width=8cm]{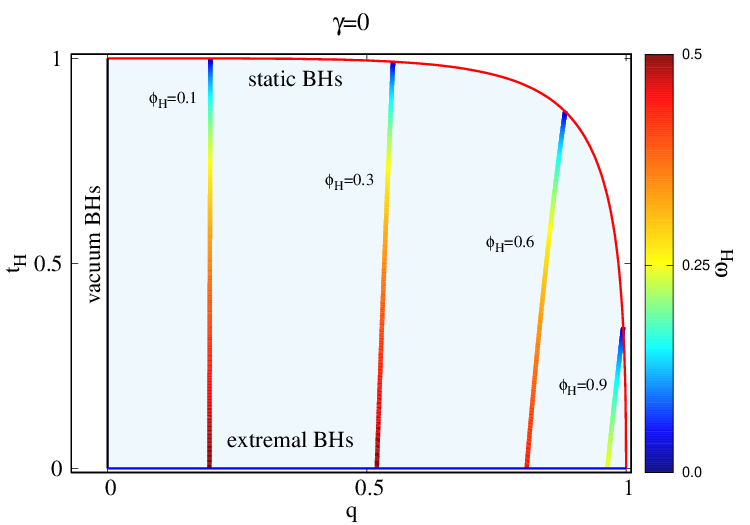}
		\end{subfigure}%
		\begin{subfigure}[b]{8cm}
			\centering
\includegraphics[width=8cm]{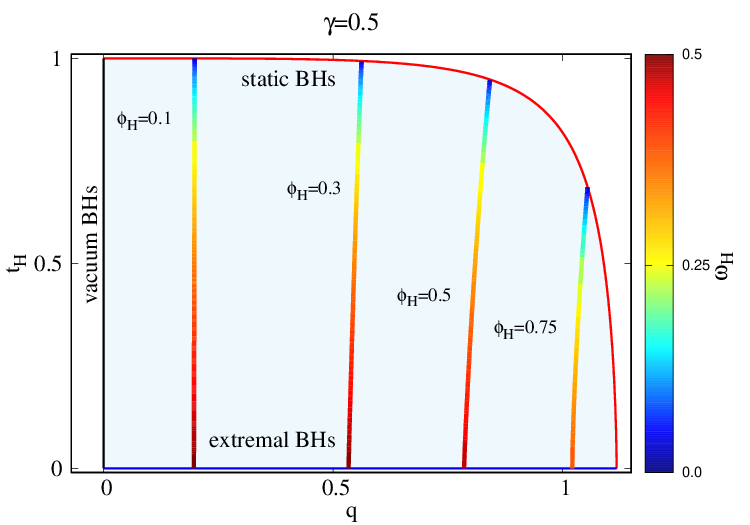}
		\end{subfigure}%
  } 
  \\
  \\
 \makebox[\linewidth][c]{%
		\begin{subfigure}[b]{8cm}
			\centering
\includegraphics[width=8cm]{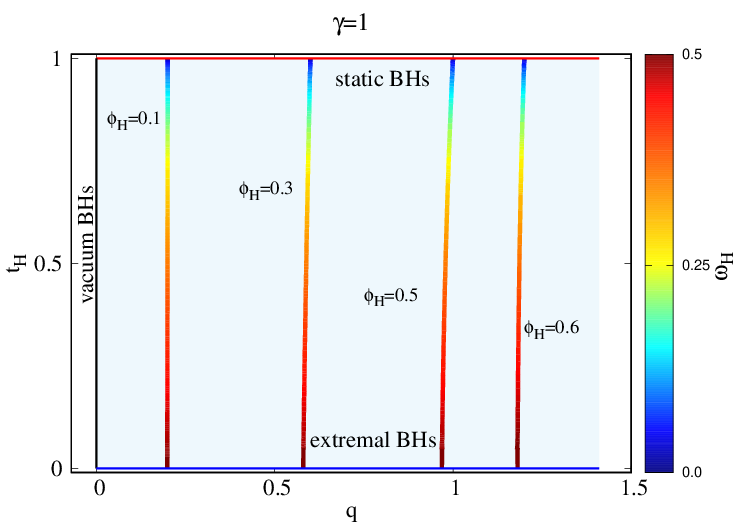}
		\end{subfigure}%
		\begin{subfigure}[b]{8cm}
			\centering
            \includegraphics[width=8cm]{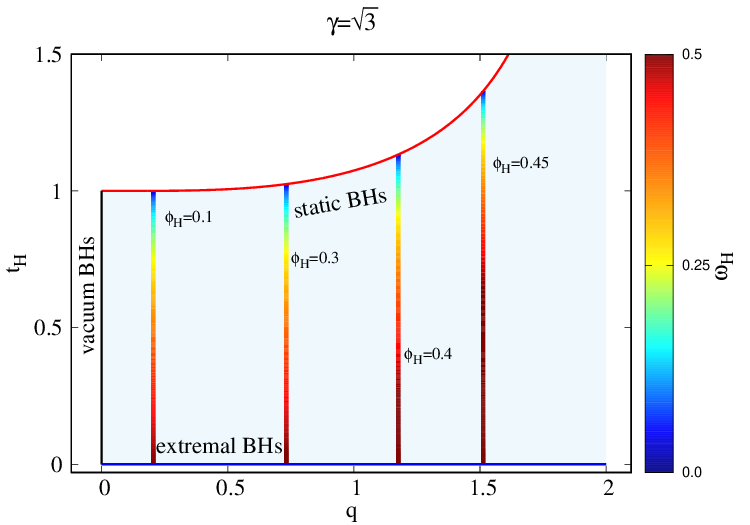}
		\end{subfigure}%
	}
 \\
	\caption{
	{\small
The domain of existence of solutions is shown in an
 electric charge-Hawking temperature  diagram
for several values of $\gamma$.
}
		\label{qtH}
  }
\end{figure}

\begin{figure}[h]
	\makebox[\linewidth][c]{%
		\begin{subfigure}[b]{8cm}
			\centering
\includegraphics[width=8cm]{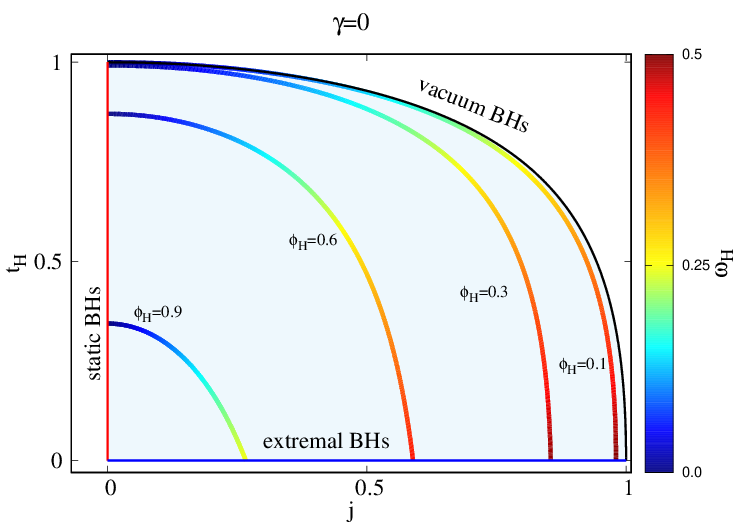}
		\end{subfigure}%
		\begin{subfigure}[b]{8cm}
			\centering
\includegraphics[width=8cm]{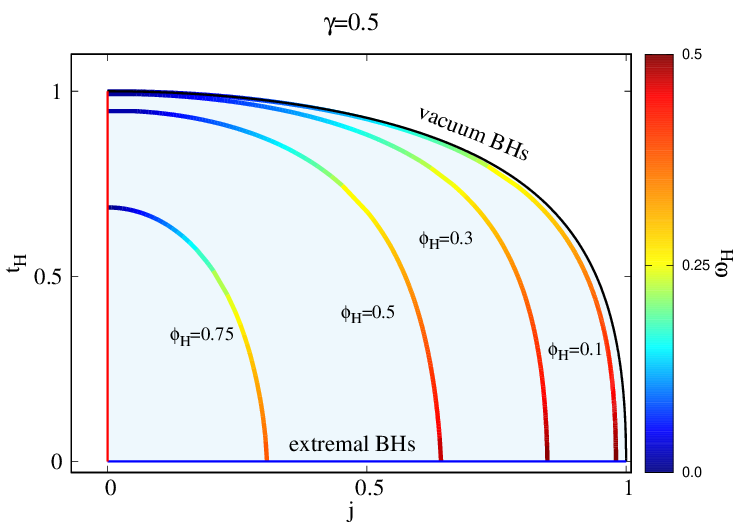}
		\end{subfigure}%
  } 
  \\
  \\
 \makebox[\linewidth][c]{%
		\begin{subfigure}[b]{8cm}
			\centering
\includegraphics[width=8cm]{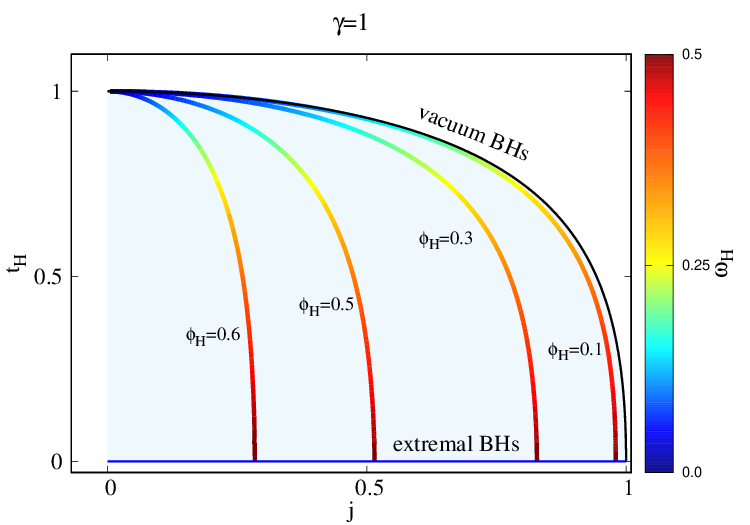}
		\end{subfigure}%
		\begin{subfigure}[b]{8cm}
			\centering
            \includegraphics[width=8cm]{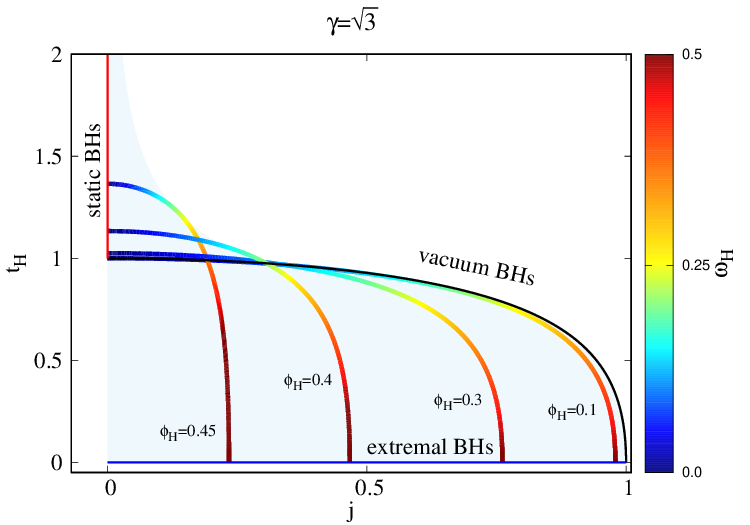} 
		\end{subfigure}%
	}
 \\
	\caption{
	{\small
Same as Figure  \ref{qtH} 
for an electric charge-Hawking temperature diagram.  
}
		\label{jtH}
  }
\end{figure}

The gyromagnetic ratio as defined in Eq. (\ref{gyro})
is shown in Figure \ref{gyr} 
as a function of (reduced)
angular momentum and electric charge
for solutions with $\gamma=1$ and $\sqrt{3}$
(a similar picture has been found for other values of $\gamma$).
Here and in Figures \ref{qj}-\ref{qj3},
 several isocurves for the electrical
chemical potential $\phi_{\mathcal{H}}$ are also shown, with the corresponding values for reduced
event horizon velocity
 shown as a color map\footnote{
For a better visualization, we restrict the range of $\omega_H$ in the color map to
 $\omega_H\leq 0.5$, although larger values
 are found in a small region close to the extremal/critical line.
}.
% and decreases with both $j$ and $q$. 
%
One can see that, 
in agreement with perturbation theory results,
the gyromagnetic ratio $g$ is smaller than $2$,
a value which is approached in the limit of small charge, only.
Also, $g$ decreases with both $j$ and $q$.

In Figure \ref{LeLp}
we show how
the ratio $L_e/L_p$  
varies
as a function of both  $j$ and $q$, for several values
of $\gamma$. 
Again, the results from the exact solution
appear  to capture the generic pattern.

In order to get a better idea on
how the dilaton coupling influences some quantities of interest,
we show
in Figure  \ref{var-gamma} 
the reduced area and temperature 
as a function of $\gamma$,
for several 
values of reduced angular momentum and a fixed 
reduced electric charge.
Based on the results there one can anticipate a KN-like behaviour of the typical
 the EMd solutions. 
As one can see,
$a_H$ and $t_H$ increase monotonically with
$\gamma$, although with relatively small changes  (at least for that value of $q=Q_e/M$).
At the same time, it seems that, for any $\gamma$,
the known behaviour of the KN solutions is recovered,
with both $a_H$ and $t_H$ decreasing with $j$.

%%%%%%%%%%%%%%%%%%%%%%%%%%%%%%%%%%%%%%%%%%%%%%%%%%%%%%
%%%%%%%%%%%%%%%%%%%%%%%%%%%%%%%%%%%%%%%%%%%%%%%% 
%%%%%%%%%%%%%%%%%%%%%%%%%%%%%%%%%%%%%%%%%%%%%%%%

%%%%%%%%%%%%%%%%%%%%%%%%%%%%%%%%%%%%%%%%%%%%%%%%%%%%%%
%%%%%%%%%%%%%%%%%%%%%%%%%%%%%%%%%%%%%%%%%%%%%%%% 

%%%%%%%%%%%%%%%%%%%%%%%%%%%%%%%%%%%%%%%%%%%%%%%%%%%%%%
%%%%%%%%%%%%%%%%%%%%%%%%%%%%%%%%%%%%%%%%%%%%%%%% 

%%%%%%%%%%%%%%%%%%
\subsection{The domain of existence and critical
behaviour}
%%%%%%%%%%%%%%%%%%%%%%%%%%%%%%%%%%%%%%%%% 

To better understand the solutions' properties
it is useful to 
consider their domain of existence.
This  is displayed in Figures 
 \ref{qj}-\ref{qj3},
 in terms of several reduced quantities, as defined 
 in Eq. (\ref{scale1}). 

In  Figure \ref{qj}  
some KN results are contrasted with those found for
EMd solutions with $\gamma=1$.
This value of the dilaton coupling constant is considered
as an illustrative case,
 %(see also Figure \ref{qj3}),
a (qualitatively) similar picture
being found for other $\gamma$.
As seen in those plots, the maximal allowed value of the angular momentum decreases with increasing
the electric charge, as expected.
We did not find any indication so far
for the violation of the Kerr bound  $j\leqslant 1$,
which appear to be satisfied for any $\gamma$.
However, as expected, the RN bound
for the electric charge
$q\leq 1$
is violated, the maximal value of
$q$ being approached in the static limit.
At the same time, the spinning BHs possess a nonzero event horizon area, 
whose maximal value is attained
by the Schwarzschild vacuum solution.

 An inspection of the diagrams
 shows that,
 for any $\gamma$,
 the domain of existence is delimited by:
 \begin{itemize}
\item  the set of static GMGHS BHs ($j=0$, red line); 
\item  the set of vacuum GR solutions -- the Kerr/Schwarzschild  BHs  ($Q_e=0$, black line).
 \end{itemize}

The remaining part of the boundary of the domain of existence is $\gamma$-dependent, 
with two different situations.

%%%%%%%%%%%%%%%%%%%%%%%%%%%%%%%%%%%%%%%%%%%%
\subsubsection{$\gamma\leq \sqrt{3}$
and extremal BHs}
%%%%%%%%%%%%%%%%%%%%%%%%%%%%%%%%%%%%%%%%% 

For values of the dilaton coupling constant up to 
the KK value, the remaining part 
 of the boundary of domain
 of existence of solutions is provided
 by
\begin{itemize}
\item  the set of extremal  BHs    (blue line).
\end{itemize}
These are limiting solutions with zero Hawking temperature\footnote{The spinning extremal BHs  have been found by directly solving 
the EMd field
equations.
There we have used
the same numerical scheme employed
 in the non-extremal case. 
The  metric Ansatz was still given by  (\ref{line_BH}),
this time, however,
with 
 $N(r)=(1-r_H/r)^2$, where $r_H\neq 0$ is an arbitrary constant;
 typically we set $r_h=0.25$),
 with 
the matter fields still given by (\ref{matter}).
 We have computed  extremal BH solutions for $\gamma= 
\{0,0.1, 0.2, 0.3, 0.4, 0.5, 1, 1.5 \} $  and  $\sqrt{3} $.}.
The existence of these  solutions
could perhaps be anticipated, based on
 known behaviour for $\gamma=0,\sqrt{3}$.
They correspond  to
 maximally spinning solutions 
 ($i.e.$ a maximal
$j$ for a given $q$).

Interestingly, 
for any $\gamma\leq \sqrt{3}$
the matter fields 
 as well as 
the Kretschmann
scalar remains finite as the zero temperature
limit is approached.
However, this does exclude the existence of
more subtle pathologies.
Indeed, we have found
that   
  the tidal forces as felt by a
timelike
observer infalling into the BH
diverges as the $extremal$ horizon is approached.
The tidal forces are
given by the components of the Riemann tensor in a frame associated with the ingoing
geodesics.
This confirms that there is a parallely propagated ($pp$) curvature singularity 
as we approach the extremality,
 the corresponding computation being given in Appendix \ref{tidal}.
This type of singularity is not rare,
other
examples of (non-vacuum) BHs with 
a  $pp$-singularity 
in near extremal limit
being reported in the literature,
see $e.g.$ 
\cite{Horowitz:1997uc,Horowitz:1997ed,Dias:2011at,Markeviciute:2018yal}  (also, this result is consistent with recent  literature indicating that extremal BHs with regular horizons are rare \cite{Horowitz:2022mly,Horowitz:2024kcx}).
Interestingly,
a study of the exact solution (\ref{KK})
shows that  this pathology
is absent for the  KK values
$\gamma=\sqrt{3}$,
with finite tidal forces in the zero-temperature limit. 

A partial
analytical understanding
of the issues encountered 
in the construction of
zero-temperature EMd BHs
with small $\gamma>0$ 
can be
achieved 
when instead of solving the full bulk equations searching for extremal solutions,
 one attempts to 
construct the corresponding 
near-horizon configurations.
 There one deals with a codimension one problem, whose solutions are easier to study.
 As discussed in Appendix  \ref{attractors},
 a $\gamma\neq 0$
 generalization of the 
 Bardeen-Horowitz
 \cite{Bardeen:1999px}
 solution
 (which corresponds to the near horizon limit
 of the extremal KN BH)
 can be found treating 
 $\gamma$ as a perturbation parameter.
 However,  no regular solution appears to exist,
 which suggests that the zero temperature bulk solutions
 may possess some pathologies.

Finally, let us remark that the extremality of a  BH  imposes a constraint  on the  global charges,
with
\begin{eqnarray}
&&
j^2+q^2=1 ~~{\rm for}~~\gamma=0,~~
{\rm and}~~
j=\frac{q(2\sqrt{1+2q^2}-q^2-2)^{3/2}}{(1-\sqrt{1+2q^2})^2}~~{\rm for}~~\gamma=\sqrt{3}.
\end{eqnarray} 
When considering
other values of $\gamma$, however, 
we have found no indication for the existence
of a simple relation between   the global charges of extremal BHs.

%%%%%%%%%%%%%%%%%%
\subsubsection{ $\gamma >\sqrt{3}$ and the issue of non-uniqueness}
%%%%%%%%%%%%%%%%%%%%%%%%%%%%%%%%%%%%%%%%% 

The situation is different for 
$\gamma>\sqrt{3}$,
in which case, we did not find any indication 
for the existence of spinning solutions 
with vanishing Hawking temperature.
 Instead, 
the  remaining part 
 of the boundary of domain
 of existence of solutions is provided by
 \begin{itemize}
\item  the set of limiting solutions (blue dotted line in the plots),
\end{itemize}
which do not appear to possess any special properties
(in our numerical scheme, they are approached for the maximal value of $\phi_\mathcal{H}$ for given $(\Omega_H, r_H)$).

%%%%%%
\begin{figure}[h] 
	\makebox[\linewidth][c]{%
		\begin{subfigure}[b]{8cm}
			\centering
\includegraphics[width=8cm]{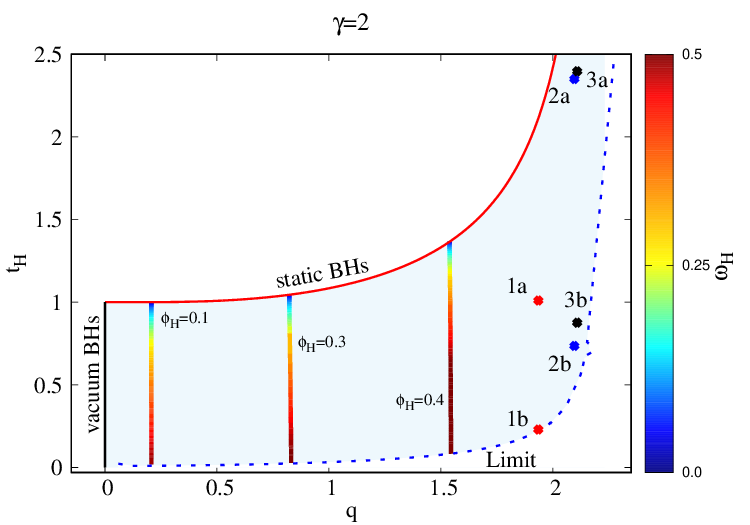}
		\end{subfigure}%
		\begin{subfigure}[b]{8cm}
			\centering 
 \includegraphics[width=8cm]{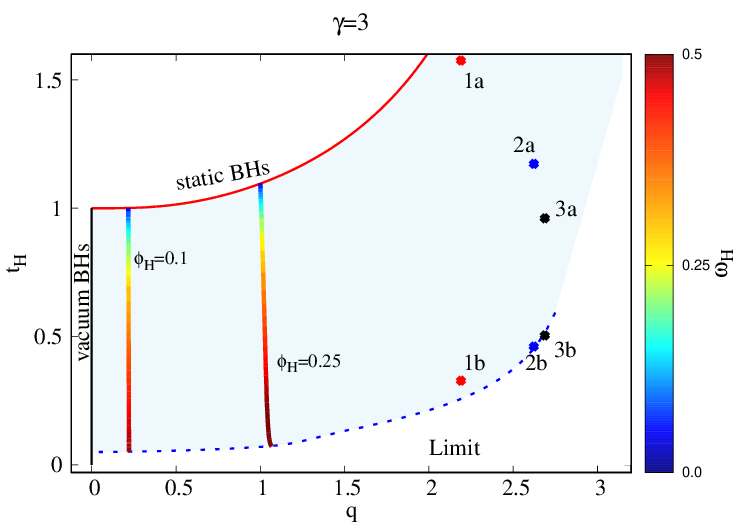}         %
		\end{subfigure}%
  } 
  \\
  \\
 \makebox[\linewidth][c]{%
		\begin{subfigure}[b]{8cm}
			\centering
\includegraphics[width=8cm]{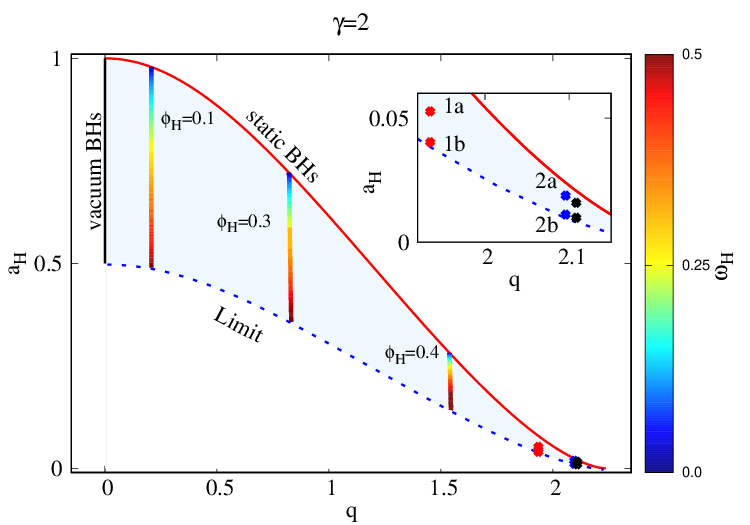}
		\end{subfigure}%
		\begin{subfigure}[b]{8cm}
			\centering
  \includegraphics[width=8cm]{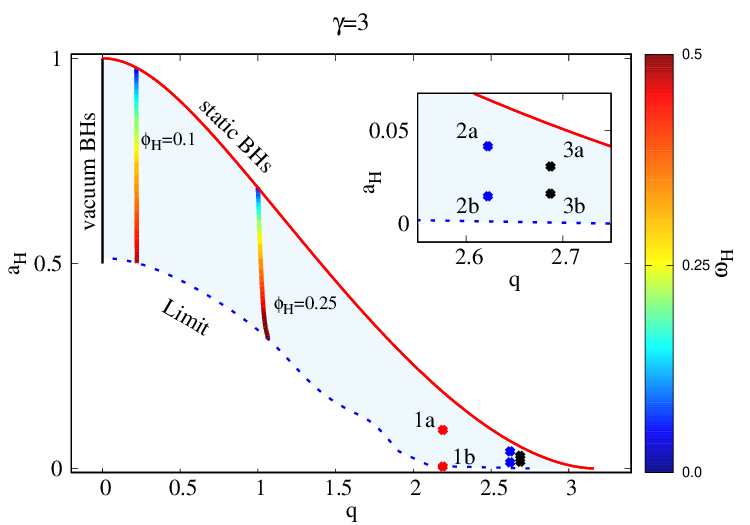}
		\end{subfigure}%
	}
  \vspace{-0.5cm}
\caption{ 
 The domain of existence of  
 solutions with $\gamma=2$
 and $\gamma=3$ is shown for Hawking temperature
 and horizon area as a function of (reduced)
 electric charge. Here and Figures  \ref{qjth3n}, \ref{qj3}, 
the dots stay for 
several representative pairs of solutions 
which possess the same global charges $M,J,Q_e$ but different horizon quantities.
}
\label{qjth3}
\end{figure}
%%%%%%%%%%%%%%%%%%%%%%%%%%%%%%%%%%%%%%%%%%

%%%%%%
\begin{figure}[h] 
	\makebox[\linewidth][c]{%
		\begin{subfigure}[b]{8cm}
			\centering
\includegraphics[width=8cm]{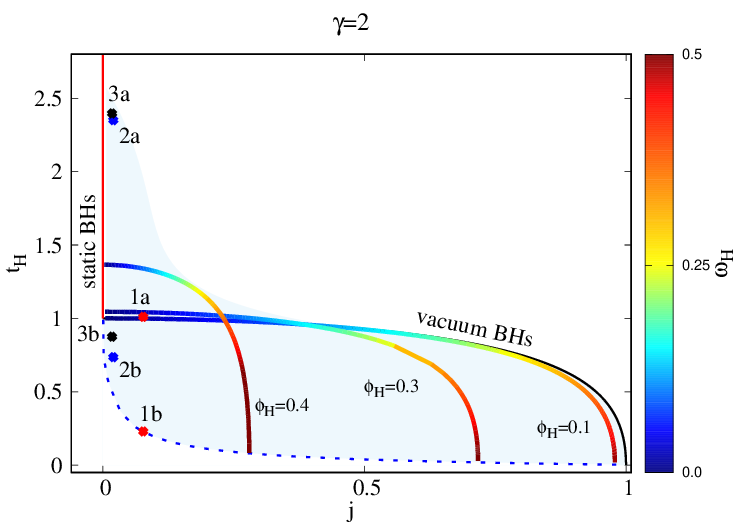}
		\end{subfigure}%
		\begin{subfigure}[b]{8cm}
			\centering 
 \includegraphics[width=8cm]{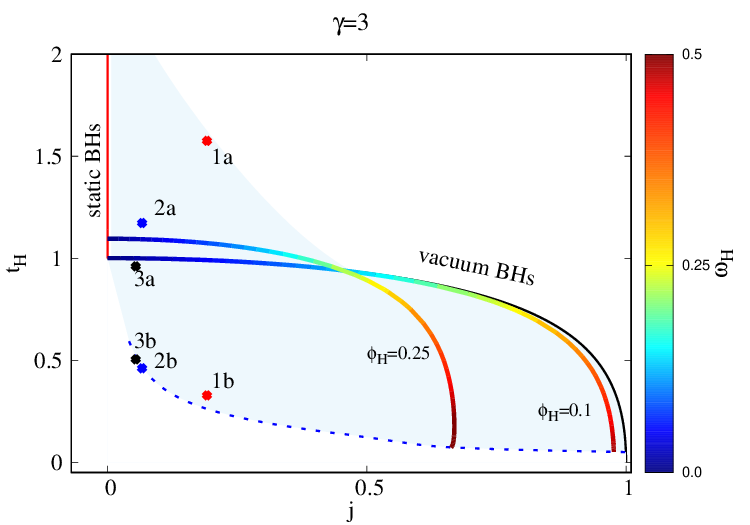}         %
		\end{subfigure}%
  } 
  \\
  \\
 \makebox[\linewidth][c]{%
		\begin{subfigure}[b]{8cm}
			\centering
\includegraphics[width=8cm]{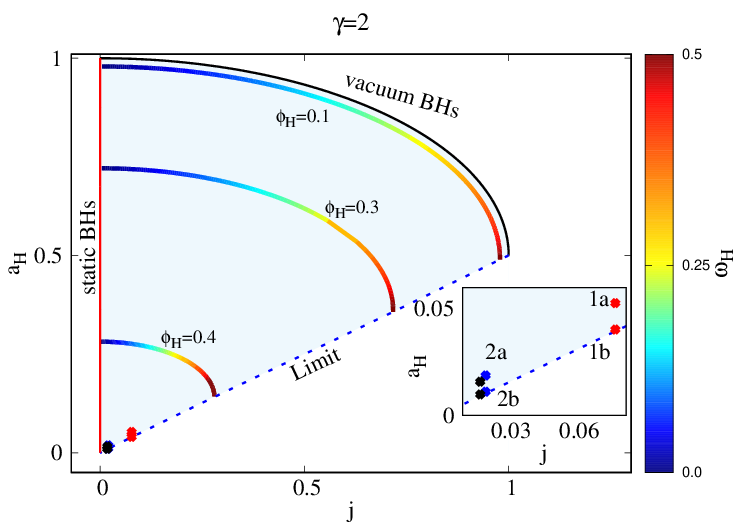}
		\end{subfigure}%
		\begin{subfigure}[b]{8cm}
			\centering
  \includegraphics[width=8cm]{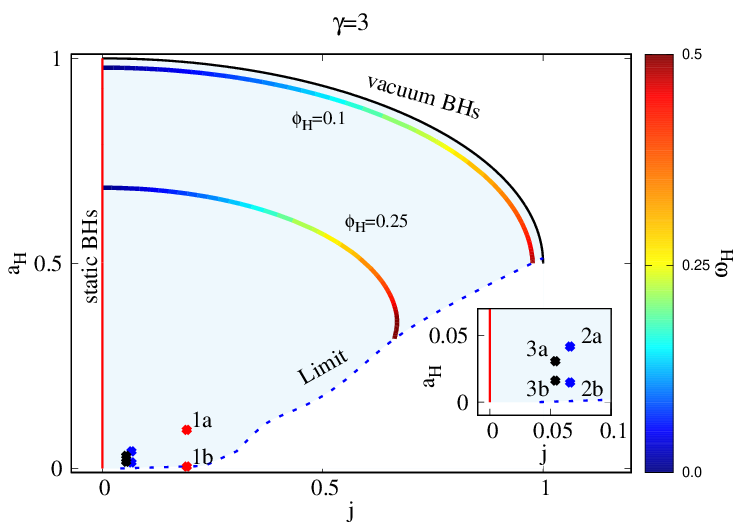}
		\end{subfigure}%
	}
  \vspace{-0.5cm}
\caption{ 
 The domain of existence of  
 solutions with $\gamma=2$
 and $\gamma=3$ is shown for Hawking temperature
 and horizon area as a function of (reduced)
  angular momentum.
}
\label{qjth3n}
\end{figure}
%%%%%%%%%%%%%%%%%%%%%%%%%%%%%%%%%%%%%%%%%%

\begin{figure}[h]
	\makebox[\linewidth][c]{%
		\begin{subfigure}[b]{8cm}
			\centering
\includegraphics[width=8cm]{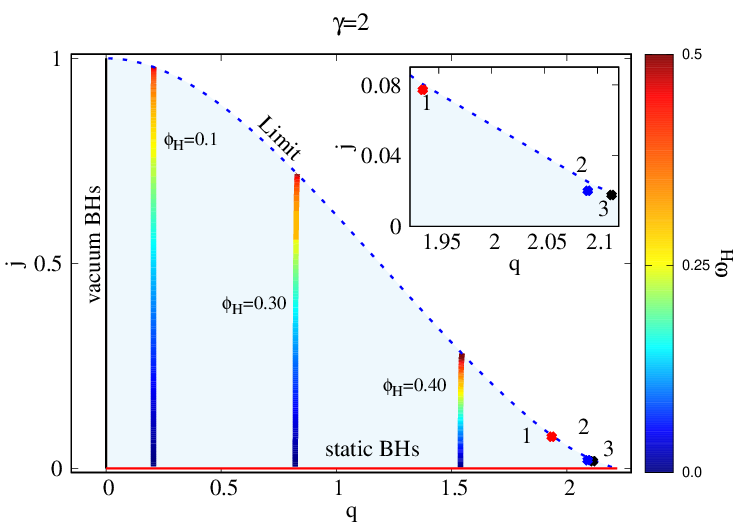}
		\end{subfigure}%
		\begin{subfigure}[b]{8cm}
			\centering
\includegraphics[width=8cm]{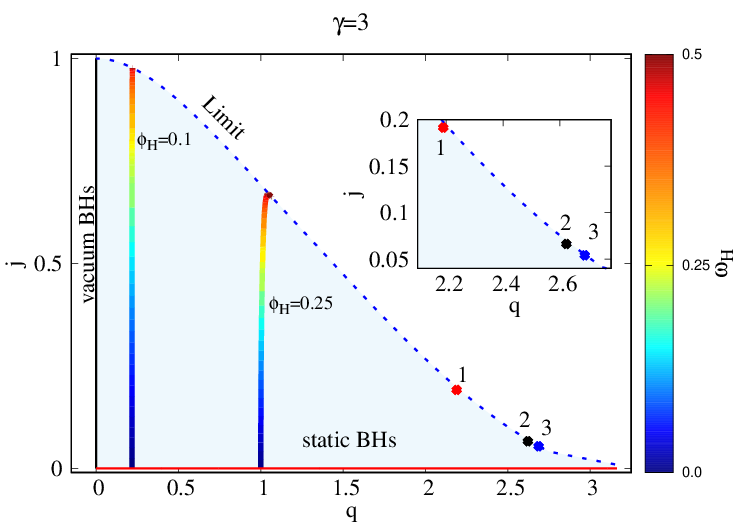}
		\end{subfigure}%
  } 
 \\
%  \vspace{-0.5cm}
\caption{  
Same as Figure 
\ref{qjth3n}
for an electric charge-angular momentum diagram.
}
\label{qj3}
\end{figure}

The diagrams for (reduced)
temperature and area are shown in Figures  
\ref{qjth3}, \ref{qjth3n}.
The results there correspond to  $\gamma=2,~3$;
however, we conjecture to hold for any  $\gamma>\sqrt{3}$.
As one can see, no zero temperature solutions 
 exist in this case, although the lower bound  
$t_H=1$,  found for  $j=0$, 
 is no longer valid in the presence of rotation.
 The corresponding electric charge-angular momentum
 diagram is shown in Figure \ref{qj3}. 

\begin{figure}[h]
	\makebox[\linewidth][c]{%
		\begin{subfigure}[b]{8cm}
			\centering
\includegraphics[width=8cm]{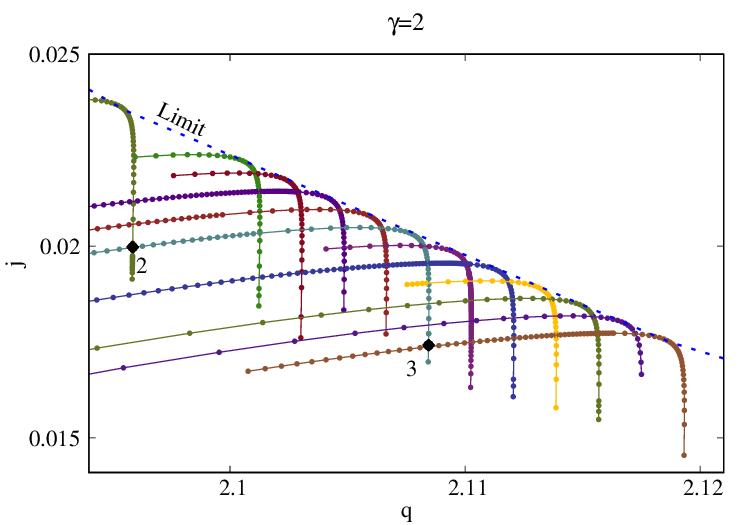}
		\end{subfigure}%
		\begin{subfigure}[b]{8cm}
			\centering
 \includegraphics[width=8cm]{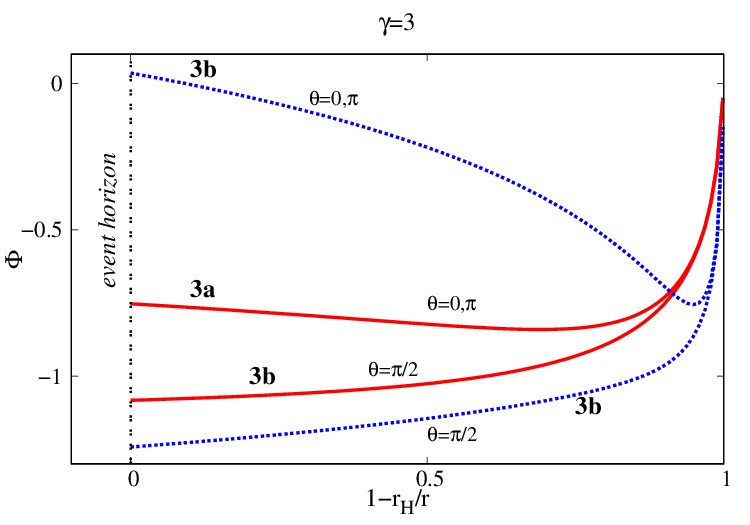}
		\end{subfigure}%
  } 
 \\
\caption{ 
{\it Left:} 
 Zoomed-in view of the left panel
 in Figure \ref{qj3}, showing multiple families of BH solutions with constant 
 values of the 
  electric potential at the horizon 
 $\phi_\mathcal{H}$
(differentiated by color).
 These curves exhibit
 turning points,
 with  a backbending towards a secondary branch;
 they also may intersect,
 which implies the non-uniqueness of solutions.  We also display the position of the solutions 2 and 3 shown in Figures  
 \ref{qjth3}-\ref{qj3}
 (left). {\it Right:} 
 The dilaton field at $\theta=0,\pi/2$
 is shown as a function of a compactified 
radial coordinate for two $\gamma=3$
solutions possessing the same global charges
(the point $3$ in the corresponding $(j,q)$-diagram  and
3a, 3b in Figures \ref{qjth3}-\ref{qjth3n}
 (right).)
}
\label{qj3n}
\end{figure}

The critical behaviour 
of EMd BHs with
$\gamma>\sqrt{3}$
is provided by the set of 
{\it singular solutions},
which is approached again
in the limit of maximal rotation.
These configurations
do not  possess
a regular horizon,
some metric functions and 
the
Kretschmann scalar 
diverging as $r\to r_H$.
As such, they
cannot be constructed directly
and are found by extrapolating the numerical results\footnote{This singular behaviour makes more difficult the accurate scanning 
of the domain of existence for 
the case
$\gamma>\sqrt{3}$,
the numerical difficulties increasing with
$\gamma$.}. In a $(q,j)$ diagram, the singular configurations are found
 very close to the limiting set, being approached
by a secondary branch of solutions\footnote{Note that the  secondary branch is  absent for smaller values of $\gamma$, in which case the families of
 solutions with constant $\phi_\mathcal{H}$
do not display a backbending,
ending in extremal configurations.} 
which possess a backbending in terms of $\phi_\mathcal{H}$,
see Figure \ref{qj3n} (left panel).

 The existence of this secondary branch appears to lead to an unexpected
 feature of the $\gamma={2}$, $3$ models: 
{\it $non-uniqueness$ of the solutions}\footnote{This feature is likely to occur for any $\gamma>\sqrt{3}$.
Moreover, we recall that Yazadjiev's uniqueness theorem in  Ref. \cite{Yazadjiev:2010bj}
holds for
$\gamma\leq \sqrt{3}$, only.
Also, 
as remarked there, 
``{\it signs for the existence of  non-uniqueness} (when $\gamma>\sqrt{3}$) "{\it seem to be found numerically in Ref. \cite{Kleihaus:2003df} }" 
(which deals, however, with rotating dyonic BHs).
}.
That is, two different configurations with the same
global charges $(M,J,Q_e)$
(or, equivalently, same reduced quantities $(j,q)$)
exist for a small region close to the critical set (which makes difficult their systematic study).
This feature can be seen 
in Figures \ref{qjth3}-\ref{qj3}
where the position of
several generic non-unique solutions is displayed.
That is, each dot in Figure \ref{qj3} corresponds to two \textit{different}
configurations with the same global charges.
Indeed, as seen in Figures \ref{qjth3}, \ref{qjth3n},
these  solutions (labeled now $a,b$)  possess different horizon area and different
temperature.
Moreover, we have verified that their (reduced)
quadrupole is also different\footnote{The computation of the quadrupole moment
%for the line-element (\ref{line_BH})
is similar to that discussed 
in Appendix A of Ref.  \cite{Herdeiro:2024pmv}}.

 %%%%%%%%%%%%%%%%%%%%%%%%
%%%%%%%%%%%%%%%%%%%%%%%%%%%%%%%%%%%%%%%%%%

 %%%%%%%%%%%%%%%%%%%%%%%%
%%%%%%%%%%%%%%%%%%%%%%%%%%%%%%%%%%%%%%%%%%

Further insight into this aspect can be found
in Figure  \ref{qj3n}  (right panel),
where we show
the profiles of the 
scalar field for two different $\gamma=3$ solutions
with the same global charges.
As one can see, the configuration 
$3b$ can be taken as an excited state,
the dilaton field
possessing 
two (symmetric) nodes on the $z$-axis ($\theta=0,\pi$).
The existence of a nodal structure
for the dilaton field
is a new feature, which is absent  for the exact solutions   
(\ref{GHS-solution1}),  (\ref{KK12}).
Furthermore, we conjecture
the existence of an infinite sequence
of excited solutions indexed by the node number
of the dilaton field.

We hope to return elsewhere with a systematic
study of these aspects.

%%%%%%%%%%%%%%%%%%%%%%%%%%%%%%%%%%%%% 
\section{Further remarks. Conclusions}
%%%%%%%%%%%%%%%%%%%%%%%%%%%%%%%%%%%%%% 
\label{Conclusions}

In this work we have constructed the spinning generalizations of the known static BHs
\cite{Gibbons:1987ps,Garfinkle:1990qj}
in the EMd model.
For a given value of the dilaton coupling constant $\gamma$,
this is a family of asymptotically 
flat, stationary, axially symmetric BHs,
that, 
in the generic case are non-singular on and outside an event horizon of spherical topology.
As with the KN BHs,
these solutions possess three global charges, 
the mass, angular momentum and electric charge,
the scalar hair being $secondary$.

While most  of the solutions' properties occur already in the well-known KN or KK cases
(where exact solutions are avaliable)
there are, however, several new features. Likely 
the most interesting aspect
revealed by our study
is that,
 for $0\leq \gamma \leq \sqrt{3}$,
the maximally
rotating solutions approach a zero temperature
extremal limit.
However, 
this limit features a  $pp$-singularity
(except for the KN and KK BHs), with 
infinite tidal forces
as measured by a freely infalling observer. 
The situation is different for 
$\gamma > \sqrt{3}$,
the critical configurations 
possessing a singular horizon.
Furthermore, an unexpected feature 
found for
$\gamma > \sqrt{3}$
is the non-uniqueness of solutions,
with (at least)
two different solutions 
%(and presumably an infinite set)
possessing the same global charges
$\{M,J,Q_e \}$.

In principle, most of the studies considered 
in the literature for the KN case
can be repeated for $\gamma \neq 0$, looking for
new feature introduced by the presence of
a dilaton field.
For example,
as possible avenues for future research we mention:
$i)$ the study of shadows (in particular for near-extremal configurations);
$ii)$  the study of geodesic motion and extreme mass ratio inspirals    and 
    $iii)$ 
the construction of solutions in the inner horizon region and clarification of the nature of pathologies there.

  Finally, it would be interesting to consider the dyonic
  generalization of the solutions in this work
  and clarify if the presence of a magnetic charge
  may cure the pathologies encountered
  in the extremal case.
  Another possibility
  would be to extend the model (\ref{action})
  by including a dilaton potential  or stringy $\alpha'$ corrections. 
  In the static case and $\gamma=1$,
  this has been shown 
\cite{Astefanesei:2019qsg,Herdeiro:2021gbw}
  to regularize the 
  GMGHS BH
  solution.
  Presumably, their spinning generalizations would also be regular.

%%%%%%%%%%%%%%%%%%%%%%%%%%%%%%%%%%%%%%%%%%%%%%%%%%%%%%%%%%%%%%%%%%%%%%%%%%%%%%%%%%%%%%%%%%%%%%%%%%%%%%
\section*{Acknowledgements}
%%%%%%%%%%%%%%%%%%%%%%%%%%%%%%%%%%%%%%%%%%%%%%%%%%%%%%%%%%%%%%%%%%%%%%%%%%%%%%%%%%%%%%%%%%%%%%%%%%%%%%

E.R. thanks K. Uzawa for useful discussions.
This work is supported by CIDMA under the FCT Multi-Annual Financing
Program for R\&D Units (UID/04106),
through the Portuguese Foundation for Science and Technology (FCT -- Fundaç\~ao para a Ci\^encia e a Tecnologia), as well as the projects: Horizon Europe staff exchange (SE) programme HORIZON-MSCA2021-SE-01 Grant No. NewFunFiCO-101086251;  2022.04560.PTDC (\url{https://doi.org/10.54499/2022.04560.PTDC}) and 2024.05617.CERN (\url{https://doi.org/10.54499/2024.05617.CERN}). E.S.C.F. is supported by the FCT grant PRT/BD/153349/2021 (\url{https://doi.org/10.54499/PRT/BD/153349/2021}) under
the IDPASC Doctoral Program. Computations have been performed at the Argus cluster at
the U. Aveiro.

%%%%%%%%%%%%%%%%%%%%%%%%%%%%%%%%%%%%%%%%%%%%%%%%%%%%%%%%%%%%%%%%%%
 \appendix

%%%%%%%%%%%%%%%%%%%%%%%%%%%%%%%%%%%%%%%
 \section{The   numerically solved equations}
 \label{equations}
%%%%%%%%%%%%%%%%%%%%%%%%%%%%%%%%%%%%%%%
\setcounter{equation}{0}
\renewcommand{\theequation}{A.\arabic{equation}}

The equations solved in practice
are a
suitable combinations of the Einstein equations (\ref{EME}) %
$
\{ 
E_\varphi^\varphi=0;~
E_t^t=0;~
E_\varphi^t=0
 \}
 $
 together with the 
 $(\varphi, t)$-Maxwell equations (\ref{ME}) and the Klein-Gordon 
equation
(\ref{DE})
such that each equation contains second derivatives for a single metric function.

%%%%%%%%%%%%%%%% ec F0 %%%%%%%%%%%
 \begin{eqnarray}
 &&
  r^2 \sin^2 \theta N
  \left(
  F_{0,rr}+\frac{1}{r^2 N}  F_{0,\theta \theta}
  \right)
  +(2r-r_H)\sin^2 \theta F_{0,r}
  +\sin\theta\cos\theta F_{0,\theta}
  \\
  \nonumber
  &&
  -\frac{1}{2 }e^{-4 F_0}r^4 \sin^4 \theta
  \left(
  W_{,r}^2+\frac{W_{,\theta}^2}{r^2 N}
  \right)
  -e^{2F_0-2\gamma \Phi}\sin^2 \theta
  \bigg[
\frac{A_\varphi \cot \theta}{r^2}
(2 A_{\varphi,\theta}+A_\varphi \cot \theta)
\\
\nonumber
&&
+N\left(
A_{\varphi,r}^2+\frac{1}{r^2 N}A_{\varphi,\theta}^2
\right)
+
e^{-4F_0}r^2
\big(
(A_{\varphi,r}-A_{\varphi}\sin\theta W_{,r})^2
+\frac{(A_{\varphi,\theta}-A_{\varphi}\sin\theta W_{,\theta})^2}{r^2N}
\big)
  \bigg]=0~,
 \end{eqnarray}
%
%%%%%%%%%%%%%%%% ec F1 %%%%%%%%%%%
 \begin{eqnarray}
 &&
  r^2 \sin^2 \theta N
  \left(
  F_{1,rr}+\frac{1}{r^2 N}  F_{1,\theta \theta}
  \right)
  +r^2 \sin^2 \theta N
  \left(
 F_{0,r}^2+ \frac{1}{r^2 N}  F_{0,\theta }^2
  \right)
  \\
  \nonumber
  &&
  -\frac{1}{2}\sin^2\theta 
  \bigg(
(2r-3r_H)F_{0,r}
-(2r- r_H)F_{1,r}
  \bigg)
  -\sin\theta \cos\theta F_{0,\theta} 
  \\
  \nonumber
  &&
  +\frac{1}{4}e^{-F_0}r^4 \sin^4 \theta
  \left(
w_{,r}^2+\frac{1}{r^2 N}W_{,\theta}^2
  \right)
  +
  r^2 \sin^2 \theta N
  \left(
\Phi_{,r}^2+\frac{1}{r^2 N}\Phi_{,\theta}^2
  \right)=0~,
  \end{eqnarray}
    %
 %%%%%%%%%%%%%%%% ec W %%%%%%%%%%%
  \begin{eqnarray}
 &&
  r^2 \sin^2 \theta N
  \left(
 W_{1,rr}+\frac{1}{r^2 N}  W_{1,\theta \theta}
  \right)
 +4 r \sin^2 \theta 
 (1-r F_{0,r})W_{,r} 
 +\frac{(3\cos\theta -4 \sin\theta F_{0,\theta}))}{N}W_{,\theta}
 \\
 \nonumber
 &&
 +4 e^{2F_0-2\gamma \Phi}\sin\theta
 \bigg[
-A_{\varphi,r}(A_{t,r}-\sin\theta A_{\varphi }W_{,r})
+
\frac{1}{r^2 N}
 (
A_{\varphi }\cot \theta +A_{\varphi,\theta}
 )
 (A_{\varphi }\sin\theta W_{,\theta}-
 A_{t,\theta})
 \bigg]=0~,
  \end{eqnarray}
%
 %%%%%%%%%%%%%%%% ec A_varphi %%%%%%%%%%%
  \begin{eqnarray}
 &&
 \nonumber
  r^2 \sin^2 \theta 
  \left(
 A_{\varphi,rr}+\frac{1}{r^2 N}  A_{\varphi,\theta \theta}
  \right)
  +r^2 \sin^2 \theta
  \left(
\frac{r_H}{r^2N}+2 F_{0,r}
  \right) A_{\varphi,r}
+(2 F_{0,\theta}+\cot \theta )\frac{\sin^2\theta}{N} A_{\varphi,\theta}
\\ 
&&
 -\frac{e^{-4F_0}r^4\sin^3 \theta}{N}
 \left(
A_{t,r}W_{ ,r}+\frac{1}{r^2N}A_{t\theta}W_{ ,\theta}
-A_{\varphi }\sin \theta 
(W_{,r}^2+\frac{1}{r^2 N}W_{,\theta}^2)
 \right)
 \\ &&
 \nonumber
 +2(\sin \theta \cos \theta  F_{0,\theta}-1)
 \frac{A_{\varphi}}{N}
 -2 \gamma \sin \theta
 \left(
r^2 \sin\theta A_{\varphi,r}\Phi_{\varphi,r}
+\frac{A_{\varphi \cos\theta+\sin\theta A_{\varphi,\theta \Phi_{ ,\theta}}}}{B}
 \right)=0~,
  \end{eqnarray}
%
 %%%%%%%%%%%%%%%% ec A_t %%%%%%%%%%%
  \begin{eqnarray}
 &&
 \nonumber
  r^2 \sin^2 \theta 
  \left(
 A_{t,rr}+\frac{1}{r^2 N}  A_{t,\theta \theta}
  \right)
  +\frac{\sin\theta \cos \theta  A_{t,\theta}}{N}
  +4 e^{2F_0 -2\gamma\Phi} A_{ \varphi}\sin^2\theta
  \bigg[
\frac{ A_{ \varphi}\cot \theta}{r^2N}
(\sin\theta A_{ \varphi}W_{,\theta}-A_{t,\theta})
\\
\nonumber
&&
- (A_{ \varphi,r}A_{t,r}+\frac{A_{ \varphi,\theta}A_{t,\theta}}{r^2N})
+A_{ \varphi } \sin\theta
(A_{ \varphi,r}W{,r}+\frac{A_{ \varphi,\theta}W_{,\theta}}{r^2N})
-2 r^2 \sin^2 \theta
(A_{ t,r}F{0,r}+\frac{A_{t,\theta}F_{0,\theta}}{r^2N})
\\
\nonumber
&&
+\frac{\sin^3 \theta}{N}
(A_{ \varphi }\cot \theta-A_{ \varphi,\theta})W_{ ,\theta} )
+2r \sin^2 \theta A_{ t,r}
+r \sin^3 \theta (2A_{ \varphi }-rA_{ \varphi,r})
W_{  ,r}
-2r^2 A_{ \varphi}\sin^3 \theta
(F_{ 0,r}W_{ ,r}
\\
%\nonumber
&&
+\frac{F_{ 0,\theta}W_{,\theta}}{r^2N})
+
2\gamma r^2 \sin^2 \theta
\left(
(-A_{ t,r}+A_{ \varphi }\sin\theta W_{ ,r})
\Phi_{,r}
+\frac{(-A_{ t,\theta}+A_{ \varphi }\sin\theta W_{ ,\theta})}{r^2N}\Phi_{,\theta}
\right)
  \bigg]=0~,
 \end{eqnarray}
%
 %%%%%%%%%%%%%%%% ec Phi %%%%%%%%%%%
  \begin{eqnarray}
 &&
% \nonumber
  r^2 \sin^2 \theta 
  \left(
 \Phi_{ ,rr}+\frac{1}{r^2 N}  \Phi_{ ,\theta \theta}
  \right)
  +\frac{2 r-r_H}{N}\sin^2\theta 
   \Phi_{ ,r }
   +\frac{\sin\theta \cos\theta \Phi_{ ,\theta}}{N}
   +e^{2F_0 -2\gamma \Phi}
   \gamma
   \bigg[
\sin^2 \theta A_{\varphi ,\theta}^2
\\
\nonumber
&&
+\frac{(A_{\varphi}\cos\theta +\sin\theta A_{\varphi,\theta})^2}{r^2N}
+
\frac{e^{-4F_0}r^2\sin^2\theta}{N}
\left(
(A_{t,r}-\sin\theta A_{\varphi}W_{ ,r})^2
+
\frac{ (A_{t,\theta}-\sin\theta A_{\varphi}W_{ ,\theta})^2}{r^2N }
\right)
   \bigg]=0~.
 \end{eqnarray}

Also,  the Einstein equations 
$E_\theta^r =0$ 
and
$E_r^r-E_\theta^\theta  =0$ 
are not solved directly\footnote{
The Einstein equation 
$E_r^r+E_\theta^\theta$
is identically zero
for the employed
 ansatz .
}, they
yielding two constraints  which are monitored in numerics.
These equations result in
 \begin{eqnarray}
 &&
 \nonumber
F_{ 0,r}^2-\frac{F_{ 0,\theta}^2}{r^2N}
-\frac{(2r-3r_H)F_{ 0,r}}{2r^2N}
-\frac{(2r-r_H)F_{ 1,r}}{2r^2N}
+\frac{\cot \theta}{ r^2N}
(F_{ 0,\theta}+F_{ 1,\theta})
-\frac{e^{-4F_0}r^2 \sin^2 \theta}{4N}
(W_{, r}^2-\frac{W_{,\theta}^2}{r^2N})
\\
%\nonumber
&&
+ \Phi_{, r}^2-\frac{\Phi_{ , \theta}^2}{r^2N} 
+\frac{e^{2F_0-2\gamma \Phi}}{r^2}
\bigg[
A_{\varphi, r}^2-\frac{(A_{\varphi } \cot \theta +A_{\varphi,\theta})^2}{r^2N}
\\
\nonumber
&&
-\frac{e^{-4F_0 }r^2}{N}
\bigg(
(A_{t, r} -A_{\varphi} \sin \theta W_{,r})^2
 -\frac{(A_{t, \theta} -A_{\varphi} \sin \theta W_{,\theta})^2}{r^2N}
\bigg)
\bigg]=0~,
 \end{eqnarray}
 \begin{eqnarray}
 &&
 \nonumber
 r \cot \theta (F_{0,r}+F_{1,r})
 +
\frac{1}{2r N}
\left(
(2r -3r_H)F_{0,\theta}
+(2r - r_H)F_{1,\theta}
\right)
-2r F_{0,r}  F_{0,\theta}
+\frac{e^{-4 F_0}r^3 \sin^2 \theta W_{ ,r}W_{ ,\theta}}{2N}
\\
&&
-2\bigg[
-r \Phi{,r}   \Phi_{ \theta}
+
e^{2F_0-2\gamma \Phi}
\bigg(
-(A_{\varphi,\theta}+A_{\varphi }\cot \theta)
\frac{A_{\varphi,r}}{r}
+\frac{e^{-4F_0 }r}{N}
\big(
A_{t,r}A_{t,\theta}
\\
\nonumber
&&
+\sin^2 \theta A_{\varphi }^2 W_{,r}W_{ ,\theta}
-\sin  \theta A_{\varphi }
(A_{t,r}W_{,\theta}+A_{t,\theta}W_{,r})
\big)
\bigg)
\bigg]=0~.
 \end{eqnarray}

%%%%%%%%%%%%%%%%%%%%%%%%%%%%%%%%%%%%%%%%%%%%%%%%%%%%%%%%%%%%%%%%%%%%%%%%%%%%%
\section{The KK BHs:
analytical $vs.$ numerical results} 
\label{appenB}

\setcounter{equation}{0}
\renewcommand{\theequation}{B.\arabic{equation}}

When $\gamma = \sqrt{3}$, the EMd model
admits the well-known KK exact solution. In this Appendix, we present the explicit form of this solution in the coordinate system
used in numerics, together with 
a discussion of the quality of the found
numerical results in this case.

When $\gamma = \sqrt{3}$, the metric functions $F_0$, $F_1$ and $W$ 
in the metric Ansatz (\ref{line_BH})
take the following expressions:
\begin{equation}
\begin{aligned} 
& e^{2F_1(r,\theta)} \;=\; \frac{\Sigma}{r^2}\,\sqrt{\,1 \;+\; \frac{(r_H+2b)\,(b+r)\,v^2}{(1-v^2)\,\Sigma}\,}, \\[6pt]
& e^{2F_0(r,\theta) }\;=\; e^{2F_1}(1-v^2)\Biggl\{
    \Bigl[\bigl(1+\tfrac{b}{r}\bigr)^2 + \tfrac{b\,(b+r_H)}{r^2}\Bigr]^2
    \;-\; b\,(b+r_H)\,\Bigl(1-\tfrac{r_H}{r}\Bigr)\,\frac{\sin^2\theta}{r^2}
    \;+\; \tfrac{(r-r_H)\,\Sigma}{r^3}\,v^2
\Biggr\}^{-1}, 
\\[6pt]
%\nonumber
& W(r,\theta) \;=\; e^{-2\,[\,F_1(r,\theta)\,-\,F_0(r,\theta)\,]}\,\sqrt{b\,(b+r_H)}\,(r_H+2b)\,\frac{\bigl(1+\tfrac{b}{r}\bigr)}{r^3\,\sqrt{1-v^2}}\,, 
\end{aligned}
\label{eq:KK-metric}
\end{equation}
with 
$\Sigma(r,\theta)=\bigl(r+b\bigr)^2 \;+\; b\,(b+r_H)\,\cos^2{\theta} $,
while the matter fields are
\begin{equation}
\begin{aligned}
%\\[6pt]
&\mathcal{A}=\frac{r v (2 b+r_H)}{2 \left(v^2 (2 b r+r r_H-\Sigma )+\Sigma \right)}\,dt + \frac{\sqrt{b} r v \left(v^2-1\right) (2 b+r_H) \sin ^2(\theta ) \sqrt{\frac{b+r_H}{1-v^2}}}{2 \left(v^2 (2 b r+r r_H-\Sigma )+\Sigma \right)}\, d\varphi\,, \\[6pt]
&\Phi=-\frac{1}{4} \sqrt{3} \log \left(1-\frac{r v^2 (2 b+r_H)}{\Sigma  \left(v^2-1\right)}\right)\,.
\end{aligned}
\label{eq:KK-metric1}
\end{equation}

Here $r_H$ and $b$ are real positive constants that, together with the 
dimensionless parameter $v\in[0,1)$, fully specify a family of solutions. 
Note that setting $v=0$ continuously recovers the (vacuum) Kerr metric.

In standard Boyer--Lindquist coordinates 
\((t,R,\theta,\varphi)\), the 
(non-extremal) KK BH
(as given by (\ref{KK}), (\ref{KK12})
with $r$ there replaced by $R$)
is specified by two parameters: 
the mass parameter $\mu$ and the rotation parameter $a$.
%with \(\,J = a\,m\,\). 
The event horizon is located at 
\(R = R_H \equiv \mu + \sqrt{\mu^2 - a^2}\). 
Also, the radial coordinate in (\ref{line_BH})
\(r\) is related to $R$ by a shift
\begin{equation}
r \;=\; R \;-\; \frac{a^2}{R_H}\,.
\label{eq:shift-r}
\end{equation}

%%%%%%%%%%%%%%%%%%%%%%%%
\begin{figure}[ht!]
\centering
\includegraphics[height=.23\textheight]{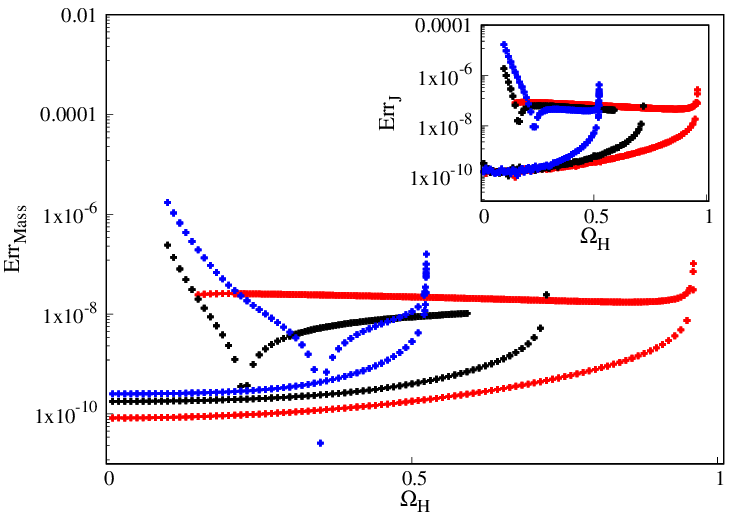}
\includegraphics[height=.23\textheight]{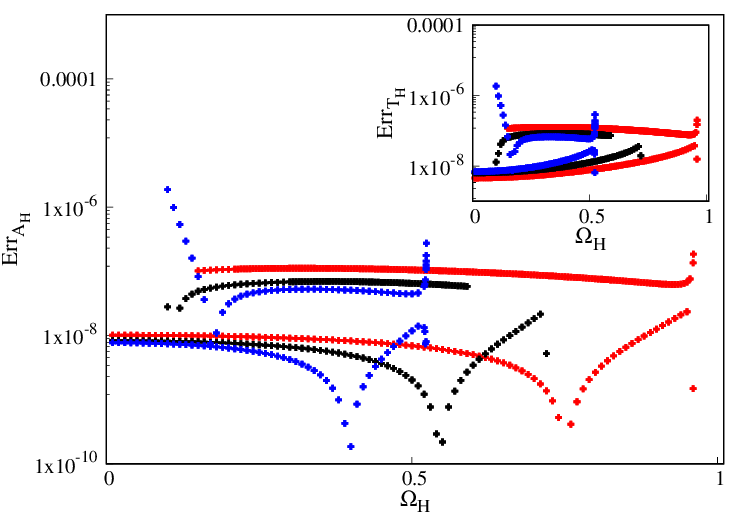}
\caption{%
A comparison between the numerically constructed KK 
solutions and  the closed-form expression 
(\ref{eq:KK-quantities}).
The plots here show $\mathrm{Err}_M$ ($\mathrm{Err}_J$ in the inset) (left) and $\mathrm{Err}_{A_H}$ ($\mathrm{Err}_{T_H}$ in the inset)(right) 
as a function of the horizon velocity, keeping the others 
input parameters fixed
($r_H=0.25$, $\phi_{\mathcal{H}}=0.45$ (blue),
 $\phi_{\mathcal{H}}=0.4$ (black) and  $\phi_{\mathcal{H}}=0.3$ (red)).
The relative errors
are computed according to (\ref{err}).
}
\label{errorKK}
\end{figure}
%%%%%%%%%%%%%%%%%%%%%

In terms of the new parameters $r_H$ and
$b=2a^2/R_H$,  the  main   quantities 
of interest are
\begin{equation}
\begin{aligned}
M \;&=\; \frac{1}{2}\,\bigl(r_H + 2b\bigr)\,\bigl(1 + \tfrac{v^2}{2\,(1-v^2)}\bigr)\,, 
~
\label{eq:KK-quantities}
J \;=\; \tfrac12\,\frac{\sqrt{\,b\,(b+r_H)\,}\,\bigl(r_H+2b\bigr)}{\sqrt{\,1-v^2\,}}\,, 
~
Q_e \;=\; \frac{v}{2\,(1-v^2)}\,(r_H + 2b)\,, 
\\[6pt]
\Omega_H \;&=\;\sqrt{\frac{1-v^2}{b+r_H}}\,\frac{\sqrt{\,b\,}}{r_H+2b}\,,
\quad
\mu_m \;=\; -\,v\,J\,, 
\quad
\phi_{\mathcal{H}} \;=\; \frac{v}{2}\,,
\\[6pt]
A_H \;&=\;\frac{4\pi\,(r_H + b)\,(r_H+2b)}{\sqrt{1-v^2}}\,, 
\qquad
T_H \;=\;\frac{r_H\,\sqrt{1-v^2}}{4\pi\,(r_H+b)\,(r_H+2b)}\,.
\end{aligned} 
\end{equation}

We have constructed numerically a set of around 700
solutions with  $\gamma = \sqrt{3}$ 
 and consider their validation against the 
 analytical expressions  
 \eqref{eq:KK-metric},
 \eqref{eq:KK-metric1},
\ref{eq:KK-quantities}.
 This can be done, for example, by computing the relative difference for some quantity $Q$
\begin{equation}
\label{err}
    \mathrm{Err}_Q \;=\; 
    \biggl\lvert 1 \;-\; \frac{Q_{\mathrm{numerical}}}{Q_{\mathrm{analytical}}}\biggr\rvert,
\end{equation}
Typical errors are illustrated in  Fig.~\ref{errorKK}, 
for $Q=\{M, J,A_H, T_H \}$.
As one can see, the numerical and analytical solutions agree to within the expected precision across a range of parameters.

%%%%%%%%%%%%%%%%%
%%%%%%%%%%%%%%%%%%%%%%
 \section{Tidal forces and a pathology of the
 $\gamma=1$ extremal BHs}
 \label{tidal}
%%%%%%%%%%%%%%%%%%%%%%%%%%%%%%%%%%%%%%%
\setcounter{equation}{0}
\renewcommand{\theequation}{C.\arabic{equation}}

%%%%%%%%%%%%%%%%%%%%%%%%%%%%%%%%%%%%%%%%%%%%%%%%%%%%%%%%%%%%%%%%%%%%%
\begin{figure}[ht!]
\begin{center}  
\includegraphics[height=.28\textheight, angle =0]{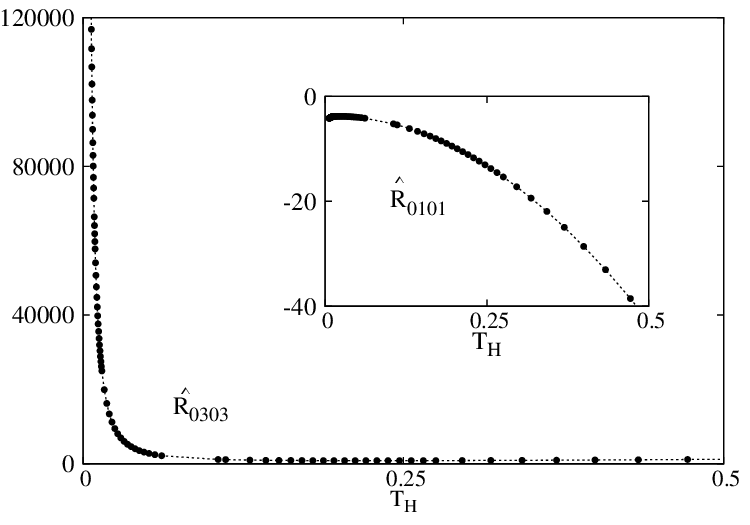} 
\end{center}
  \vspace{-0.5cm}
\caption{ 
 The components 
 $\hat{R}_{0101}$
 and
$\hat{R}_{0303}$
 of the Riemann tensor in the PPON frame
 are shown as a function of  Hawking 
 temperature for a family of 
 $\gamma=1$ EMd BHs with
 $\Omega_H=0.87$
 and 
  $\phi_{\mathcal{H}}=0.3$.
}
\label{pathology}
\end{figure}
%%%%%%%%%%%%%%%%%%%%%%%%%%%%%%%%%%%%%%%%%%

As originally found in Ref. \cite{Horowitz:1997uc},
the 
static charged, near extremal BHs
in EMd model
can exhibit 
diverging tidal forces as measured by a freely infalling observer, 
while having all curvature scalars finite at the horizon.

 In order to test whether the near extremal spinning solutions also possess 
a tidal force singularity, we need to analyse the geodesic motion in these backgrounds. 
We start by considering the metric Ansatz
 as given  by (\ref{line_BH}), and look for radial, 
timelike ingoing geodesics parametrised by the proper time $\tau$
 and with the tangent vector $\dot{x}^\mu=\mathrm{d}x^\mu/\mathrm{d}\tau$
 (with $x^0=t,x^1=r,x^2=\theta$ and $x^3=\varphi$). 
The Killing vector fields $\partial_t$  and $\partial_\varphi$ 
give us two conserved quantities
\begin{eqnarray}
E=-g_{t \mu}\dot{x}^\mu, \qquad L =g_{\varphi \mu}\dot{x}^\mu.
\end{eqnarray}
In what follows we consider radial static geodesics  in the equatorial plane
($i.e.$ $\dot{\theta}=0$), and $L=0$. 
Using the normalization condition ${\dot{x}^\mu\dot{x}_\mu=-1}$ we obtain
\begin{eqnarray}
\label{eq-geodesic}
\dot{r}=e^{-F_1}\sqrt{e^{-2F_0}E^2-N},
~~
\dot{t} = \frac{e^{-2F_0}E}{N},~~
\dot{\varphi}=\dot{t}W .
\end{eqnarray}

In order to compute the curvature measured by a freely 
falling observer along the radial timelike ingoing geodesic, 
we change into a parallely propagated orthonormal frame (PPON). 
In the PPON, we require $(\hat{e}_0)_a=\dot{x}_a$.
This results in
\begin{eqnarray}
\nonumber
&&
(\hat{e}_0)_\mu=
\frac{e^{ F_1}\sqrt{e^{-2F_0}E^2-N}}{N}\partial_\mu r
-
 E  \partial_\mu t,
~~
(\hat{e}_1)_\mu=
\frac{e^{-F_0+F_1} E}{N}\partial_\mu r
-\sqrt{E^2-e^{ 2F_0}N }\partial_\mu t,~~
\\
\nonumber
&&
(\hat{e}_2)_\mu= 
e^{F_1} r \partial_\mu \theta,~~
(\hat{e}_3)_\mu= 
e^{-F_0} r\sin \theta \partial_\mu \varphi
-e^{-F_0} r W \sin \theta \partial_\mu t,
\end{eqnarray}
which satisfies the orthonormality condition
\begin{equation}
\nonumber
g^{\mu \nu }(\hat{e}_a)_\mu 
(\hat{e}_b)_ \nu=\eta_{a b}\,.
\end{equation}
 The components of the Riemann tensor in the PPON frame are related to the components in the coordinate frame by
\begin{equation}
\hat{R}_{abcd}=R_{\alpha\beta\gamma\delta}(\tilde{e}_a)^\alpha
(\tilde{e}_b)^\beta
(\tilde{e}_c)^\gamma
(\tilde{e}_d)^\delta~.
\end{equation}
As one can see in Fig. \ref{pathology}, 
the components $\hat{R}_{0303}$
is diverging as the extremal limit is approached
(this holds as well for 
$\hat{R}_{0202}$), therefore exhibiting a parallely propagated curvature singularity\footnote{It is interesting to remark that not 
all components of  $\hat{R}_{abcd}$
are diverging, see $e.g.$ the inset in Fig. \ref{pathology}.}
(the results in that figure are for a particle  initially at rest
at asymptotic infinity $i.e.$ $E=1$ in (\ref{eq-geodesic})).

%%%%%%%%%%%%%%%%%%%%%%%%%%%%%%%%%%%%%%%
 \section{The near-horizon extremal solutions:
 a perturbative result}
 \label{attractors}
%%%%%%%%%%%%%%%%%%%%%%%%%%%%%%%%%%%%%%%
\setcounter{equation}{0}
\renewcommand{\theequation}{D.\arabic{equation}}
 
Following $e.g.$~\cite{Bardeen:1999px,Astefanesei:2006dd},
one considers the following line element
 with an isometry group $SO(2,1)\times U(1)$:
\begin{eqnarray}
\label{metric-nh}
 ds^2=v_1(\theta)  \left ( -r^2 dt^2+\frac{dr^2}{r^2}+\beta^2 d\theta^2 \right)
 + v_2 (\theta)\sin^2(\theta) \left(d\varphi- K r dt \right)^2 ,
\end{eqnarray}
together with the matter fields
\begin{eqnarray} 
A=A_\varphi (\theta) (d\varphi- K r dt)+e r dt,~~\Phi \equiv \phi(\theta)~.
\end{eqnarray}   
In these relations,
$0\leq r<\infty$, while $\theta,\varphi$ and $t$ have the usual range;
also, $\beta,~K,~e$
 are real parameters.
 
The above Ansatz would
 describe the
neighbourhood of the event horizon of an extremal EMd BH 
(and will 
be an attractor for the full bulk solutions
\cite{Astefanesei:2006dd}).
However, the metric functions
$v_1$
and 
$v_2$
are not independent and
one can show that the Einstein equations imply the simple relation
\begin{eqnarray}
v_2(\theta)=\frac{U^2}{v_1(\theta)},~~
{\rm and}~~\beta=1,
\end{eqnarray}
with $U$ a constant.

 Also, the  equations satisfied by $v_1$,
  $A_\varphi$ and $\phi$
  can be derived from the reduced Lagrangian
\begin{eqnarray} 
\label{Leff}
\nonumber
{\cal L}_{eff}=
-\frac{\sin \theta v_1'^2}{2v_1^2}
+\frac{2\cos \theta v_1'}{v_1}
+\frac{U^2 K^2 \sin^3 \theta}{2v_1^2}
-2\sin \theta
\bigg(
\phi'^2
+e^{-2\gamma \phi}
v_1 (\frac{A_\varphi'^2}{U^2   \sin^2 \theta} -\frac{(K A_\varphi-e)^2}{v_1^2})
\bigg)~,
\end{eqnarray} 
where a prime denotes a derivative $w.r.t.$ the angular variable $\theta$.

The solution for $\gamma=0$ 
corresponds to the near horizon limit
of the extremal KN BH and
reads
\cite{Bardeen:1999px}
 \begin{eqnarray} 
\label{KN1n}
v_1(\theta)= r_0^2\left(1-\frac{a^2}{r_0^2} \sin^2 \theta\right),~~
A_\varphi(\theta)= \frac{1}{1-\frac{a^2}{r_0^2} \sin^2 \theta}  
  \frac{ q Q_e M}{r_0^2}\sin^2 \theta~,~~\phi=0~,
\end{eqnarray} 
in terms of two parameters $Q_e $ and $a$.
Also,
\begin{eqnarray} 
\label{KN2}
e = \frac{ Q_e^3  }{ r_0^2},~~ 
U =r_0^2,~~K=\frac{2aM}{r_0^2}~,
~~{\rm with}~~
 r_0=\sqrt{2a^2+Q_e^2 },~~
    M=\sqrt{ a^2+Q_e^2 }.
\end{eqnarray} 
In the above expressions,
 $Q_e$ is the electric charge
and $a=J/ M$. 

One can  try to construct a perturbative solution in $\gamma$
around the KN attractor.
There one writes
\begin{eqnarray} 
&&
\nonumber
\phi (\theta)=\gamma \phi_1 (\theta)+\dots,~~
v_1 (\theta)=v_{KN}(\theta)+\gamma v_{11} (\theta)+\dots,~~
A_\varphi (\theta)=A_{\varphi,KN}(\theta)+\gamma A_{\varphi,1} (\theta)+\dots,~~
\\
\label{pert}
&&
U=U_{KN}+\gamma U_1+\dots,~~
e=e_{KN}+\gamma e_1+\dots,
\end{eqnarray} 
the $\gamma=0$ case (index $KN$) corresponding to  
(\ref{KN1n}).
Then a straightforward computation leads to
\begin{eqnarray} 
&& 
\nonumber
 \phi_1 (\theta)=
 \frac{Q_e^2}{Q_e^2+2a^2}
 \phi_{10}
 + \phi_{11}\log (\tan\frac{\theta}{2}) 
 -\frac{Q_e^2+ a^2}{Q_e^2+a^2
 +a^2 \cos^2 \theta}
 \\
 &&
 {~~~~~~}
 +\frac{Q_e^2 }{2(Q_e^2+2a^2)}
 \log (\frac{ Q_e^2+a^2
 +a^2 \cos^2 \theta}{\sin^2 \theta})
\end{eqnarray} 
which, for any choice of the integration 
constants $\phi_{10},\phi_{11}$,  is divergent at $\theta=0$
and/or $\theta=\pi$.
This suggest that the bulk (extremal) solution with small $\gamma$
would possess some pathologies.

One may argue that this is an artifact of the first order perturbation theory. 
However, 
at least for $\gamma=1$ (the only value we have considered more systematic in this context),
we have failed
to find nonperturbative (numerical) attractor solutions.

For completeness, we include here the near-horizon solution for
the KK value $\gamma=\sqrt{3}$.
 The solution in this case is more complicated than 
 in the pure EM case, with
\begin{eqnarray} 
&&
\nonumber
v_1(\theta)=a^2 (1+ \cos^2 \theta) \sqrt{1+\frac{2x^2}{(1-x^2) (1+\cos^2 \theta)}},~~
A_\varphi(\theta)=-a\frac{ x \sqrt{1-x^2} \sin^2 \theta}{1+x^2+(1-x^2) \cos^2 \theta},
\\
&&
\phi(\theta)=-\frac{\sqrt{3}}{4} \log\left( 1+\frac{2x^2}{(1-x^2) (1+\cos^2 \theta)}\right),
\end{eqnarray} 
and
\begin{eqnarray} 
q=0,~~U =\frac{2a^2}{\sqrt{1-x^2}}~,
\end{eqnarray} 
again in terms of two free parameters $a$ and $0\leq x\leq 1$.
 The fact that 
 no pathologies are present for
 $\gamma=\sqrt{3}$ 
 shows the limitation
 of the perturbative approach. 
Also, note that 
different from the $\gamma=0$
case,  the static limit is  singular for $\gamma=\sqrt{3}$.

%%%%%%%%%%%%%%%%%%%%%%%%%%%%%%%%%%%%%%%%%%%%%%%%%%%%%%%%%

\bibliographystyle{JHEP}
\bibliography{biblio}

@article{Markeviciute:2018yal,
    author = "Markeviciute, Julija and Santos, Jorge E.",
    title = "{Evidence for the existence of a novel class of supersymmetric black holes with AdS$_5\times$S$^5$ asymptotics}",
    eprint = "1806.01849",
    archivePrefix = "arXiv",
    primaryClass = "hep-th",
    doi = "10.1088/1361-6382/aaf680",
    journal = "Class. Quant. Grav.",
    volume = "36",
    number = "2",
    pages = "02LT01",
    year = "2019"
}

@article{Astefanesei:2006dd,
    author = "Astefanesei, Dumitru and Goldstein, Kevin and Jena, Rudra P. and Sen, Ashoke and Trivedi, Sandip P.",
    title = "{Rotating attractors}",
    eprint = "hep-th/0606244",
    archivePrefix = "arXiv",
    reportNumber = "TIFR-TH-06-15, HRI-P-06-06-002",
    doi = "10.1088/1126-6708/2006/10/058",
    journal = "JHEP",
    volume = "10",
    pages = "058",
    year = "2006"
}

@article{Bardeen:1999px,
    author = "Bardeen, James M. and Horowitz, Gary T.",
    title = "{The Extreme Kerr throat geometry: A Vacuum analog of AdS(2) x S**2}",
    eprint = "hep-th/9905099",
    archivePrefix = "arXiv",
    reportNumber = "NSF-ITP-99-29",
    doi = "10.1103/PhysRevD.60.104030",
    journal = "Phys. Rev. D",
    volume = "60",
    pages = "104030",
    year = "1999"
}

@article{Dias:2011at,
    author = "Dias, Oscar J. C. and Horowitz, Gary T. and Santos, Jorge E.",
    title = "{Black holes with only one Killing field}",
    eprint = "1105.4167",
    archivePrefix = "arXiv",
    primaryClass = "hep-th",
    doi = "10.1007/JHEP07(2011)115",
    journal = "JHEP",
    volume = "07",
    pages = "115",
    year = "2011"
}

@article{kaluza,
    author = "Kaluza, Th.",
    title = {{Zum Unit\"atsproblem der Physik}},
    eprint = "1803.08616",
    archivePrefix = "arXiv",
    primaryClass = "physics.hist-ph",
    reportNumber = "HUPD-8401",
    doi = "10.1142/S0218271818700017",
    journal = "Sitzungsber. Preuss. Akad. Wiss. Berlin (Math. Phys. )",
    volume = "1921",
    pages = "966--972",
    year = "1921"
}

@article{Klein:1926tv,
    author = "Klein, Oskar",
    editor = "Taylor, J. C.",
    title = "{Quantum Theory and Five-Dimensional Theory of Relativity. (In German and English)}",
    doi = "10.1007/BF01397481",
    journal = "Z. Phys.",
    volume = "37",
    pages = "895--906",
    year = "1926"
}

@article{Horowitz:1997uc,
    author = "Horowitz, Gary T. and Ross, Simon F.",
    title = "{Naked black holes}",
    eprint = "hep-th/9704058",
    archivePrefix = "arXiv",
    reportNumber = "UCSBTH-97-05",
    doi = "10.1103/PhysRevD.56.2180",
    journal = "Phys. Rev. D",
    volume = "56",
    pages = "2180--2187",
    year = "1997"
}

@article{Horowitz:1997ed,
    author = "Horowitz, Gary T. and Ross, Simon F.",
    title = "{Properties of naked black holes}",
    eprint = "hep-th/9709050",
    archivePrefix = "arXiv",
    reportNumber = "UCSBTH-97-20",
    doi = "10.1103/PhysRevD.57.1098",
    journal = "Phys. Rev. D",
    volume = "57",
    pages = "1098--1107",
    year = "1998"
}

@book{chandrasekhar1998mathematical,
  title={The mathematical theory of black holes},
  author={Chandrasekhar, S.},
  isbn={9780198503705},
  lccn={93181092},
  series={International series of monographs on physics},
  year={1998},
  address={New York},
  publisher={Clarendon Press}
}

@book{wald1984general,
  title={General Relativity},
  author={Wald, R.M.},
  isbn={9780226870328},
  lccn={83017969},
  year={1984},
  address={Chicago},
  publisher={University of Chicago Press}
}

@book{heusler_1996, place={Cambridge}, series={Cambridge Lecture Notes in Physics}, title={Black Hole Uniqueness Theorems}, DOI={10.1017/CBO9780511661396}, publisher={Cambridge University Press}, author={Heusler, Markus}, year={1996}, collection={Cambridge Lecture Notes in Physics}}

@article{Carter:1970ea,
    author = "Carter, Brandon",
    title = "{The commutation property of a stationary, axisymmetric system}",
    doi = "10.1007/BF01647092",
    journal = "Commun. Math. Phys.",
    volume = "17",
    pages = "233--238",
    year = "1970"
}

@article{Forgacs:1979zs,
    author = "Forgacs, P. and Manton, N. S.",
    title = "{Space-Time Symmetries in Gauge Theories}",
    reportNumber = "LPTENS 79/3",
    doi = "10.1007/BF01200108",
    journal = "Commun. Math. Phys.",
    volume = "72",
    pages = "15",
    year = "1980"
}

@article{Kundt:1966zz,
    author = "Kundt, Wolfgang and Trumper, M.",
    title = "{Orthogonal decomposition of axi-symmetric stationary spacetimes}",
    doi = "10.1007/BF01325677",
    journal = "Z. Phys.",
    volume = "192",
    pages = "419--422",
    year = "1966"
}

@article{Carter:1969zz,
    author = "Carter, Brandon",
    title = "{Killing horizons and orthogonally transitive groups in space-time}",
    doi = "10.1063/1.1664763",
    journal = "J. Math. Phys.",
    volume = "10",
    pages = "70--81",
    year = "1969"
}

@article{Carter:2009nex,
    author = "Carter, Brandon",
    title = "{Republication of: Black hole equilibrium states}",
    doi = "10.1007/s10714-009-0888-5",
    journal = "Gen. Rel. Grav.",
    volume = "41",
    number = "12",
    pages = "2873--2938",
    year = "2009"
}

@article{Yazadjiev:2010bj,
    author = "Yazadjiev, Stoytcho S.",
    title = "{A Classification (uniqueness) theorem for rotating black holes in 4D Einstein-Maxwell-dilaton theory}",
    eprint = "1009.2442",
    archivePrefix = "arXiv",
    primaryClass = "hep-th",
    doi = "10.1103/PhysRevD.82.124050",
    journal = "Phys. Rev. D",
    volume = "82",
    pages = "124050",
    year = "2010"
}

@article{Carter:1992dbj,
    author = "Carter, Brandon",
    editor = "De Sabbata, V. and Zhang, Zheng-Jiu",
    title = "{Mechanics and equilibrium geometry of black holes, membranes, and strings}",
    eprint = "hep-th/0411259",
    archivePrefix = "arXiv",
    journal = "NATO Sci. Ser. C",
    volume = "364",
    pages = "283--357",
    year = "1992"
}

@article{Heusler:1996ft,
    author = "Heusler, Markus",
    editor = "Straumann, N. and Jetzer, P. and Lavrelashvili, George V.",
    title = "{No hair theorems and black holes with hair}",
    eprint = "gr-qc/9610019",
    archivePrefix = "arXiv",
    reportNumber = "ZU-TH-28-96",
    journal = "Helv. Phys. Acta",
    volume = "69",
    number = "4",
    pages = "501--528",
    year = "1996"
}

@article{Herdeiro:2019oqp,
    author = "Herdeiro, Carlos A. R. and Oliveira, Jo\~ao M. S.",
    title = "{On the inexistence of solitons in Einstein\textendash{}Maxwell-scalar models}",
    eprint = "1902.07721",
    archivePrefix = "arXiv",
    primaryClass = "gr-qc",
    doi = "10.1088/1361-6382/ab1859",
    journal = "Class. Quant. Grav.",
    volume = "36",
    number = "10",
    pages = "105015",
    year = "2019"
}

@article{Simon,
author = {Simon,Walter },
title = {The multipole expansion of stationary {E}instein–{M}axwell fields},
journal = {Journal of Mathematical Physics},
volume = {25},
number = {4},
pages = {1035-1038},
year = {1984},
NOTE = {DOI:10.1063/1.526271},
}

@Article{10.21468/SciPostPhys.15.4.154,
	title={{Gravitational multipoles in general stationary spacetimes}},
	author={Daniel R. Mayerson},
	journal={SciPost Phys.},
	volume={15},
	pages={154},
	year={2023},
	publisher={SciPost},
	doi={10.21468/SciPostPhys.15.4.154},
	url={https://scipost.org/10.21468/SciPostPhys.15.4.154},
}

@article{Garfinkle:1990qj,
    author = "Garfinkle, David and Horowitz, Gary T. and Strominger, Andrew",
    title = "{Charged black holes in string theory}",
    reportNumber = "UCSB-TH-90-66",
    doi = "10.1103/PhysRevD.43.3140",
    journal = "Phys. Rev. D",
    volume = "43",
    pages = "3140",
    year = "1991",
    note = "[Erratum: Phys.Rev.D 45, 3888 (1992)]"
}

@article{Horne:1992zy,
    author = "Horne, James H. and Horowitz, Gary T.",
    title = "{Rotating dilaton black holes}",
    eprint = "hep-th/9203083",
    archivePrefix = "arXiv",
    reportNumber = "UCSBTH-92-11",
    doi = "10.1103/PhysRevD.46.1340",
    journal = "Phys. Rev. D",
    volume = "46",
    pages = "1340--1346",
    year = "1992"
}

@article{Frolov:1987rj,
    author = "Frolov, Valeri P. and Zelnikov, A. I. and Bleyer, U.",
    title = "{Charged Rotating Black Hole From Five-dimensional Point of View}",
    journal = "Annalen Phys.",
    volume = "44",
    pages = "371--377",
    year = "1987"
}

@article{RASHEED1995379,
title = {The rotating dyonic black holes of Kaluza-Klein theory},
journal = {Nuclear Physics B},
volume = {454},
number = {1},
pages = {379-401},
year = {1995},
issn = {0550-3213},
doi = {https://doi.org/10.1016/0550-3213(95)00396-A},
url = {https://www.sciencedirect.com/science/article/pii/055032139500396A},
author = {Dean Rasheed},
}

@article{Casadio:1996sj,
    author = "Casadio, R. and Harms, B. and Leblanc, Y. and Cox, Paul H.",
    title = "{New perturbative solutions of the Kerr-Newman dilatonic black hole field equations}",
    eprint = "hep-th/9606069",
    archivePrefix = "arXiv",
    reportNumber = "UAHEP-958",
    doi = "10.1103/PhysRevD.55.814",
    journal = "Phys. Rev. D",
    volume = "55",
    pages = "814--825",
    year = "1997"
}

@article{Kleihaus:2003df,
    author = "Kleihaus, Burkhard and Kunz, Jutta and Navarro-Lerida, Francisco",
    title = "{Stationary black holes with static and counter rotating horizons}",
    eprint = "gr-qc/0309082",
    archivePrefix = "arXiv",
    doi = "10.1103/PhysRevD.69.081501",
    journal = "Phys. Rev. D",
    volume = "69",
    pages = "081501",
    year = "2004"
}

@article{Kleihaus:2003sh,
    author = "Kleihaus, Burkhard and Kunz, Jutta and Navarro-Lerida, Francisco",
    title = "{Rotating dilaton black holes with hair}",
    eprint = "gr-qc/0306058",
    archivePrefix = "arXiv",
    doi = "10.1103/PhysRevD.69.064028",
    journal = "Phys. Rev. D",
    volume = "69",
    pages = "064028",
    year = "2004"
}

@article{Pacilio:2018gom,
    author = "Pacilio, Costantino",
    title = "{Scalar charge of black holes in Einstein-Maxwell-dilaton theory}",
    eprint = "1806.10238",
    archivePrefix = "arXiv",
    primaryClass = "gr-qc",
    doi = "10.1103/PhysRevD.98.064055",
    journal = "Phys. Rev. D",
    volume = "98",
    number = "6",
    pages = "064055",
    year = "2018"
}

@article{Heusler:1993cj,
    author = "Heusler, Markus and Straumann, Norbert",
    title = "{The First law of black hole physics for a class of nonlinear matter models}",
    reportNumber = "ZU-TH-1-93-REV, MPA-723-REV",
    doi = "10.1088/0264-9381/10/7/008",
    journal = "Class. Quant. Grav.",
    volume = "10",
    pages = "1299--1322",
    year = "1993"
}

@article{Heusler:1993ke,
    author = "Heusler, M. and Straumann, N.",
    title = "{Mass variation formulae for Einstein Yang-Mills Higgs and Einstein dilaton black holes}",
    doi = "10.1016/0370-2693(93)90158-E",
    journal = "Phys. Lett. B",
    volume = "315",
    pages = "55--66",
    year = "1993"
}

@article{PhysRevLett.30.71,
  title = {Mass Formula for Kerr Black Holes},
  author = {Smarr, Larry},
  journal = {Phys. Rev. Lett.},
  volume = {30},
  issue = {2},
  pages = {71--73},
  numpages = {0},
  year = {1973},
  month = {Jan},
  publisher = {American Physical Society},
  doi = {10.1103/PhysRevLett.30.71},
  url = {https://link.aps.org/doi/10.1103/PhysRevLett.30.71}
}

@article{Shiraishi:1992np,
    author = "Shiraishi, Kiyoshi",
    title = "{Spinning a charged dilaton black hole}",
    eprint = "1511.08543",
    archivePrefix = "arXiv",
    primaryClass = "gr-qc",
    reportNumber = "AJC-HEP-6",
    doi = "10.1016/0375-9601(92)90712-U",
    journal = "Phys. Lett. A",
    volume = "166",
    pages = "298--302",
    year = "1992"
}

@article{Elgood:2020svt,
    author = "Elgood, Zachary and Meessen, Patrick and Ort\'\i{}n, Tom\'as",
    title = "{The first law of black hole mechanics in the Einstein-Maxwell theory revisited}",
    eprint = "2006.02792",
    archivePrefix = "arXiv",
    primaryClass = "hep-th",
    reportNumber = "IFT-UAM/CSIC-20-068",
    doi = "10.1007/JHEP09(2020)026",
    journal = "JHEP",
    volume = "09",
    pages = "026",
    year = "2020"
}

@article{Herdeiro_2015,
	doi = {10.1088/0264-9381/32/14/144001},
	url = {https://doi.org/10.1088/0264-9381/32/14/144001},
	year = 2015,
	month = {jun},
	publisher = {{IOP} Publishing},
	volume = {32},
	number = {14},
	pages = {144001},
	author = {Carlos Herdeiro and Eugen Radu},
	title = {Construction and physical properties of Kerr black holes with scalar hair},
	journal = {Classical and Quantum Gravity},
}

@phdthesis{Delgado:2022pwo,
    author = "Delgado, Jorge F. M.",
    title = "{Spinning Black Holes with Scalar Hair and Horizonless Compact Objects within and beyond General Relativity}",
    eprint = "2204.02419",
    archivePrefix = "arXiv",
    primaryClass = "gr-qc",
    school = "Aveiro U.",
    year = "2022"
}

@article{Ortin:2022uxa,
    author = "Ortin, Tomas and Pere\~niguez, David",
    title = "{Magnetic charges and Wald entropy}",
    eprint = "2207.12008",
    archivePrefix = "arXiv",
    primaryClass = "hep-th",
    reportNumber = "IFT-UAM/CSIC-22-40",
    doi = "10.1007/JHEP11(2022)081",
    journal = "JHEP",
    volume = "11",
    pages = "081",
    year = "2022"
}

@article{Hajian:2022lgy,
    author = "Hajian, Kamal and Sheikh-Jabbari, M. M. and Tekin, Bayram",
    title = "{Gauge invariant derivation of zeroth and first laws of black hole thermodynamics}",
    eprint = "2209.00563",
    archivePrefix = "arXiv",
    primaryClass = "hep-th",
    doi = "10.1103/PhysRevD.106.104030",
    journal = "Phys. Rev. D",
    volume = "106",
    number = "10",
    pages = "104030",
    year = "2022"
}

@article{Bardeen:1973gs,
    author = "Bardeen, James M. and Carter, B. and Hawking, S. W.",
    title = "{The Four laws of black hole mechanics}",
    doi = "10.1007/BF01645742",
    journal = "Commun. Math. Phys.",
    volume = "31",
    pages = "161--170",
    year = "1973"
}

@article{Gauntlett:1998fz,
    author = "Gauntlett, Jerome P. and Myers, Robert C. and Townsend, Paul K.",
    title = "{Black holes of D = 5 supergravity}",
    eprint = "hep-th/9810204",
    archivePrefix = "arXiv",
    reportNumber = "QMW-PH-98-38, DAMTP-1998-132, MCGILL-98-29",
    doi = "10.1088/0264-9381/16/1/001",
    journal = "Class. Quant. Grav.",
    volume = "16",
    pages = "1--21",
    year = "1999"
}

@article{Copsey:2005se,
    author = "Copsey, Keith and Horowitz, Gary T.",
    title = "{The Role of dipole charges in black hole thermodynamics}",
    eprint = "hep-th/0505278",
    archivePrefix = "arXiv",
    doi = "10.1103/PhysRevD.73.024015",
    journal = "Phys. Rev. D",
    volume = "73",
    pages = "024015",
    year = "2006"
}

@article{Prabhu:2015vua,
    author = "Prabhu, Kartik",
    title = "{The First Law of Black Hole Mechanics for Fields with Internal Gauge Freedom}",
    eprint = "1511.00388",
    archivePrefix = "arXiv",
    primaryClass = "gr-qc",
    doi = "10.1088/1361-6382/aa536b",
    journal = "Class. Quant. Grav.",
    volume = "34",
    number = "3",
    pages = "035011",
    year = "2017"
}

@book{Frolov:1998wf,
    editor = "Frolov, V. P. and Novikov, I. D.",
    title = "{Black hole physics: Basic concepts and new developments}",
    doi = "10.1007/978-94-011-5139-9",
    year = "1998"
}

@article{SCHONAUER1989279,
title = {Efficient vectorizable PDE solvers},
journal = {Journal of Computational and Applied Mathematics},
volume = {27},
number = {1},
pages = {279-297},
year = {1989},
note = {Special Issue on Parallel Algorithms for Numerical Linear Algebra},
issn = {0377-0427},
doi = {https://doi.org/10.1016/0377-0427(89)90371-3},
url = {https://www.sciencedirect.com/science/article/pii/0377042789903713},
author = {W. Schönauer and R. Wei\ss}
}

@incollection{SCHONAUER1990279,
title = {Efficient vectorizable PDE solvers},
editor = {Henk A. {van der Vorst} and Paul {van Dooren}},
series = {Advances in Parallel Computing},
publisher = {North-Holland},
volume = {1},
pages = {279-297},
year = {1990},
booktitle = {Parallel Algorithms for Numerical Linear Algebra},
issn = {0927-5452},
doi = {https://doi.org/10.1016/B978-0-444-88621-7.50019-0},
url = {https://www.sciencedirect.com/science/article/pii/B9780444886217500190},
author = {W. Schönauer and R. Wei\ss}
}

@article{SCHONAUER2001473,
title = {How We solve PDEs},
journal = {Journal of Computational and Applied Mathematics},
volume = {131},
number = {1},
pages = {473-492},
year = {2001},
issn = {0377-0427},
doi = {https://doi.org/10.1016/S0377-0427(00)00255-7},
url = {https://www.sciencedirect.com/science/article/pii/S0377042700002557},
author = {Willi Schönauer and Torsten Adolph}
}

@book{Ortin:2015hya,
    author = "Ortin, Tomas",
    title = "{Gravity and Strings}",
    edition = "2nd ed.",
    doi = "10.1017/CBO9781139019750",
    isbn = "978-0-521-76813-9, 978-0-521-76813-9, 978-1-316-23579-9",
    publisher = "Cambridge University Press",
    series = "Cambridge Monographs on Mathematical Physics",
    month = "7",
    year = "2015"
}

@article{Duff:1986hr,
    author = "Duff, M. J. and Nilsson, B. E. W. and Pope, C. N.",
    title = "{Kaluza-Klein Supergravity}",
    doi = "10.1016/0370-1573(86)90163-8",
    journal = "Phys. Rept.",
    volume = "130",
    pages = "1--142",
    year = "1986"
}

@article{Overduin:1997sri,
    author = "Overduin, J. M. and Wesson, P. S.",
    title = "{Kaluza-Klein gravity}",
    eprint = "gr-qc/9805018",
    archivePrefix = "arXiv",
    doi = "10.1016/S0370-1573(96)00046-4",
    journal = "Phys. Rept.",
    volume = "283",
    pages = "303--380",
    year = "1997"
}

@article{Newman:1965my,
    author = "Newman, E T. and Couch, R. and Chinnapared, K. and Exton, A. and Prakash, A. and Torrence, R.",
    title = "{Metric of a Rotating, Charged Mass}",
    doi = "10.1063/1.1704351",
    journal = "J. Math. Phys.",
    volume = "6",
    pages = "918--919",
    year = "1965"
}

@article{Gibbons:1987ps,
    author = "Gibbons, G. W. and Maeda, Kei-ichi",
    title = "{Black Holes and Membranes in Higher Dimensional Theories with Dilaton Fields}",
    reportNumber = "UTAP-48-87, LPTENS-87-10",
    doi = "10.1016/0550-3213(88)90006-5",
    journal = "Nucl. Phys. B",
    volume = "298",
    pages = "741--775",
    year = "1988"
}

@article{Ashtekar:2000hw,
    author = "Ashtekar, Abhay and Fairhurst, Stephen and Krishnan, Badri",
    title = "{Isolated horizons: Hamiltonian evolution and the first law}",
    eprint = "gr-qc/0005083",
    archivePrefix = "arXiv",
    doi = "10.1103/PhysRevD.62.104025",
    journal = "Phys. Rev. D",
    volume = "62",
    pages = "104025",
    year = "2000"
}

@phdthesis{Pacilio:2018kdk,
    author = "Pacilio, Costantino",
    title = "{Black holes beyond general relativity: theoretical and phenomenological developments}",
    school = "SISSA, Trieste",
    year = "2018"
}

@article{Compere:2007vx,
    author = "Compere, Geoffrey",
    title = "{Note on the First Law with p-form potentials}",
    eprint = "hep-th/0703004",
    archivePrefix = "arXiv",
    reportNumber = "ULB-TH-07-10",
    doi = "10.1103/PhysRevD.75.124020",
    journal = "Phys. Rev. D",
    volume = "75",
    pages = "124020",
    year = "2007"
}

@article{Rakhmanov:1993yd,
    author = "Rakhmanov, Malik",
    title = "{Dilaton black holes with electric charge}",
    eprint = "hep-th/9310174",
    archivePrefix = "arXiv",
    reportNumber = "CALT-68-1885",
    doi = "10.1103/PhysRevD.50.5155",
    journal = "Phys. Rev. D",
    volume = "50",
    pages = "5155--5163",
    year = "1994"
}

@article{Bokulic:2023oxw,
    author = "Bokuli\'c, Ana and Smoli\'c, Ivica",
    title = "{Generalizations and challenges for the spacetime block-diagonalization}",
    eprint = "2303.00764",
    archivePrefix = "arXiv",
    primaryClass = "gr-qc",
    reportNumber = "ZTF-EP-23-01",
    doi = "10.1088/1361-6382/ace589",
    journal = "Class. Quant. Grav.",
    volume = "40",
    number = "16",
    pages = "165010",
    year = "2023",
    note = "[Erratum: Class.Quant.Grav. 41, 029501 (2024)]"
}

@article{Coleman:1991ku,
    author = "Coleman, Sidney R. and Preskill, John and Wilczek, Frank",
    title = "{Quantum hair on black holes}",
    eprint = "hep-th/9201059",
    archivePrefix = "arXiv",
    reportNumber = "IASSNS-HEP-91-64, CALT-68-1764, HUTP-92-A003",
    doi = "10.1016/0550-3213(92)90008-Y",
    journal = "Nucl. Phys. B",
    volume = "378",
    pages = "175--246",
    year = "1992"
}

@article{Fernandes:2022gde,
    author = "Fernandes, Pedro G. S. and Mulryne, David J.",
    title = "{A new approach and code for spinning black holes in modified gravity}",
    eprint = "2212.07293",
    archivePrefix = "arXiv",
    primaryClass = "gr-qc",
    doi = "10.1088/1361-6382/ace232",
    journal = "Class. Quant. Grav.",
    volume = "40",
    number = "16",
    pages = "165001",
    year = "2023"
}

@article{Galtsov:1995zm,
    author = "Galtsov, D. V. and Garcia, A. A. and Kechkin, O. V.",
    title = "{Symmetries of the stationary Einstein-Maxwell dilaton - axion theory}",
    doi = "10.1063/1.531212",
    journal = "J. Math. Phys.",
    volume = "36",
    pages = "5023--5041",
    year = "1995"
}

@article{Wells:1998gc,
    author = "Wells, Clive G.",
    title = "{Extending the black hole uniqueness theorems. 2. Superstring black holes}",
    eprint = "gr-qc/9808045",
    archivePrefix = "arXiv",
    reportNumber = "DAMTP-98-106, DAMTP-1998-106",
    month = "8",
    year = "1998"
}

@article{Herdeiro:2014goa,
    author = "Herdeiro, Carlos A. R. and Radu, Eugen",
    title = "{Kerr black holes with scalar hair}",
    eprint = "1403.2757",
    archivePrefix = "arXiv",
    primaryClass = "gr-qc",
    doi = "10.1103/PhysRevLett.112.221101",
    journal = "Phys. Rev. Lett.",
    volume = "112",
    pages = "221101",
    year = "2014"
}

@article{Herdeiro:2016tmi,
    author = "Herdeiro, Carlos and Radu, Eugen and R\'unarsson, Helgi",
    title = "{Kerr black holes with Proca hair}",
    eprint = "1603.02687",
    archivePrefix = "arXiv",
    primaryClass = "gr-qc",
    doi = "10.1088/0264-9381/33/15/154001",
    journal = "Class. Quant. Grav.",
    volume = "33",
    number = "15",
    pages = "154001",
    year = "2016"
}

@article{Delgado:2020rev,
    author = "Delgado, Jorge F. M. and Herdeiro, Carlos A. R. and Radu, Eugen",
    title = "{Spinning black holes in shift-symmetric Horndeski theory}",
    eprint = "2002.05012",
    archivePrefix = "arXiv",
    primaryClass = "gr-qc",
    doi = "10.1007/JHEP04(2020)180",
    journal = "JHEP",
    volume = "04",
    pages = "180",
    year = "2020"
}

@article{Herdeiro:2024pmv,
    author = "Herdeiro, Carlos and Radu, Eugen and dos Santos Costa Filho, Etevaldo",
    title = "{Spinning Proca-Higgs balls, stars and hairy black holes}",
    eprint = "2406.03552",
    archivePrefix = "arXiv",
    primaryClass = "gr-qc",
    doi = "10.1088/1475-7516/2024/07/081",
    journal = "JCAP",
    volume = "07",
    pages = "081",
    year = "2024"
}

@article{Herdeiro:2021gbw,
    author = "Herdeiro, Carlos and Radu, Eugen and Uzawa, Kunihito",
    title = "{De-singularizing the extremal GMGHS black hole via higher derivatives corrections}",
    eprint = "2103.00884",
    archivePrefix = "arXiv",
    primaryClass = "hep-th",
    doi = "10.1016/j.physletb.2021.136357",
    journal = "Phys. Lett. B",
    volume = "818",
    pages = "136357",
    year = "2021"
}

@article{Astefanesei:2019qsg,
    author = "Astefanesei, Dumitru and Bl\'azquez-Salcedo, Jose Luis and Herdeiro, Carlos and Radu, Eugen and Sanchis-Gual, Nicolas",
    title = "{Dynamically and thermodynamically stable black holes in Einstein-Maxwell-dilaton gravity}",
    eprint = "1912.02192",
    archivePrefix = "arXiv",
    primaryClass = "gr-qc",
    doi = "10.1007/JHEP07(2020)063",
    journal = "JHEP",
    volume = "07",
    pages = "063",
    year = "2020"
}

@article{Maeda:2018hqu,
    author = "Maeda, Hideki and Martinez, Cristian",
    title = "{Energy conditions in arbitrary dimensions}",
    eprint = "1810.02487",
    archivePrefix = "arXiv",
    primaryClass = "gr-qc",
    doi = "10.1093/ptep/ptaa009",
    journal = "PTEP",
    volume = "2020",
    number = "4",
    pages = "043E02",
    year = "2020"
}

@book{Hawking:1973uf,
    author = "Hawking, Stephen W. and Ellis, George F. R.",
    title = "{The Large Scale Structure of Space-Time}",
    doi = "10.1017/9781009253161",
    isbn = "978-1-009-25316-1, 978-1-009-25315-4, 978-0-521-20016-5, 978-0-521-09906-6, 978-0-511-82630-6, 978-0-521-09906-6",
    publisher = "Cambridge University Press",
    series = "Cambridge Monographs on Mathematical Physics",
    month = "2",
    year = "2023"
}

@article{Herdeiro:2014jaa,
    author = "Herdeiro, Carlos and Radu, Eugen",
    title = "{Ergosurfaces for Kerr black holes with scalar hair}",
    eprint = "1406.1225",
    archivePrefix = "arXiv",
    primaryClass = "gr-qc",
    doi = "10.1103/PhysRevD.89.124018",
    journal = "Phys. Rev. D",
    volume = "89",
    number = "12",
    pages = "124018",
    year = "2014"
}

@article{Chinea:2002jz,
    author = "Chinea, F. J. and Navarro-Lerida, F.",
    title = "{Stationary axisymmetric SU(2) Einstein-Yang-Mills fields with restricted circularity conditions are Abelian}",
    eprint = "gr-qc/0201082",
    archivePrefix = "arXiv",
    doi = "10.1103/PhysRevD.65.064010",
    journal = "Phys. Rev. D",
    volume = "65",
    pages = "064010",
    year = "2002"
}

@article{Ballesteros:2023iqb,
    author = "Ballesteros, Romina and G\'omez-Fayr\'en, Carmen and Ort\'\i{}n, Tom\'as and Zatti, Matteo",
    title = "{On scalar charges and black hole thermodynamics}",
    eprint = "2302.11630",
    archivePrefix = "arXiv",
    primaryClass = "hep-th",
    reportNumber = "IFT-UAM/CSIC-23-018",
    doi = "10.1007/JHEP05(2023)158",
    journal = "JHEP",
    volume = "05",
    pages = "158",
    year = "2023"
}

@article{Ashtekar:1999sn,
    author = "Ashtekar, Abhay and Corichi, Alejandro",
    title = "{Laws governing isolated horizons: Inclusion of dilaton couplings}",
    eprint = "gr-qc/9910068",
    archivePrefix = "arXiv",
    reportNumber = "CGPG-99-10-2",
    doi = "10.1088/0264-9381/17/6/301",
    journal = "Class. Quant. Grav.",
    volume = "17",
    pages = "1317--1332",
    year = "2000"
}

@article{Horowitz:2022mly,
    author = "Horowitz, Gary T. and Kolanowski, Maciej and Santos, Jorge E.",
    title = "{Almost all extremal black holes in AdS are singular}",
    eprint = "2210.02473",
    archivePrefix = "arXiv",
    primaryClass = "hep-th",
    doi = "10.1007/JHEP01(2023)162",
    journal = "JHEP",
    volume = "01",
    pages = "162",
    year = "2023"
}

@article{Horowitz:2024kcx,
    author = "Horowitz, Gary T. and Santos, Jorge E.",
    title = "{Smooth extremal horizons are the exception, not the rule}",
    eprint = "2411.07295",
    archivePrefix = "arXiv",
    primaryClass = "hep-th",
    doi = "10.1007/JHEP02(2025)169",
    journal = "JHEP",
    volume = "02",
    pages = "169",
    year = "2025"
}

@article{Deshpande:2024itz,
    author = "Deshpande, Rhucha and Lunin, Oleg",
    title = "{Multi-charged geometries with cosmological constant}",
    eprint = "2408.08254",
    archivePrefix = "arXiv",
    primaryClass = "hep-th",
    doi = "10.1007/JHEP03(2025)131",
    journal = "JHEP",
    volume = "03",
    pages = "131",
    year = "2025"
}

@article{Deshpande:2024vbn,
    author = "Deshpande, Rhucha and Lunin, Oleg",
    title = "{Rotating Einstein-Maxwell black holes in odd dimensions}",
    eprint = "2411.01795",
    archivePrefix = "arXiv",
    primaryClass = "hep-th",
    doi = "10.1007/JHEP06(2025)066",
    journal = "JHEP",
    volume = "06",
    pages = "066",
    year = "2025"
}

@article{Kunz:2006jd,
    author = "Kunz, Jutta and Maison, Dieter and Navarro-Lerida, Francisco and Viebahn, Jan",
    title = "{Rotating Einstein-Maxwell-dilaton black holes in D dimensions}",
    eprint = "hep-th/0606005",
    archivePrefix = "arXiv",
    doi = "10.1016/j.physletb.2006.06.024",
    journal = "Phys. Lett. B",
    volume = "639",
    pages = "95--100",
    year = "2006"
}

@article{Kleihaus:2016auo,
    author = "Kleihaus, Burkhard and Kunz, Jutta and Radu, Eugen",
    title = "{Charged, rotating black objects in Einstein-Maxwell-dilaton theory in $D\ge 5$}",
    eprint = "1605.05756",
    archivePrefix = "arXiv",
    primaryClass = "gr-qc",
    doi = "10.3390/e18050187",
    journal = "Entropy",
    volume = "18",
    pages = "187",
    year = "2016"
}

@article{Blazquez-Salcedo:2013wka,
    author = "Blazquez-Salcedo, Jose Luis and Kunz, Jutta and Navarro-Lerida, Francisco",
    title = "{Properties of rotating Einstein-Maxwell-Dilaton black holes in odd dimensions}",
    eprint = "1311.0062",
    archivePrefix = "arXiv",
    primaryClass = "gr-qc",
    doi = "10.1103/PhysRevD.89.024038",
    journal = "Phys. Rev. D",
    volume = "89",
    number = "2",
    pages = "024038",
    year = "2014"
}

\end{document}